\documentclass[11pt]{article}

\pdfoutput=1
\usepackage{yfonts}
\usepackage{color}
\usepackage{mhchem}
\usepackage{xcolor}
\usepackage{cite}
\usepackage{sectsty}

\usepackage{environ}
\NewEnviron{myalign}{%
\begin{align}
\scalebox{1.15}{$\BODY$}
\end{align}
}
\NewEnviron{smalign}{%
\begin{align}
\scalebox{0.85}{$\BODY$}
\end{align}
}
\usepackage{yfonts}
\usepackage{color}
\usepackage{mhchem}
\usepackage{xcolor}
\usepackage{mathrsfs}
\usepackage[mathscr]{eucal}
\usepackage{cite}
\usepackage{hyperref}
\hypersetup{colorlinks=true,linkcolor=darkgray,anchorcolor=black,citecolor=red}
\usepackage[toc,page]{appendix}
\usepackage{amsfonts}
\usepackage{bbold}
\usepackage{textcomp}
\usepackage[DIV13]{typearea}
\usepackage{amsmath, amsthm, amssymb, mathtools,empheq,latexsym,dsfont}
\usepackage{bbm}
\usepackage{slashed, simplewick}
\usepackage[utf8]{inputenc}
\usepackage{graphicx,placeins}
\usepackage{makeidx}
\usepackage[font=small,labelfont=bf]{caption}
\usepackage{nicefrac}
\usepackage{subfigure}
\usepackage{array, bigdelim,multirow,multicol}
\usepackage[integrals]{wasysym}
\usepackage{ulem}
\usepackage{fancybox}
\usepackage{bm}
\usepackage{float}
\usepackage{rotating}
\usepackage{colortbl}
\usepackage{booktabs}
\usepackage[top=2cm,textwidth=16.6cm,textheight=22.75cm]{geometry}
\usepackage{doi}
\graphicspath{{immagini/}}

\def\otau{{\tau_0}}
\def\otaus{{\tau_0}^*}
\def\ogamma{{\gamma_0}}

\definecolor{Gray}{gray}{0.92}
\newcommand{\diag}{\mathtt{diag}}

\newcommand{\nc}{\newcommand}
\nc{\Dfb}{\mbox{$\raisebox{2mm}{\boldmath ${}^\leftrightarrow$}\hspace{-4mm} D$}}

\def\dd{\displaystyle}
\nc{\btb}{\begin{tabular}}    \nc{\etb}{\end{tabular}}

\renewcommand{\arg}{{\rm Arg}}

\newcommand{\be}{\begin{equation}}
\newcommand{\ee}{\end{equation}}
\newcommand{\bea}{\begin{eqnarray}}
\newcommand{\eea}{\end{eqnarray}}

\makeatletter
\renewcommand*{\@fnsymbol}[1]{\ensuremath{\ifcase#1\or *\or  \mathsection\or \ddagger\or
\dagger\or \mathparagraph\or \|\or **\or \dagger\dagger
\or \ddagger\ddagger \else\@ctrerr\fi}}
\makeatother

\makeatletter
\@addtoreset{equation}{section}
\makeatother

\DeclareMathOperator{\im}{Im}

\begin{document}
 \unitlength = 1mm

\setlength{\extrarowheight}{0.2 cm}

\title{
\begin{flushright}
\begin{minipage}{0.2\linewidth}
\normalsize
\end{minipage}
\end{flushright}
 {\Large\bf Fermion masses, critical behavior and universality}\\[1.3cm]}
\date{}

\author{
Ferruccio~Feruglio$^{1}$
\thanks{E-mail: {\tt feruglio@pd.infn.it}}
\\*[20pt]
\centerline{
\begin{minipage}{\linewidth}
\begin{center}
$^1${\small
INFN, Sezione di Padova, Via Marzolo~8, I-35131 Padua, Italy}\\*[10pt]
\end{center}
\end{minipage}}
\\[10mm]}
\maketitle
\thispagestyle{empty}

\centerline{\large\bf Abstract}
\begin{quote}
\indent
We look for signals of critical behavior in the Yukawa sector. By reviewing a set of models for the fermion masses,
we select those where a symmetry-breaking order parameter sits at a transition point between 
a disordered phase and an ordered one. Many models based on ordinary flavor symmetries are formulated 
in terms of small corrections to a symmetric limit, which can hardly be interpreted unambiguously as a sign of near-criticality. Different is the case of nonlinearly realized flavor symmetries
when the system is always in the broken phase. 
By inspecting a large number of modular and CP invariant models of lepton masses, 
we find that most of them cluster around the fixed point $\tau= i$,
where the system enjoys enhanced symmetry. Since a priori all values of the modulus $\tau$ are equally
acceptable to describe the fermion spectrum,
we regard this preference as a hint of near-criticality.
We analyze in detail these models in the vicinity
of all fixed points, showing that only one possibility provides a good description of neutrino masses and mixing angles.
Near the fixed points the models exhibit a universal behavior.
Mass ratios and mixing angles scale
with appropriate powers of the order parameter, independently of the details of the theory, a feature reminiscent of systems belonging
to the same universality class in second-order phase transitions. 
The observations of this work are inspired 
by the role near-criticality might play in solving the naturalness problem and 
are motivated by the fascinating possibility that most of the free parameters of the Standard Model
could find a common explanation.
\end{quote}

\newpage
\tableofcontents

\newpage
\section{Introduction}
The American photographer Wilson Bentley was fascinated by the beauty of snow crystals, their regularity, symmetry and elegance. Over forty-six years, he captured more than 5,000 snow crystal images. In his book ``{\it Snow Crystals}'' (1931) he noticed that no two snowflakes are perfectly identical, though they share a common pattern: a distinctive six-fold symmetry. This feature had long since caught the attention of scientists. Already in 1611, Johannes Kepler published the monograph "The Six-Cornered Snowflake" where he tried to explain how
the complex symmetrical structure of ice crystals could emerge out of the air. Today we know that the formation of snow crystals is a complex out-of-equilibrium phenomenon, taking place during the liquid-solid phase transition when the fluid is kept at a temperature slightly below its freezing point~\cite{libbrecht}. The hexagonal symmetry is due to the geometry of the initial ice crystals. During its growth in the out-of-equilibrium phase, the snowflake is exposed to sharply fluctuating conditions. 
Thermal variations and collisions with dust particles in the air generate random nucleation sites. The local temperature and humidity affect the rate of growth of the snowflake making it highly unlikely that two identical snowflakes will form. 
While the ultimate shape of a snowflake is not predictable, the universal feature of all ice crystals resides in their
common origin during the critical transition, which produces a hexagonal seed.

We seem to face a similar situation in particle physics today. While the precise value of most of the parameters
of the Standard Model (SM) escape our comprehension, there are indications that they might reflect a critical behavior of the system. 
Hints of criticality came from the discovery of the Higgs. Thanks to the accurate knowledge of the top mass and the strong coupling constant,
it has been realized that the SM electroweak vacuum, in the absence of new physics below the Planck scale $M_P$, lies very close to the boundary between stability and metastability. The Higgs quartic coupling becomes negative at very high energies, the precise value of the transition point depending on the values of the top and Higgs 
masses and of the strong coupling constant. Moreover, the Higgs quartic coupling remains small in a very wide
range of high energies~\cite{Isidori:2001bm,Froggatt:2001pa,Elias-Miro:2011sqh,Degrassi:2012ry,Masina:2012tz,Buttazzo:2013uya}. 

Another aspect of the Higgs potential is the extreme smallness of the Higgs VEV in Planck units. This is directly related to the hierarchy problem. In the SM the Higgs VEV measures the separation between the symmetric and the spontaneously broken phase. If we believe that a field theory description should remain valid at energy scales much larger than the electroweak scale, perhaps even up to the Planck scale $M_P$, we have to explain the appearance of a tiny dimensionless parameter given by the Higgs VEV evaluated in units of $M_P$. 
The Higgs quadratic coupling appears to be tuned to set the SM near the phase transition \cite{Giudice:2006sn,Giudice:2008bi,Wetterich:2011aa,Giudice:2017pzm,Craig:2022uua}. Why is the system near criticality? These hints of near-criticality are somehow similar to that of gauge coupling unification: they can be mere accidents or indications of some fundamental aspect of Nature.

Also in cosmology we can find hints of near-criticality. The smallness of the cosmological constant sets the universe
at the border between an expanding phase and a collapsing one. At a different level,
the evolution of the early universe can be interpreted in terms of a small deviation from a de Sitter geometry
that, if realized exactly, would provide an indefinite exponential expansion. 
At the same time, the closeness to an exact symmetrical phase reveals helpful in classifying models
of inflation according to their universal properties.
Many models can be shown to belong to few universality classes, characterized by a small number of parameters \cite{Ketov:2012jt,Kehagias:2013mya,Roest:2013fha,Binetruy:2014zya,Ketov:2019toi}.

Among the SM parameters that still lack explanation, those describing the fermion spectrum
are especially intriguing. To describe all observable quantities we need up to 22 independent parameters,
to be compared with the two characterizing the pure Higgs sector in the SM, at the lowest order.
While several regularities and approximate empirical relations among them have long been remarked,
strongly suggesting the existence of an organizing principle, no convincing fundamental rationale
for the flavor puzzle has been established so far~\cite{Feruglio:2015jfa}. Perhaps we will never be able to accurately
predict fermion masses, mixing angles and CP-violating phases in terms of a small number of
fundamental parameters, much as we are unable to predict the exact shape of a single snow crystal.
Nevertheless, some universal features of the spectrum, having 
its origin during a critical transition, might still shed some light on this fascinating puzzle.

The aim of this work is to start exploring the relevance of critical phenomena in flavor physics.
A phase transition occurs when a control parameter is tuned and, crossing a threshold, we observe a change in the organization of a system. Typical control parameters in statistical mechanics are the temperature and pressure of the system or some external factor, such as a magnetic field in a ferromagnetic material, that can be varied by the observer. In particle physics
we can vary the energy scale or, in a gedanken experiment, the coupling constants and the background fields of the theory under examination. The change in the system organization typically, but not exclusively, consists of a transition between two different symmetry patterns distinguished by the value of an order parameter. An added value of critical phenomena is the universal behavior of seemingly different physical systems in the vicinity of a critical point, allowing for a classification independent of the specific microscopic realization.

Establishing whether the Yukawa sector is close to a phase transition in the above sense is an impossible task today, 
and the purpose of this note is much more modest. 
Indeed, unlike in the Higgs system, there is no baseline theory of fermion masses and mixing angles. 
On the contrary, myriads of models capture some features of the problem. 
Most of them are based on flavor symmetries and, in a simple-minded approach, we
will tentatively require for near-criticality the closeness of the vacuum preferred by the data
to the transition point from a symmetric phase to a non-symmetric one. Already this viewpoint
presents several difficulties, that we comment on more extensively  in Section~\ref{cri}.
Any realistic candidate flavor symmetry cannot be exact~
\cite{Feruglio:2019ybq}, and near-criticality might be misinterpreted as synonymous with approximate symmetry, 
broken by small order parameters gauging the distance of the system from the critical point. 
Though this is a common feature of many models
of fermion masses, we will see that there are important classes of models that do not possess this property, a priori.
To this purpose, a survey of the main types of models will be given in Section~\ref{survey}. While in general
the requirement of approximate symmetry reveals not sufficient to identify a critical behavior, in specific contexts it is highly suggestive of the proximity to a phase transition.

In particular, in an interesting class of models the flavor symmetry is nonlinearly realized.
For example, this happens with modular invariance and its generalizations, playing a key role in string theory compactifications~\cite{Giveon:1994fu}.
While linear flavor symmetries take advantage of the existence of a special vacuum configuration
where the symmetry is unbroken, by definition such a configuration simply does not exist in the nonlinear case.
The symmetry is always in the broken phase and a priori there is no lamppost helping the search for a realistic
vacuum state. Everywhere in the space
of vacua the flavor symmetry is completely broken but in loci of zero measure, where a residual symmetry group
possibly survives. All vacua are equally viable candidates to describe the fermion spectrum, without 
prejudice against or in favor of any of them. 
By comparing the theory with the data, it might happen that the vacuum maximizing the agreement between the observed fermion spectrum and the theory falls close to one of the loci enjoying a residual symmetry.
We interpret this case as a significant indication for the vicinity of the system to a phase transition.

Having in mind this viewpoint, in Sections~\ref{modularsec},~\ref{micp} and~\ref{lmm}, we will analyze the full set of models of lepton masses based on modular invariance~\cite{Feruglio:2017spp}. 
Hundred of such models have been proposed in recent years \cite{petcov},
differing in the detailed implementation: choice of the level $N$ of the finite modular group, of the modular weights and representations of the matter fields. Additional important freedom is related to the kinetic terms of the theory,
often chosen to be minimal and flavor-universal in existing constructions, but allowed to evade this assumption in the general case~\cite{Chen:2019ewa}. 

In all these models the vacuum is parametrized by a complex (VEV of a) field $\tau$,
the modulus, living in the upper half of the complex plane. Anywhere in this domain modular invariance is fully broken, but at three inequivalent fixed points, where a finite residual symmetry is preserved. In a bottom-up approach no region of the moduli space is favored a priori and $\tau$ is treated as a free parameter, varied to
maximize the agreement between data and theory. Depending on $N$, on the modular weights, and the specific kinetic terms, any point of the moduli space might provide the best fit to the data.
It is remarkable that, in a large class of CP and modular invariant models, data exhibit a significant preference 
for $\tau$ near the fixed point $\tau=i$, where the theory is invariant under the nonlinear transformation $\tau\to -1/\tau$. 
We consider this preference as a hint of near-criticality for the Yukawa sector. We also show that the behavior of the system in the vicinity of 
the fixed points $\tau=i$ and $\tau=-1/2+i~ \sqrt{3}/2 $ is independent of the level $N$, of the weights of the matter
fields and of the assumed form of the kinetic terms of the theory. We prove that all models 
having lepton doublets in an irreducible representation of the relevant finite modular group fall in the same universality class of which we identify the main properties, extending and completing the discussion of ref. \cite{Feruglio:2022kea}.

For models manifesting near-criticality, we look for a classification based on universal features. 
We focus on aspects that are independent, as much as possible, of the details of the model like
unknown order-one Lagrangian parameters or the transformation properties of 
matter fields. Natural candidates are the scaling laws of physical quantities with respect to variations of
the order parameters. 
We will see that in many interesting models the order parameters fall in different representations
of the relevant symmetry group, much as in the case of continuous phase transitions between crystals \cite{landau,birman}. Due to the large variety of possible combinations, we will not attempt a general classification
of flavor models, but we will select a sample of models large enough to illustrate this point. 

In Section \ref{road} we briefly mention 
few mechanisms suggested in the literature, able to drive the system close to criticality during the cosmological evolution.
In a final Section we draw our conclusions.
\section{Critical behavior in flavor physics}
\label{cri}
\begin{figure}[h!]
\centering
\includegraphics[width=0.5\linewidth]{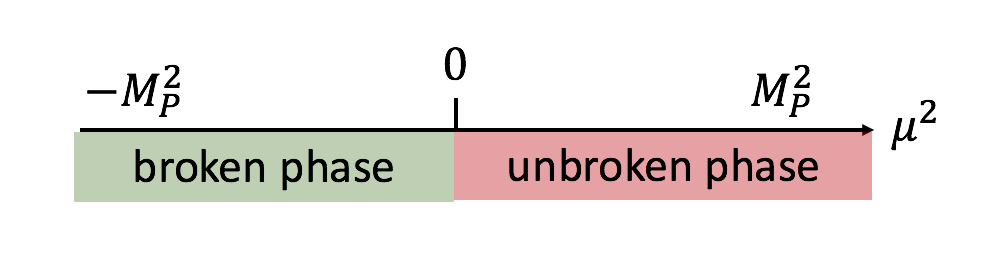}~~~~~~~~~~~~~~~~~~~~~~~~~~~~~~~~
\includegraphics[width=0.25\linewidth]{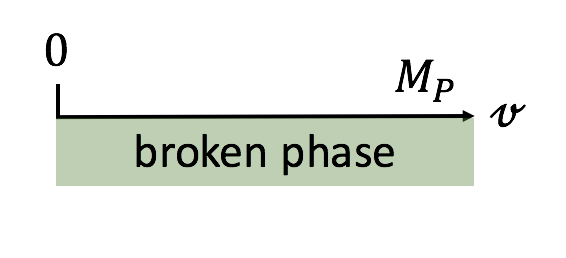}
\caption{Left: electroweak phase as a function of the control parameter $\mu^2$. Right: order parameter of the electroweak phase transition.}
\label{higgs1}
\end{figure}

To investigate the critical properties of the fermion sector, it is useful to recall how this concept applies to the hierarchy
problem. When $\mu^2$, the quadratic parameter of the Higgs sector here identified with the relevant control parameter, is positive(negative), the electroweak symmetry is unbroken(broken). Adopting the Planck mass as a fundamental unit, $\mu^2$ is expected to be of order one, while
the observed value of $\mu^2$ is very close to the critical
point separating the unbroken electroweak phase from the broken one (see Fig. \ref{higgs1}, left panel).
The hierarchy problem is translated into the question: why $\mu^2$ is nearly critical~\cite{Giudice:2006sn,Giudice:2008bi}?
Slightly varying $\mu^2$ around this value produces a phase transition. 
\vskip 0.2 cm
\noindent
We cannot hope to identify a similar picture in the Yukawa sector. 
First of all, it would be practically impossible to determine for each model the set of control parameters playing the role of $\mu^2$. 
We can overcome this difficulty by shifting our focus from the control parameters to the 
order parameter. In the Higgs system, this would mean ignoring the 
parameter $\mu^2$ and formulating near-criticality as the closeness to zero of the electroweak
VEV $v$ evaluated in units of $M_P$ (see Fig. \ref{higgs1}, right panel). 

Neglecting control parameters is not a harmless procedure for the purpose we have in mind. Near-criticality requires the control parameter $\xi$ to lie close to a critical value $\xi_c$ separating two phases. 
An order parameter $u$ is designed to vanish when $\xi>\xi_c$ and to be different from zero when $\xi<\xi_c$.
To infer the closeness to criticality from the value of $u$, we need to know the
functional dependence $u(\xi)$. When the phase transition is of the second order, $u(\xi)$ is continuous at $\xi=\xi_c$
and a hint of criticality can be deduced from the smallness of $u$. When the transition is of the first order,
$u(\xi)$ jumps discontinuously from zero to a non-vanishing value and its value does not necessarily provide a clue
of near-criticality. For example, in the liquid-gas first-order transition we cannot establish how close we are to the border separating the two phases by only measuring the density of the system.
In the isotropic-nematic phase transition of uniaxial liquid crystals, the order parameter $u$
can be estimated by minimizing the function:
\begin{myalign}
V=\dd\frac{\xi}{2} u^2-\dd\frac{1}{3}u^3+\dd\frac{\lambda}{4}u^4~~~~~~~~~~~~~~~~~~(\lambda>0)~~~,
\end{myalign}  
where we have conveniently rescaled all the variables. For $\xi>1/(4\lambda)$ there is a single minimum
at $u=0$, and a second minimum appears at $u_+=(1+\sqrt{1-4\lambda\xi})/(2\lambda)$ when $\xi\le 1/(4\lambda)$. We have $V(0)=V(u_+)$ at the
critical value $\xi_c=2/(9\lambda)$. The presence of two separate degenerate minima signals instability and near
$\xi=\xi_c$ a first-order phase transition drives $u=0$ to $u=2/(3\lambda)$. A small $u$ is not necessarily related to
$\xi\approx \xi_c$ and we cannot use the order parameter to recognize near-criticality. While first-order phase
transitions are of interest and possibly also relevant to flavor physics, by tracking the order parameter alone we are
led to exclude them from our analysis. Second-order phase transitions map a disordered phase
($u=0$) to an ordered one ($u\ne 0$). The order parameter reflects the symmetry content
of the system, higher in the disordered phase and lower in the ordered one.

We are led to assume that possible phase transitions occurring in the Yukawa sector involve flavor symmetries. Here we face another obstacle since, contrary to the well-established electroweak symmetry, there is no evidence for a fundamental flavor symmetry. Even assuming such a symmetry exists, there is no consensus about the choice of the symmetry group. 
Thus we can only proceed by inspecting some realistic classes of models. An empirical property that applies to any models built along these lines is the fact that the flavor symmetry is realized in the broken phase,
thus requiring a symmetry-breaking sector. In the fermion sector we can look for models where the order parameters of flavor symmetry breaking are also close to zero, suggesting near-criticality of the system~\footnote{Strictly speaking, when the relevant symmetry is a local one, the VEV of a scalar field with nontrivial transformation properties (like the Higgs field in the electroweak theory) is not a true order parameter~\cite{Elitzur:1975im,Frohlich:1981yi}.
Nevertheless, at least in perturbation theory, this quantity evaluated in a fixed gauge is useful to classify the phases of the system~\cite{Olynyk:1984pz,Beekman:2019pmi}. Here we will adopt this simple-minded viewpoint.
}.

An objection comes immediately to mind. Many models of fermion masses are built as small deviations from a symmetry limit. A flavor symmetry broken by order-one relative effects seems to be completely useless as well as out of control.
For this reason, near-criticality seems to be present in models based on flavor symmetries by construction,
and not as a possible prominent property. In the next Section we see that this is not necessarily the case.
Indeed, while models of quark masses and mixing angles generally reflect a symmetric pattern in first
approximation, entire classes of models for neutrino masses and lepton mixing angles do not lie close to a nontrivial symmetric
limit. Moreover, when flavor symmetries are nonlinearly realized the choice of the vacuum is not as straightforward as in the linearly
realized case. There is no origin in moduli space where the symmetry is unbroken. On the contrary, at every point in moduli space the symmetry is in the broken phase. There can be loci of zero measure enjoying residual symmetries and the identification of a symmetry-breaking order parameter is less obvious. 

\begin{figure}[h!]
\centering
\includegraphics[width=0.5\linewidth]{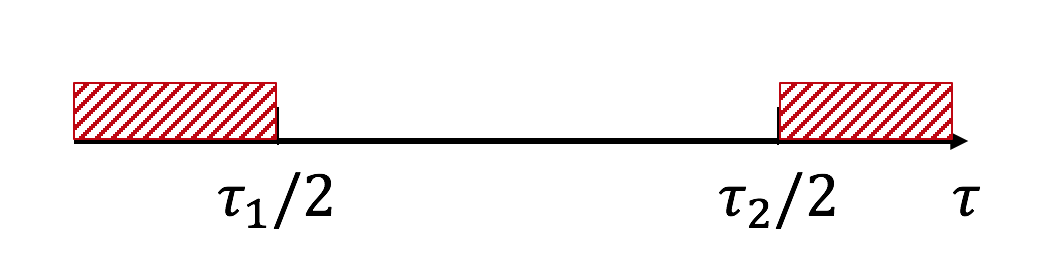}
\caption{Independent vacua of the theory invariant under $G$ can be restricted between $\tau_1/2$ and $\tau_2/2$.}
\label{toyfun}
\end{figure}

As a toy example, consider a theory whose vacuum is described by a real scalar field
$\tau$ and the flavor group $G$ is generated by the two parity symmetries $\mathbb{Z_2}$ and $\mathbb{Z_2}'$ acting on $\tau$ as:
\begin{myalign}
\label{zz}
\begin{array}{rl}
\tau &\xrightarrow{\mathbb{Z_2}} \tau_1-\tau\\
\tau &\xrightarrow{\mathbb{Z_2}'} \tau_2-\tau~~~,
\end{array}
\end{myalign} 
where $\tau_{1,2}$ are of order one in units of the fundamental scale of the theory and $\tau_1<\tau_2$. We assume
the symmetry is gauged, so that the vacua described by $\tau$ and $\tau_{1,2}-\tau$ are indistinguishable. 
The full gauge symmetry $G$ is nonlinearly realized (as usual, by linear we mean linear and homogeneous). We can always find a new field variable
such that either $\mathbb{Z_2}$ or $\mathbb{Z_2}'$ is linearly realized. For instance $\mathbb{Z_2}$ acts on $u=\tau-\tau_1/2$ as $u\to -u$. However, no coordinate choice exists for which the whole action of $G$ is linear. 
The absolute value of $\tau$ has no physical meaning and, by exploiting the gauge invariance under $G$, we can restrict $\tau$ to the region between $\tau_1/2$ and $\tau_2/2$, see fig. \ref{toyfun}. Arbitrarily large values of $\tau$ can all be mapped here.
\vskip 0.2 cm
\noindent
If we include a generic fermion sector and we require invariance under $G$, we end up with $\tau$-dependent Yukawa couplings.
Given a dataset of fermion masses and mixing angles, the choice of $\tau$ offering the best fit is not obvious.
If the vacuum of the theory lies close to $\tau_1/2$, the parity $\mathbb{Z_2}$ is an approximate symmetry of the system. However, the order parameter for the $\mathbb{Z_2}$ symmetry breaking
is not $\tau$, but the deviation of $\tau$ from $\tau_1/2$.

Nonlinearly realized symmetries can play an important role in the solution of the flavor puzzle. In realistic string theory compactifications, the moduli space is shaped by discrete nonlinear transformations of a group $G$, removing an intrinsic redundancy in the description of the vacua and representing gauge symmetries that
all sectors of the theory, including the flavor one, are bound to respect. 
The vacuum of the theory is described by a moduli space $\mathcal{M}$. 
The independent vacua are parametrized by scalar fields $\tau$ and Yukawa couplings are functions of $\tau$, restricted by the requirement of invariance of the theory under $G$~\cite{Hamidi:1986vh,Dixon:1986qv,Lauer:1989ax,Lauer:1990tm,
Kobayashi:2018rad,Kobayashi:2018bff,
Baur:2019kwi,
Baur:2019iai,
Kobayashi:2019uyt,Kikuchi:2020frp,
Nilles:2020tdp,
Kikuchi:2020nxn,
Baur:2020jwc,Ishiguro:2020nuf,Nilles:2020gvu,Baur:2020yjl,Ding:2021iqp,Nilles:2021glx,Ding:2020zxw,Ishiguro:2021ccl,Baur:2022hma}.
The $\tau$ vacuum expectation values (VEVs) do not directly correspond to the order parameters we are interested in.
Unlike the case of linearly realized symmetries, the size of these VEVs has no absolute meaning since it does not remain unchanged under a nonlinear transformation~\footnote{Under a linear transformation
a scalar multiplet $\tau$ undergoes a unitary transformation $\tau\to U(g)\tau$ and its size $||\tau||$ is unchanged.}.
In general there is no point in field space, like the origin for linearly realized symmetries, left invariant by the
whole symmetry group $G$. 
At the generic point of the moduli space the gauge flavor symmetry $G$ is completely broken, except
at special points $\tau_0$, where a subgroup $H$ of $G$ is preserved. The value of $\tau_0$ can be of order one in units of the fundamental scale, but the field $\tau$ is not a good
order parameter for the breaking of $G$ into $H$ and appropriate new coordinates should be defined to characterize
the breaking. In Section~\ref{modularsec} we show how to define such coordinates and characterize near-criticality in the general case.
\section{Model survey}
\label{survey}
Are hints of near-criticality also suggested by the observed pattern of fermion masses? 
In this Section, by reviewing several models of fermion masses, we start bringing to light possible indications of
critical behavior in this new context.
We focus on second-order phase transitions, mapping a disordered phase to an ordered one.
A key element of our analysis is the moduli space ${\cal M}$, whose elements $\tau$ parametrize the 
vacua of the theory. For definiteness, we can view $\tau$ as (VEVs of) a set of dimensionless and gauge-invariant scalar fields. Canonical dimensions can be recovered by rescaling the fields $\tau$ by an appropriate mass parameter. The requirement of
gauge invariance of $\tau$ is more restrictive and can be relaxed in a more general framework. 
The fields $\tau$ transform non-trivially under a flavor symmetry group $G$ and allow to define 
an order parameters $u(\tau)$ for the breaking of $G$ or of one of its subgroups.
The group $G$ is completely general, covering both the case of traditional flavor symmetries, where $\tau$ are assigned to linear representations of $G$, and modular symmetries, where $G$ is a discrete gauge symmetry whose action on $\tau$ is nonlinear. We allow for both global and local groups $G$. The corresponding systems are physically distinguishable but, for the sole purpose of studying the properties of the fermion mass spectrum, 
we can treat them on an equal footing. Acting with $G$ on a given element $\tau$ of ${\cal M}$ we obtain an orbit.
In this class of theories points on the same orbit identify the same vacuum and the inequivalent vacua are described by the domain ${\cal M}/G$ of ${\cal M}$~\footnote{Here we identify the moduli space with ${\cal M}$,
whereas in the string and mathematical literature, the moduli space describes the inequivalent vacua and is represented by the quotient ${\cal M}/G$.}.

In this setup, fermion mass matrices $m_{ij}(\lambda;\tau)$ depend on both the vacuum parameters $\tau$ and a set of Lagrangian parameters $\lambda$.
There can be additional discrete parameters describing the transformation properties of the fields under $G$,
like charges or representations. The phase of the system is meant to be completely specified in terms of $\tau$. 
At this stage we are not interested in the dynamics driving the system to a given phase.
Thus we omit from $\lambda$ the control parameters that determine the vacuum $\tau$ itself, like those occurring in the scalar potential.
In general, in most of the space ${\cal M}$ spanned by the fields $\tau$, the symmetry is completely broken, but there can be "critical" points $\tau_0$ where some nontrivial subgroup $H$ of $G$ is preserved. In fact, along an orbit of the group
passing through $\tau_0$ we have the same residual symmetry and $\tau_0$ stands merely for a representative point of the orbit. We start by adopting the following tentative necessary condition for near-criticality.

\begin{quote}
{\it For near-criticality to occur, the value of $\tau$ reproducing the observed pattern of masses and mixing angles should lie close to one critical region. If the little group of the region is $H$, the transition involved is between the $H$-symmetric and the ordered phase.} 
\end{quote}
We stress the "kinematical" character of this definition. There is no
reference to the dynamics leading the system to approach the critical point $\tau_0$.

For many flavor models, with a linearly realized symmetry, the above condition is easily met. The values of $\tau$
reproducing correctly the data are typically smaller or even much smaller than one,
and the point $\tau_0=0$, where the symmetry $G$ is unbroken, often defines a reasonable first approximation 
of the observed masses and mixing angles. In other words, the data are reproduced by small perturbations
around the symmetric point $\tau_0=0$. This is the case of a vast class of models describing the quark sector.
For this reason, it is useful to inspect more closely some examples, to identify features that can
better characterize a near-critical system.

The simplest class of models is that relying on a continuous abelian Froggatt-Nielsen (FN) flavor symmetry~\cite{Froggatt:1978nt}: $G=U(1)_F$.
The most economic symmetry-breaking sector consists of a complex scalar field $\tau$ carrying, conventionally,
a negative unit of the FN charge $F$ and consistent with the data when $|\tau|<1$. Hence this system is close to the
critical point $\tau_0=0$, where $U(1)_F$ is unbroken. The quarks are often assigned non-negative charges $F(X)$, $(X=q,u^c,d^c)$. Quark electroweak doublets $q_i$ have charges that can be ordered as $F(q_1)\ge F(q_2)\ge F(q_3)\ge 0$.
We get the approximate predictions:
\begin{myalign}
\left\vert(V_{CKM})_{ij}\right\vert\approx\dd\frac{|\tau|^{F(q_i)}}{|\tau|^{F(q_j)}}\le 1~~~~~~~~~~~~(i\le j)~~~,
\end{myalign} 
which imply
\begin{myalign}
V_{ud}\approx V_{cs}\approx V_{tb}\approx 1~~~,~~~~~~~~~V_{ub}\approx V_{us}\times V_{cb}~~~.
\end{myalign} 
This prediction is confirmed at the level of the order of magnitudes and is completely independent of the specific
choice of the charges $F(X)\ge 0$. We interpret this as a universal property of this class of models.
Other features, such as the quark mass ratios or the individual elements of the mixing matrix, depend on
the actual value of the FN charges. In this example we have a single order parameter that can be chosen, for example, as
$u=|\tau|$ or $u=|\tau|^2$. The domain ${\cal M}/G$ can be represented by a straight line where $\tau$ is real and non-negative.
The disordered phase at $\tau_0=0$ can be approached along a single direction.

When ${\cal M}$ is spanned by fields $\tau$ transforming in a reducible representation of $G$, 
it is more difficult to identify the universal features of the mass spectrum since there are many elements that
compete in establishing the final result. 
To correctly reproduce the data, in general it is not sufficient to break the flavor group by a generic choice of $\tau$. The size and the orientation of $\tau$ in flavor space should be carefully chosen to achieve a realistic pattern of masses and mixing angles. As a result, higher predictability often comes at the expense of a complicated symmetry-breaking sector. We look for properties of the system that follow directly from the symmetry-breaking pattern and are as much as possible independent of details of the model such as the Lagrangian parameters $\lambda$, 
at least in a convenient portion of the parameter space.

To illustrate this case,
an instructive example is the supersymmetric version of the model of quark masses in ref. \cite{Dudas:2013pja}, (see also ref. \cite{Falkowski:2015zwa,Linster:2018avp}) where $G=U(2)$. We collect chiral multiplets and representations
in table 1.

\begin{myalign}
\nonumber
\begin{array}{c|cccccc|cc}
\hline
&q_a&u^c_a&d^c_a&q_3&u^c_3&d^c_3&\tau_a&\tau_3\\
\hline
SU(2)_F&2&2&2&1&1&1&2&1\\
\hline
U(1)_F&+1&+1&+1&0&0&+1&-1&-1\\
\hline
\end{array}
\end{myalign}
\vskip 0.2 cm
\begin{center}{{\bf Table 1} Multiplets and transformation properties of the supersymmetric model in ref. \cite{Dudas:2013pja,Falkowski:2015zwa,Linster:2018avp}, $(a=1,2)$.}
\end{center}
\vskip 0.5 cm
The space ${\cal M}$ is spanned by the three complex fields $(\tau_1,\tau_2,\tau_3)$ and consists of several layers,
in each of which $G$ is broken down to a specific subgroup. We can view each layer as a union of orbits with 
isomorphic residual symmetries. Once the parameters $\lambda$ are fixed, points lying on the same orbit 
describe the same physics, and the variables $(\tau_1,\tau_2,\tau_3)$ provide a redundant parametrization
of the quantities we are interested in. For this reason, it is convenient to project ${\cal M}$ to the region ${\cal M}/G$ where each orbit is described by a single point. By a $U(2)$ transformation it is always possible to reach points in ${\cal M}$ of the type $(\tau_1,\tau_2,\tau_3)=
(\tau_1,0,\tau_3)$, where $(\tau_1,\tau_3)$ are real and non-negative. At the same time, this region cannot be reduced any more,
since two distinct points $(\tau_1,0,\tau_3)$ and $(\tau_1',0,\tau_3')$ (with nonnegative entries) cannot be related by a 
$U(2)$ transformation. Thus, the most general vacuum of the system depends on the two 
real non-negative parameters $(\tau_1,\tau_3)$~\footnote{As an alternative description, we can move to the orbit space ${\cal M}_I$, spanned by the independent invariant polynomial built out of $\tau$, in this case $I_1=|\tau_1|^2+|\tau_2|^2\ge 0$ and $I_2=|\tau_3|^2\ge 0$. In the interior of ${\cal M}_I$, $U(2)$ is completely broken. Non-trivial residual symmetries are achieved at the boundary of ${\cal M}_I$: $SU(2)$ is unbroken along $I_1=0$, while a subgroup $U(1)'\ne U(1)$ is preserved at $I_2=0$. Where the two previous boundaries meet, at $I_1=I_2=0$, $U(2)$ is unbroken.
We can view $\tau_1\ge 0$ and $\tau_3\ge 0$ as representative of $I_1$ and $I_2$, respectively.}.
In terms of $(\tau_1,\tau_3)$ the Yukawa couplings read~\footnote{In the $(\bar L R)$ convention.}:

\begin{myalign}
Y_u=
\left(
\begin{array}{ccc}
0&u_{12}\tau_3^{2}&0\\
-u_{12}\tau_3^{2}&u_{22}\tau_1^2 &u_{23}\tau_1\\
0&u_{32}\tau_1&u_{33}
\end{array}
\right)~~~,
\end{myalign}
\begin{myalign}
Y_d=
\left(
\begin{array}{ccc}
0&d_{12}\tau_3^{2}&0\\
-d_{12}\tau_3^{2}&d_{22}\tau_1^2&d_{23}\tau_1\tau_3\\
0&d_{32}\tau_1&d_{33}\tau_3
\end{array}
\right)
~~~,
\end{myalign}
where $(u_{ij},d_{ij})$ are complex coefficients, here expected to be of order one.
Quark masses and mixing angles depend on a large set of parameters: the complex coefficients $(u_{ij},d_{ij})$ and the vacuum $(\tau_1,\tau_3)$. We look for predictions that are 
characteristic of the symmetry breaking preferred by the data. 
For $\tan\beta$ of ${\cal O}(1)$, data favor the vacuum $(\tau_1,\tau_3)\approx (0.05,0.02)$, which appears
to be close both to the origin, where the whole $U(2)$ is unbroken and to the critical region $(\tau_1\ne 0,\tau_3=0)$ where
$U(2)$ is broken down to a $U(1)'\ne U(1)$ subgroup, see fig. 3. This last interpretation has the advantage of "explaining"
the mild hierarchy between $\tau_3$ and $\tau_1$. Adopting this point of view we can study the behavior
of the system in a neighborhood of $(0<\tau_1<1,\tau_3=0)$, by keeping $\tau_1$ fixed and expanding the
quantities of interests in powers of $\tau_3$, assumed to be smaller than $\tau_1$. 
Unless we know the coefficients $(u_{ij},d_{ij})$, we cannot determine the physical quantities with a good degree of accuracy. Nevertheless, near the critical region $(0<\tau_1<1,\tau_3=0)$, we can identify general scaling laws.
In particular, up to unknown order-one coefficients, we find~\cite{Falkowski:2015zwa,Linster:2018avp}:
\begin{myalign}
\label{uni1}
\begin{array}{lll}
y_u\approx \tau_1^2 \left(\dd\frac{\tau_3}{\tau_1}\right)^4~~~,&y_c\approx \tau_1^2~~~,&y_t\approx 1~~~,\\
y_d\approx \tau_1^2 \left(\dd\frac{\tau_3}{\tau_1}\right)^4~~~,&y_s\approx \tau_1^2 \left(\dd\frac{\tau_3}{\tau_1}\right)~~~,&y_b\approx \tau_1~~~,\\
V_{ub}\approx \tau_1 \left(\dd\frac{\tau_3}{\tau_1}\right)^2~~~,&V_{cb}\approx \tau_1~~~,&V_{us}\approx\left(\dd\frac{\tau_3}{\tau_1}\right)^2~~~.
\end{array}
\end{myalign} 
This scaling behavior relies on the choice of charges in table 1 (which in turns 
determine the near-critical vacuum), but is independent of the exact values of the coefficients $(u_{ij},d_{ij})$.
Assuming the coefficients $(u_{ij},d_{ij})$ to be of order one, an immediate consequence of these scaling relations are approximate equalities such as:
\begin{myalign}
\label{un_equalities}
\begin{array}{ll}
V_{us}V_{cb}\approx V_{ub}~~~,&
V_{us}\approx \left(\dd\frac{m_d}{m_s}\right)^{2/3}~~~,\\
V_{cb}\approx \left(\dd\frac{m_c}{m_t}\right)^{1/2}~~~,&
\dd\frac{V_{ub}}{V_{cb}}\approx \left(\dd\frac{m_u}{m_c}\right)^{1/2}~~~.
\end{array}
\end{myalign} 
The system sits close to the critical orbit with a $U(1)'$ residual symmetry, which results in several scaling properties
that do not depend on the details of the model and represent a quantitative test of universality.
Notice that in this discussion we are not looking for a dynamical explanation of the hierarchy
$\tau_3/\tau_1<1$. In our interpretation, it is the vicinity of the vacuum to the critical point $(\tau_1\ne 0,\tau_3=0)$ that provides a justification for such a hierarchy, much as the criticality of the Higgs system legitimates the smallness of the ratio between the electroweak VEV and the Planck scale.

\begin{figure}[h!]
\centering
\includegraphics[width=0.5\linewidth]{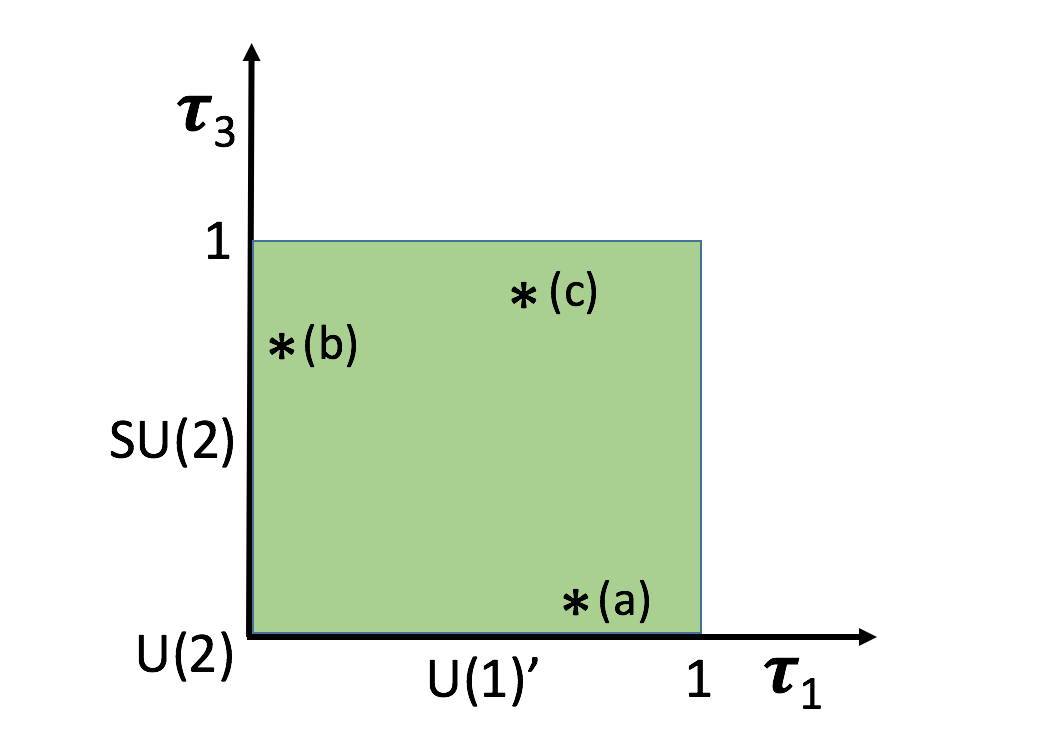}
\caption{Region ${\cal M}/G$ for the $U(2)$ model. The green area highlights the domain where an expansion in powers
of $\tau_{1.3}$ is meaningful. The stars marked $(a)$, $b$ and $(c)$ would suggest vicinity to phase transitions where the unbroken groups are $U(1)'$, $SU(2)$ and $U(2)$, respectively.}
\label{U2}
\end{figure}
\vskip 0.2 cm
\noindent
The region ${\cal M}/G$ is drawn in fig. \ref{U2}. As for most of the linearly-realized flavor symmetries, the $U(2)$ model
favors a priori the region in green, where $\tau_{1,3}$ are small parameters and the operators of the lowest dimension dominate the effective field theory. 
The case illustrated above is qualitatively described by the point marked $(a)$ in the plot, close to the line where $U(1)'$ is unbroken. As an order parameter for the $U(1)'$ breaking we can take, for example, $u_a=|\tau_3|^2$. The hypothetical point $(b)$ would suggest the vicinity to a phase transition where the unbroken group is
$SU(2)$, for which we could adopt $u_b=|\tau_1|^2+|\tau_2|^2$ as order parameter. More complicated cases can occur,
like that represented by the point $(c)$, for which we cannot decide whether it is significantly closer to the $U(1)'$ phase 
or the $SU(2)$ one. Here a single order parameter is insufficient to fully characterize the broken phase. In general, the number of order parameters needed to completely specify the broken phase is equal to the dimension
of the domain ${\mathcal M}/G$. Moreover, several different phases can be accessible to the system. In the example under discussion, beyond the unbroken phase and the completely broken one, there are also
the $SU(2)$ and the $U(1)'$ phases. For generic $G$ and ${\mathcal M}$ the phase diagram can be very rich, and many transitions are possible. When $\tau$ is localized near a critical region where some nontrivial subgroup $H$ of $G$ is preserved, we can refer near-criticality to the transition from the exact $H$ phase to the broken one.
We provide a general discussion of the possible transitions as well as the related scaling laws in Appendix~\ref{fmcu}.

We might be led to think that equalities like those in eq. (\ref{un_equalities})
are common features of all Yukawa systems approaching a critical point. 
We provide a counterexample by discussing
the extreme case associated to the choice $G=U(3)_q\times U(3)_{u^c}\times U(3)_{d^c}$, the group of minimal
flavor violation (MFV)~\cite{DAmbrosio:2002vsn}, and $\tau\equiv{\cal Y}_u\oplus {\cal Y}_d=(\bar 3,3,1)\oplus (\bar 3,1,3)$, $q=(3,1,1)$, $u^c=(1,\bar 3,1)$, $d^c=(1,1,\bar 3)$.
Though the description of quark masses, mixing angles and phases goes beyond the scope of MFV, 
it is a fact that the data favor a region of ${\cal M}$ close to:
\begin{myalign}
\label{region}
{\cal Y}_u\propto
\left(
\begin{array}{ccc}
0&0&0\\
0&0&0\\
0&0&1
\end{array}
\right)~~~,~~~~~
{\cal Y}_d\propto
\left(
\begin{array}{ccc}
0&0&0\\
0&0&0\\
0&0&1
\end{array}
\right)~~~,
\end{myalign} 
where $G$ is broken down to $H=U(2)_q\times U(2)_{u^c}\times U(2)_{d^c}\times U(1)$. Moreover, an interesting property of any $G$-invariant scalar potential depending on $\tau$ is that it admits extrema of this type~\cite{Alonso:2013nca}.
Thus our tentative criterium for near-criticality is satisfied and the pattern of eq. (\ref{region}) is a good first-order approximation for a model of quark masses.
The subgroup $H$ is not an exact symmetry and it is broken down to the baryon number and the weak hypercharge groups by the entries of six irreducible representations $u_i$ $(i=1,...,6)$: $(\bar 2,2,1)_0$, $(\bar 2,1,2)_0$, $(1,2,1)_{-1}$
$(1,1,2)_{-1}$, $(\bar 2,1,1)_{+1}$ and $(\bar 2,1,1)_{+1}$.~\footnote{The index refers to the $U(1)$ charge.}
To discuss the physical properties, we move to the restricted region ${\cal M}/G$, where each orbit is represented
by a single point. Indeed, by performing a $G$ transformation,
it is always possible to reach a region in ${\cal M}$ where $\tau$ is parametrized as:
\begin{myalign}
{\cal Y}_d=\hat {\cal Y}_d~~~,~~~~~{\cal Y}_u=\hat{\cal Y}_u V_{CKM}~~~,
\end{myalign} 
where $\hat {\cal Y}_{u,d}$ are diagonal non-negative matrices and $V_{CKM}$ is the unitary quark mixing matrix, depending
on three angles and one phase. Thus the most general vacuum of the system depends on ten 
real parameters describing the independent types of orbits. Two of them are fixed by the choice in eq. (\ref{region}) and the remaining eight describe the independent degrees of freedom of the representations $u_i$. 
Unlike the previous case, we have many vacuum parameters. Moreover, these parameters are spread over many different orders of magnitude
and each observable has an independent scaling behavior, not allowing to identify equalities like those in eq. (\ref{un_equalities})~\footnote{Apart from the trivial scaling of the top and bottom masses.}. While in this specific example the conclusion could already be anticipated from the premises,
we can easily figure out other examples where a large number of independent orbits
practically forbids significant relations among the observables, even assuming 
all dimensionless Lagrangian parameters to be of the same order of magnitude.

Although in the quark sector, featuring a hierarchical mass spectrum and small mixing angles, it is relatively easy
to identify a nontrivial symmetric pattern of masses reasonably close to observation, this is not the case 
for the lepton sector. In a large class of models based on discrete symmetries,
the observed lepton mixing matrix is approximately reproduced through breaking terms that
force the charged lepton mass matrix and the neutrino mass matrix to have different residual symmetries, whose intersection is trivial. 
The system obeys our criterium for near-criticality, but the hypothetical phase transition is realized in a very peculiar way. The system is close to the critical region where the full group $G$ is unbroken and, in such a symmetric phase, lepton mass ratios and mixing angles are completely undetermined. Several independent order parameters are needed to characterize the broken phase and a crucial ingredient for the correct output is their relative orientation in the space ${\mathcal M}/G$. 

We illustrate this behavior in one example.
For instance, the model of ref.~\cite{Altarelli:2005yx}, tailored to reproduce an approximate tribimaximal mixing, is based on the group $G=A_4\times \mathbb{Z}_3$, where $A_4$ is the group of even permutations of four objects, generated by two
elements: $S$ and $T$.
The relevant part of the manifold ${\cal M}$
is spanned by two $A_4$ triplets $(\varphi_S,\varphi_T)$ and one singlet $\xi$: $\tau=\varphi_T\oplus \varphi_S\oplus \xi$. Representations are arranged such that $\varphi_T$ mainly couples to charged leptons and
$\varphi_S\oplus \xi$ mainly couples to neutrinos. A reliable description of the data relies entirely on a special vacuum alignment among the different components of $\tau$. 
In an appropriate basis for $S$ and $T$, tribimaximal mixing arises from:
\begin{myalign}
\label{va}
\varphi_T=(v_T,0,0)~~~,~~~~~\varphi_S=(v_S,v_S,v_S)~~~,~~~~~\xi_S=v~~~~~~~~~~~~~~(v_T\approx v_S\approx v\ll1)~~~.
\end{myalign} 
Today exact tribimaximal mixing is experimentally ruled out, but this picture can be easily made compatible with data
by adding small corrections.
The multiplet $\varphi_S\oplus \xi$ breaks $G$ down to the $\mathbb{Z}_2^S$ subgroup of $A_4$ generated by $S$, while 
$\varphi_T$ breaks $G$ down to $\mathbb{Z}_3^T\times \mathbb{Z}_3$, $\mathbb{Z}_3^T$ being the $A_4$ subgroup generated by $T$. The net result is that $G$ is completely broken by the VEVs in eq. (\ref{va}). Given the smallness of the parameters $v_{S,T},v$, the system is close 
to the origin of ${\cal M}$ where $G$ is unbroken. There are as many independent order parameters as entries of the complex multiplets $\varphi_S$ ,$\varphi_T$ and $\xi$, but only the special alignment in eq. (\ref{va}) drives the system close to the data. Moreover,
 in the symmetric phase lepton mass ratios, mixing angles and phases are completely undetermined.

It is a logical possibility that in the symmetric phase all relevant observables are completely undetermined and the observed pattern entirely arises from a clever alignment of the order parameters. However, we can ask whether
a realistic description of the lepton spectrum is compatible with a nontrivial symmetric limit, providing
a seed that persists at some level in the broken phase, as is the case in most models describing the quark sector.
To answer this question, ref.~\cite{Reyimuaji:2018xvs} presents a comprehensive discussion, applying to flavor groups $G$ of any type, 
continuous or discrete, global or local, abelian or non-abelian.
The analysis is performed under the assumption that the light neutrino masses are Majorana
and the symmetry directly constrains the light neutrino mass matrix~
\footnote{The latter assumption is not innocent: if light neutrinos originate from
the seesaw mechanism a necessary, but not sufficient, condition for its validity is that the neutrino mass matrix is non-singular in the symmetric limit.}.
An additional assumption is that, when the observed lepton masses and mixing angles are close to a symmetric point, the symmetric limit is such that:
\begin{enumerate}
\item[i)] the mixing matrix is not fully undetermined;
\item[ii)] both the $\theta_{23}$ and $\theta_{12}$ angles are allowed to be non-vanishing;
\item[iii)] the non-vanishing charged lepton masses are not forced to be degenerate.
\end{enumerate}
The conditions i), ii) and iii) reflect the closeness of the examined configuration to the data. Under these rather general assumptions, ref.~\cite{Reyimuaji:2018xvs} shows that the only possible unbroken symmetry compatible with normal ordering of neutrino masses requires left-handed lepton doublets to consist of three trivial singlets (either invariant or flipping the sign under $G$). This means that both neutrino masses and lepton mixing angles are unconstrained. Each entry $(m_\nu)_{ij}$ of the neutrino mass matrix is an independent Lagrangian parameter. No field $\tau$ is needed to describe neutrinos (it might be needed to explain the hierarchy of charged lepton masses). If all elements $(m_\nu)_{ij}$ are comparable in size,
neutrino masses are expected to be nearly degenerate and lepton mixing angles all of the same order,
as in the anarchy proposal~\cite{Hall:1999sn,Haba:2000be,deGouvea:2003xe,Espinosa:2003qz,deGouvea:2012ac}. 
Near-criticality goes undetected in the neutrino sector, which sits exactly at the unbroken phase.

The main lesson from this short review is that most of the models based on linearly-realized flavor symmetries obey the condition of near-criticality by construction. Mass matrices $m_{ij}(\tau)$ are dominated by the first few terms
in the power expansion around $\tau=0$, where the flavor group $G$ is unbroken. The condition $|\tau|\ll 1$ is a necessary one to justify such an expansion. Even if near-criticality formally applies, it does not represent a useful tool to
discriminate models in this class or to reliably detect a phase transition. 
Nonetheless, near-criticality is realized in a variety of different ways. In the quark sector, the symmetric phase 
typically provides a decent first approximation of the data, storing a seed that persists in the broken phase.
Sometimes this seed resides in a subgroup $H$ of the full flavor group $G$ and the transition from $H$ down to the
broken phase is described by a single order parameter. Mass ratios and mixing angles exhibit characteristic scaling relations, independent of the order-one parameters entering the mass matrices.
On the contrary, in the lepton sector it seems difficult to 
identify a nontrivial symmetric limit. Either the mass spectrum is completely undetermined in the symmetric phase
and the transition requires a special alignment of multiple order parameters, or a large part of the systems sits
permanently in the exact phase where masses and mixing angles are unconstrained.
It is interesting to compare this picture to the one arising in the case of nonlinearly realized flavor symmetries,
which will be discussed in the next Section.
\section{Critical behavior of nonlinearly realized flavor symmetries}
\label{modularsec}
In this Section we discuss nonlinearly realized flavor symmetries.
We focus on the case where the vacua of the theory are redundantly described by a space ${\cal M}$.
The redundancy is removed by the action of a discrete gauge symmetry group $G$ and the physically
inequivalent vacua are the elements of the domain ${\cal M}/G$. This is the typical framework describing
the moduli space in string theory compactifications, where $G$ embodies the network of duality transformations.
As anticipated in Section~\ref{cri}, there is no origin
in ${\cal M}$ where $G$ is unbroken. The group is broken everywhere in ${\cal M}/G$, but there can be points 
$\tau_0$ that remain fixed under some subgroup $H$ of $G$. Fermion mass matrices $m_{ij}(\tau)$ 
can be expressed as combinations of functions $Y_k(\tau)$ with appropriate transformations properties under $G$:
\begin{myalign}
m_{ij}(\tau)=\sum_k c^k_{ij} Y_k(\tau)~~~.
\end{myalign}
In the example of Section~\ref{cri}, where $G$ is generated by the two parity symmetries $\mathbb{Z_2}$ and  
$\mathbb{Z_2}'$ of eq. (\ref{zz}), matter fields $\psi$
can be assigned to representations of the type $(\sigma_1^\psi,\sigma_2^\psi)$ under $(\mathbb{Z_2},\mathbb{Z_2}')$, where $\sigma_i$ are equal to $\pm 1$. Yukawa couplings entering fermion bilinears are functions of $\tau$ transforming 
in representations $(\sigma_1,\sigma_2)$. They can be expanded as linear combinations of basis functions $Y^k_{\sigma_1,\sigma_2}(\tau)$ ($k$ integer):
\begin{myalign}
\begin{array}{rcl}
Y^k_{+,+}&=&\cos\dd\frac{2k\pi(\tau-\tau_1/2)}{(\tau_2-\tau_1)}\\
Y^k_{+,-}&=&\cos\dd\frac{(2k+1)\pi(\tau-\tau_1/2)}{(\tau_2-\tau_1)}\\
Y^k_{-,+}&=&\sin\dd\frac{(2k+1)\pi(\tau-\tau_1/2)}{(\tau_2-\tau_1)}\\
Y^k_{-,-}&=&\sin\dd\frac{(2k+2)\pi(\tau-\tau_1/2)}{(\tau_2-\tau_1)}~~~.
\end{array}
\end{myalign}
Due to the nonlinear and gauge character of the $G$ transformations, the size of $\tau$ has no absolute meaning. 
At the same time, $\tau$ is not a good order parameter. The gauge group $G$ is fully broken everywhere, but at the two extremes
of the interval in fig. \ref{toyfun}, $\tau_1/2$ and $\tau_2/2$, where $\mathbb{Z_2}$  
and $\mathbb{Z_2}'$ are residual symmetries, respectively.
The value of $\tau$ favored by the data
is less obvious than in the case of linearly realized flavor symmetries. A value of $\tau$ near the center of the 
interval in fig. \ref{toyfun} would be perfectly acceptable from the viewpoint of the effective theory.
A value of $\tau$ close to the extremes of the interval would be remarkable since, unlike in linearly realized
flavor symmetries, these points are not privileged, a priory. We are led to regard the preference for $\tau$ near 
$\tau_1/2$ (or $\tau_2/2$) as a significant indication of the near-criticality of the system. In this case the fermion spectrum
might have its origin in a transition from the unbroken $\mathbb{Z_2}$  
(or $\mathbb{Z_2}'$) phase to the broken one. One of the problems of this type of transition is the identification of
a good order parameter. In this toy model, $u=\tau-\tau_1/2$ and $u=\tau-\tau_2/2$ are order parameters for the
transitions involving $\mathbb{Z_2}$ and $\mathbb{Z_2}'$, respectively. 

We discuss here how it is possible to define an order parameter $u(\tau)$ in the general case, when
$G$ is nonlinearly realized in the space ${\cal M}$ and the value of $\tau$ preferred by the data
lies close to a critical point $\tau_0$, fixed by a nontrivial subgroup $H_0$ of $G$:
\begin{myalign}
\label{htau0}
h\tau_0=\tau_0~~~~~~~~~~~~~~~~(h\in H_0)~~~.
\end{myalign} 
We choose a coordinate system $u(\tau)$ in ${\cal M}$ such that $H_0$ is linearly realized and $u(\tau_0)=0$.
This is always possible if $H_0$ is a continuous compact group \cite{Coleman:1969sm}. 
If $G$ is discrete and $H_0$
is a finite group, we can easily adapt the proof of ref.~\cite{Coleman:1969sm}. First, observe that when $\tau'=\tau-\tau_0$ is very small
we can expand $h_i\tau'$ in powers of $\tau'$ around $\tau'=0$:
\begin{myalign}
h_i\tau'=D(h_i)\tau'+{\cal O}(\tau'^2)~~~~~~~~~~~~~~~~(h_i\in H_0)~~~.
\end{myalign} 
The constant term ${\cal O}(\tau'^0)$ is absent due to eq. (\ref{htau0}) and it is not difficult to prove that $D(h_i)$ provides a linear representation of $H_0$.
We define:
\begin{myalign}
\label{defu}
u\equiv\dd\frac{1}{d(H_0)}\sum_i D^{-1}(h_i) h_i\tau'~~~,
\end{myalign} 
$d(H_0)$ being the dimension of the group $H_0$. Acting with an element $h_0$ of $H_0$ on $u$ we get:
\begin{myalign}
\begin{array}{ll}
h_0 u&=\dd\frac{1}{d(H_0)}\sum_i D^{-1}(h_i) h_i h_0\tau'\\
&=\dd\frac{1}{d(H_0)}\sum_i D^{-1}(h_i h_0 h_0^{-1}) (h_i h_0)\tau'\\
&=\dd\frac{1}{d(H_0)}\sum_i 
D(h_0) D^{-1}(h_i h_0) (h_i h_0)\tau'\\
&=D(h_0)u~~~.
\end{array}
\end{myalign} 
From eq. (\ref{defu}) we see that $u(\tau_0)=0$ and we can adopt 
$u(\tau)$ as order parameter for the breaking of $H_0$. Moreover, the symmetry $H_0$ is
linearly realized on the manifold ${\cal M}$ spanned by the fields $u$, and we can follow the same steps 
of our discussion in Section~\ref{survey}.
We now move to a realistic realization of this framework, applied to a set of models for lepton masses.
\subsection{Critical behavior of modular invariant flavor models}
We focus on modular 
invariance as candidate flavor symmetry. Modular invariance is an intrinsic property of the moduli space in string theory compactifications. It is a discrete gauge symmetry, crucial
for the correct identification 
of the vacuum in the theory~\cite{Feruglio:2019ybq}. 
We are interested in modular invariant models of fermion masses, with ${\cal N}=1$ rigid supersymmetry, the extension
to ${\cal N}=1$ supergravity being straightforward. 
The restriction to the supersymmetric case is a technical requirement that, forcing Yukawa couplings to be holomorphic
functions of the modulus $\tau$, allows to make use of explicit analytical expressions. 
While supersymmetry-breaking effects are a necessary ingredient of any realistic construction, it has been shown that, in a large portion of the parameter space, they have a negligible impact on the prediction of the fermion spectrum~\cite{Criado:2018thu}. For this reason we neglect them here.
The flavor group is the modular group $G=SL(2,\mathbb{Z})$, consisting of $2\times 2$ matrices $\gamma$
of the form
\begin{myalign}
\gamma=
\left(
\begin{array}{cc}
a&b\\
c&d
\end{array}
\right)~~~,
\end{myalign} 
where $a$, $b$, $c$ and $d$ are integers and $ad-bc=1$. 
The modular group is generated by two elements:
\begin{myalign}
S=\left(\begin{array}{cc}0&1\\-1&0\end{array}\right)~~~,~~~~~~~~~~~~~~~~~~~~T=\left(\begin{array}{cc}1&1\\0&1\end{array}\right)~~~,
\end{myalign} 
satisfying the equalities
\begin{myalign}
S^4=(ST)^3=\mathbb{1}~~~~~~~~~~~~~~~S^2T=TS^2~~~,
\end{myalign} 
that can be adopted as the abstract relations defining the group.
The field content of the theory includes a set of chiral supermultiplets $(\tau,\varphi)$~\footnote{We denote by $(\tau,\varphi)$ both the chiral superfields and their scalar components.}. The scalar component of the modulus $\tau$, parametrizing the vacuum of the theory, is a complex field living in the upper half complex plane ${\cal M}$ and can be thought as a label for the (conformally equivalent) metrics on a torus compactification.
The chiral superfields $\varphi$ describe the matter multiplets. Under $SL(2,\mathbb{Z})$, $\tau$ and $\varphi$ transform as
\begin{myalign}
\label{tran}
\begin{array}{l}
\tau\xrightarrow{\gamma}\dd\frac{a\tau+b}{c\tau+d}~~~~~~~~~~~~~~~~~~~~~~~~~~~~~~~~~~~~~~~~~
\\[10 pt]
\varphi \xrightarrow{\gamma} j(\gamma,\tau)^{-k_\varphi}\rho^N_\varphi(\gamma)\varphi~~~~~~~~~~~~~~~~~~~~~~~j(\gamma,\tau)\equiv(c\tau+d)~~~.
\end{array}
\end{myalign} 
The elements $S$ and $T$ act on the modulus as $\tau\xrightarrow{S} -1/\tau$, $\tau\xrightarrow{T} \tau+1$. 
The automorphy factor $j(\gamma,\tau)=c\tau+d$ satisfies the cocycle condition
\begin{myalign}
\label{cocy}
j(\gamma_1\gamma_2,\tau)=j(\gamma_1,\gamma_2\tau)j(\gamma_2,\tau)
\end{myalign} 
 and guarantees that the transformation
in eq. ({\ref{tran}) is a nonlinear realization of $SL(2,\mathbb{Z})$.
In a toroidal compactification of two extra dimensions, $\tau$ and $\gamma\tau$ describe the same (metrics on a) torus.
In this sense we can view $G=SL(2,\mathbb{Z})$ as a discrete gauge symmetry.
The transformation law of the matter multiplets $\varphi$ is characterized by 
\begin{itemize}
\item[1.]
the weight $k_\varphi$, here assumed to be integer;
\item[2.]
a finite copy of the modular group, $SL(2,\mathbb{Z}_N)$ $(N=1,2,3,4,...)$. The positive integer $N$ is the level of the representation and is common to all matter multiplets. To include two distinct levels $N_1$ and $N_2$ in the same
construction, it is sufficient to adopt as level $N$ the least common multiple of $N_1$ and $N_2$. 
For the first few levels
$N=2,3,4,5$, the finite modular group $SL(2,\mathbb{Z}_N)$ is isomorphic to the double covering of the permutation groups
$S_3$, $A_4$, $S_4$, $A_5$, respectively;
\item[3.] a unitary representation $\rho^N_\varphi$ of $SL(2,\mathbb{Z}_N)$.
\end{itemize}
In eqs. (\ref{tran}) we use a matrix notation: if the representation $\rho_\varphi(\gamma)$ is reducible, we write it in a block diagonal form and 
the weight $k_\varphi$ is a vector with independent integer entries for each block.
Therefore $j(\gamma,\tau)^{-k_\varphi}$ is a diagonal matrix that commutes
with $\rho_\varphi(\gamma)$. When the representations
are irreducible, a unique weight is allowed and $j(\gamma,\tau)^{-k_\varphi}$
is an overall factor. Thus a realization of the modular group in the field
space is characterized by the triplet $(k_\varphi,N,\rho^N_\varphi)$ for each irreducible multiplet $\varphi$, $N$ being fixed. Here $(k_\varphi,N,\rho^N_\varphi)$ are considered free parameters of the theory, while in a fundamental theory such as string theory the field content and the $SL(2,\mathbb{Z})$ action are determined by the compactification.

The modular group can be combined with CP, whose
most general action on $\tau$ and $\varphi$, up to modular transformations, is given by~\footnote{We denote conjugations of fields(numbers) with a bar(asterisk).}~\cite{Baur:2019kwi,Novichkov:2019sqv,Kobayashi:2019uyt,Ding:2021iqp}:
\begin{myalign}
\label{CP}
\tau\xrightarrow{CP}-\bar\tau~~~~~~~~~~~~~
\varphi \xrightarrow{CP} X_{CP}~\bar\varphi~~~,
\end{myalign} 
where $X_{CP}$ is a matrix obeying, for each $\gamma\in SL(2,\mathbb{Z})$, the consistency condition:
 \begin{myalign}
 X_{CP}~\rho^{N*}_\varphi(\gamma)~ X_{CP}^{-1}=\rho^N_\varphi(\gamma')~~~~~~~~~~~\gamma'\in SL(2,\mathbb{Z})~~~.
\end{myalign} 
The unitary matrices $\rho^{N}_\varphi(S)$ and $\rho^{N}_\varphi(T)$ are always symmetric in the cases discussed here.
It follows that it is not restrictive to choose $X_{CP}=\mathbb{1}$.
When the theory is CP invariant, CP violation can only arise from spontaneous breaking.

The relevant part of the classical action reads:
 \begin{myalign}
\label{one}
{\mathscr S}=\int d^4 x d^2\theta d^2\bar\theta~ K(\bar\tau,\bar\varphi,\tau,\varphi)+\int d^4 x d^2\theta~ w(\tau,\varphi)+\int d^4 x d^2\bar\theta~ \bar w(\bar\tau,\bar\varphi)~,
\end{myalign} 
where $K(\bar\tau,\bar\varphi,\tau,\varphi)$ is the K\"ahler potential, a real gauge-invariant function describing the kinetic terms and $w(\tau,\varphi)$ is the superpotential, a holomorphic gauge-invariant function describing Yukawa interactions. 
Modular(CP) invariance requires ${\mathscr S}$ to remain unchanged under the transformations of eq. (\ref{tran}(\ref{CP})). This imposes a strong restriction on the superpotential $w(\tau,\varphi)$. By expanding $w(\tau,\varphi)$
in powers of the matter fields $\varphi$, each $\tau$-dependent coefficient of the expansion is a 
modular form of level $N$ and given weight.
These span a finite-dimensional linear space, and few independent parameters are typically sufficient to
characterize the whole superpotential. Modular invariance is not so effective in constraining the K\"ahler potential.
While minimal K\"ahler potentials are generally adopted in model building, non-minimal and flavor-dependent ones
are allowed in the general case. When analyzing the theory predictions in Sections~\ref{micp} and~\ref{lmm}, we will always consider the most general 
$K(\bar\tau,\bar\varphi,\tau,\varphi)$ allowed by modular invariance.

By exploiting modular invariance, we can restrict the modulus $\tau$ to
the fundamental domain ${\cal F}=\{\tau\in {\cal M}\vert~~~~ |\tau|\ge 1, |{\Re}(\tau)|\le 1/2\}$, such that each point of ${\cal M}$ can be mapped into ${\cal F}$ by a $SL(2,\mathbb{Z})$ transformation, but no two points of the interior of ${\cal F}$ are related by $SL(2,\mathbb{Z})$ transformations, see fig.~\ref{lines}. The fundamental domain ${\cal F}$ corresponds to 
the region ${\cal M}/G$ of our general discussion.
In a generic point of ${\cal F}$ the discrete symmetry $SL(2,\mathbb{Z})$ and $CP$ are completely broken, that is $\gamma \tau = \tau$ and $-\bar\tau=\tau$ (or their combination $-\bar\tau=\gamma \tau$) have no solution for $\gamma\in SL(2,\mathbb{Z})$. There are special points and lines of ${\cal F}$ where a part of the flavor symmetry (including CP) is preserved. Fixed points of ${\cal F}$ under $SL(2,\mathbb{Z})$ are $\otau =( i,\omega, i\infty)$, where
$\omega=-1/2 + i \sqrt{3}/2$. They are left invariant by the finite subgroups generated by $\ogamma=(S, ST,T)$, respectively~\footnote{The point $+1/2 + i \sqrt{3}/2$, equivalent to $-1/2 + i \sqrt{3}/2$, is invariant under $TS$, $\tau\xrightarrow{TS} (\tau-1)/\tau$.}: 
\begin{myalign}
\ogamma~\otau=\otau~~~.
\end{myalign} 
The elements $(S, ST,T)$ have order 4, 3 and $\infty$, respectively.
Moreover, any point of ${\cal F}$ is left invariant by the action of the element $S^2$, which can have a nontrivial
action only in field space. For this reason at $\otau =( i,\omega, i\infty)$ the theory is invariant under the discrete group
$G_0=(\mathbb{Z}_4^S,\mathbb{Z}_2^{ST}\times\mathbb{Z}_2^{S^2},\mathbb{Z}^{T}\times\mathbb{Z}_2^{S^2})$, respectively.
If the action ${\mathscr S}$ of the theory is also $CP$-invariant, $CP$ is spontaneously broken anywhere in ${\cal F}$, except
along the line $\Re{\tau}=0$ and, up to modular transformations, on the boundary of ${\cal F}$, that is $|\tau|=1$ and $|\Re(\tau)|=1/2$. Hence at the three points $\otau=( i,\omega, i\infty)$ also CP is preserved, see fig. \ref{lines}.

\begin{figure}[h!]
\centering
\includegraphics[width=0.5\linewidth]{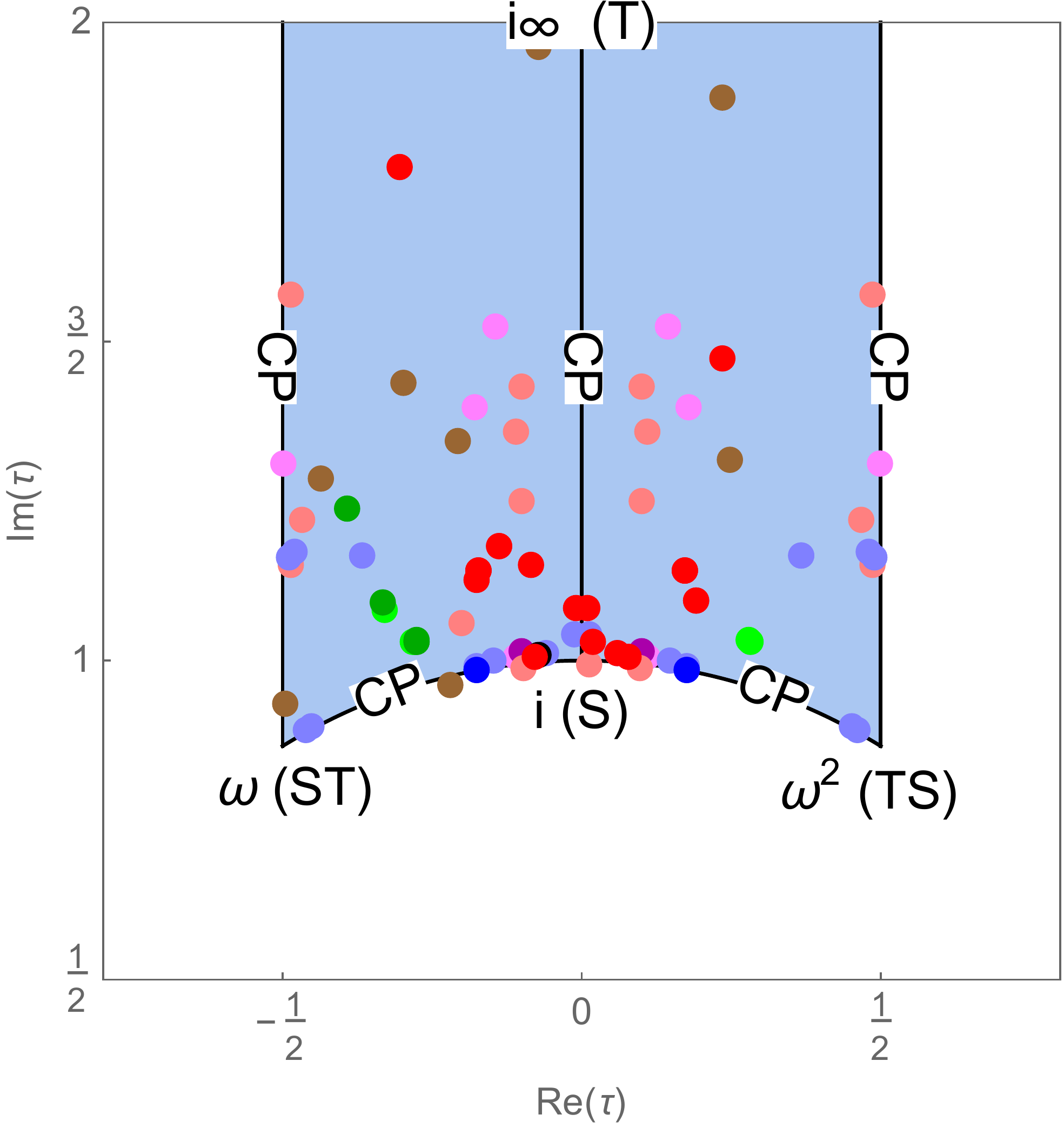}\includegraphics[width=0.5\linewidth]{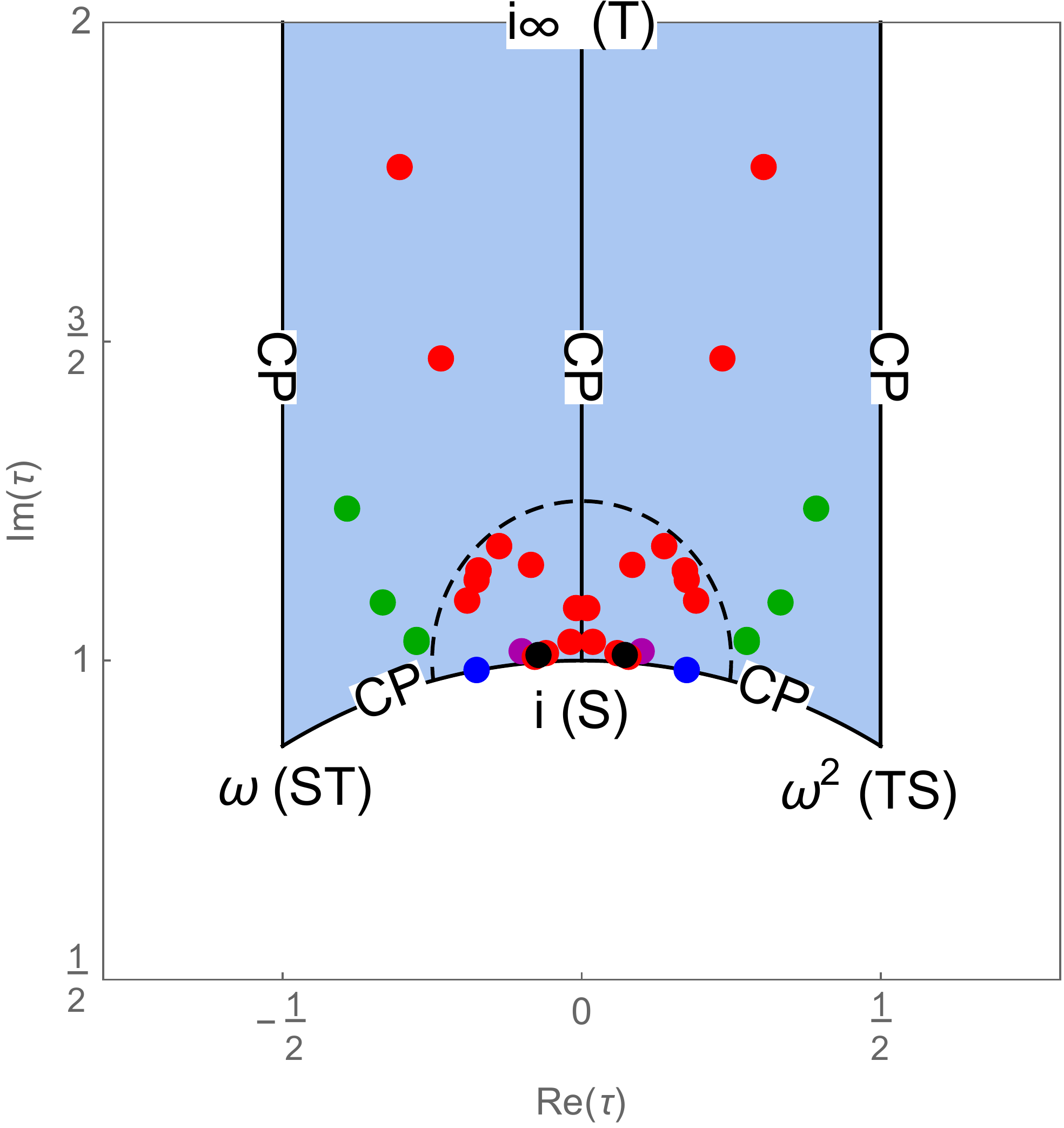}
\caption{Fundamental domain ${\cal F}$ (light blue region) and fixed points (see text). Dots are the best-fit values of $\tau$ in models of ref. \cite{Criado:2018thu,Ding:2019zxk} ($\Gamma_3$ - light red), 
\cite{Yao:2020qyy,Okada:2020brs} ($\Gamma_3 \& CP$ - red),
\cite{Novichkov:2018ovf} ($\Gamma_4$ - light magenta), 
\cite{Novichkov:2019sqv} ($\Gamma_4 \& CP$ - magenta), 
\cite{Liu:2020akv} ($\Gamma_4'$ - light blue), 
\cite{Liu:2020akv} ($\Gamma_4' \& CP$ - blue),
 \cite{Wang:2021mkw} ($\Gamma_5' \& CP$ - black), 
 \cite{Li:2021buv} ($\Gamma_6'$ - light green), 
 \cite{Li:2021buv} ($\Gamma_6' \& CP$ - green), 
 \cite{Ding:2020msi} ($\Gamma_7$ - brown). 
 We use the notation $\Gamma_N'=SL(2,\mathbb{Z}_N)$ and $\Gamma_N=SL(2,\mathbb{Z}_N)/\{\pm \mathbb{1}\}$.
 In the left panel all models are displayed. The right panel includes only CP invariant models, for which the full pair of points
 $\tau$ and $-\bar\tau$ is shown. The dashed line
 represents the contour $|\tau-i|=0.25$.}
\label{lines}
\end{figure}

\begin{figure}[h!]
\centering
\includegraphics[width=0.75\linewidth]{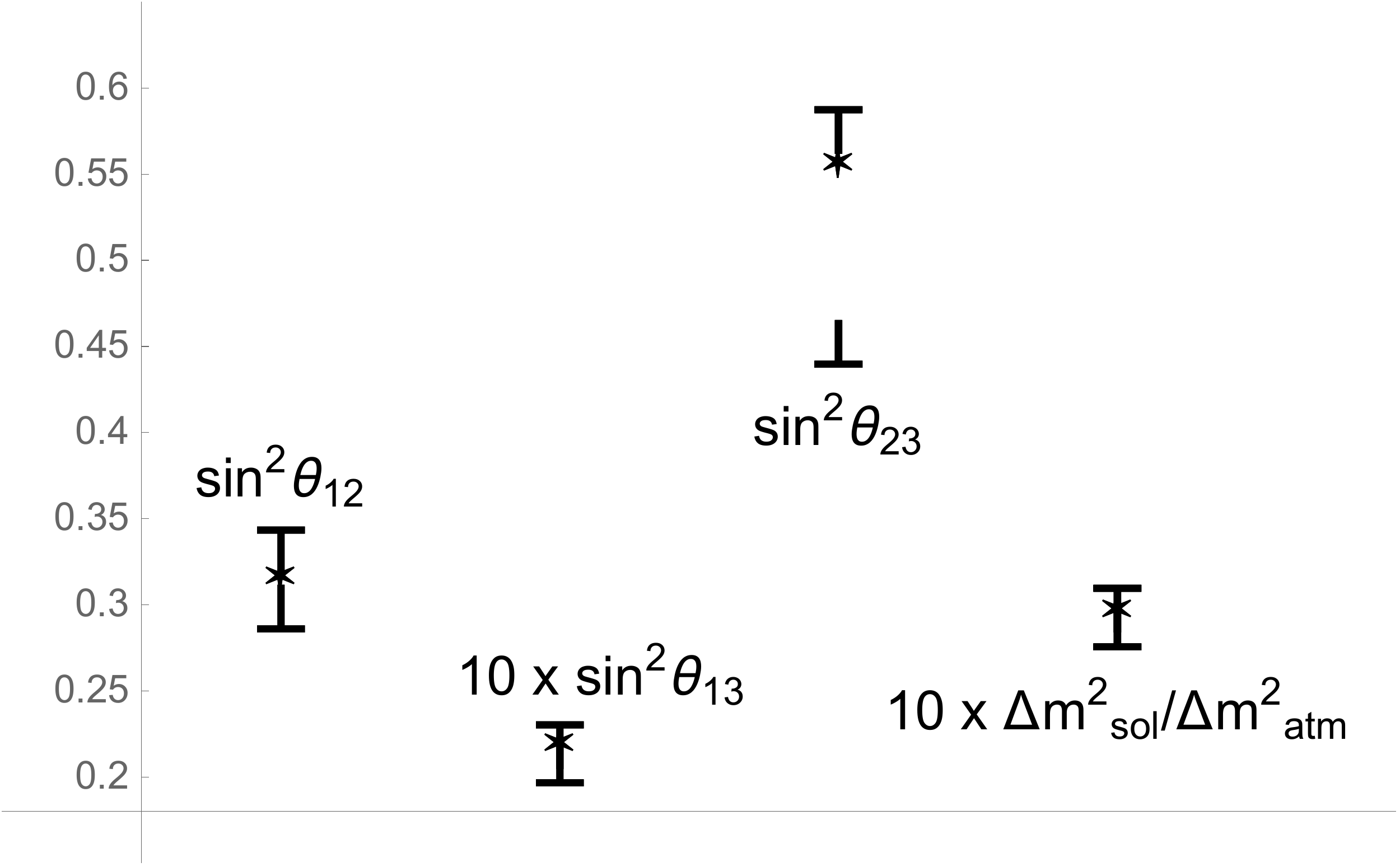}
\caption{Mixing angles and $\Delta m^2_{sol}/\Delta m^2_{atm}$ of 14 pairs of CP and modular invariant models featuring normal ordering and $|\tau-i|<0.25$ (see text). Shown in the plot are the intervals covered by the model predictions. A star indicates the average over the 14 models.}
\label{angles}
\end{figure}
\subsection{Snow-plots in the fundamental domain ${\cal F}$}
\label{snowplot}
Do data favor any critical region in ${\cal F}$?
Of course we do not have a "standard model" in this case and we should rely on examples that have been accumulated in model building, mostly in the lepton sector. Many independent models of lepton masses, mixing angles and phases, based on different choices of level, matter weights and representations have been proposed. We might feel uneasy since no baseline model springs up among them. But we can turn this weakness into an opportunity by looking for some 
common features of these successful models, which might shed light
on a fundamental organizing principle. To this purpose, in ref. \cite{Feruglio:2022kea} more than 100 models of lepton masses, which reproduce accurately
the data, were selected. In these models SU(2) lepton doublets are mostly assigned to a three-dimensional irreducible representation $\rho^N_l$ of the finite group $SL(2,\mathbb{Z}_N)$, for several choices of $N$: 3, 4, 5, 6 and 7. 
This assignment has the advantage of minimizing the number of Lagrangian free parameters. Kinetic terms are assumed to be flavor universal in all models, an assumption that we will relax later on.
The total lepton number $L$ is violated and the neutrino mass matrix originates from the dimension-5 Weinberg operator, either directly or via the seesaw mechanism. The predictions depend on the value of $\tau$ and 5 or 6 Lagrangian parameters. Three parameters are in a one-to-one correspondence with the charged lepton masses. Once these have been fixed, all the remaining observables, neutrino masses, lepton mixing angles and phases are described by the residual
Lagrangian parameters (two or three) and by the $\tau$ VEV, here denoted simply by $\tau$. Some of these models are invariant under CP, which is spontaneously broken by the value of the $\tau$. 
 
In all cases, $\tau$ is treated as an extra free parameter, varied to maximize the agreement between data and theoretical predictions. There is no prejudice about the value of $\tau$, nor about possible dynamical mechanisms that can determine $\tau$ or favor some region of ${\cal F}$. In fig. \ref{lines} we plot 103 best-fit points in the fundamental domain of $SL(2,\mathbb{Z})$.
All these models lead to an excellent description of neutrino masses and mixing angles, and predict nontrivial CP phases. 
The best-fit points for $\tau$ are not equally distributed over the fundamental domain ${\cal F}$. They are accumulated
along the boundary of ${\cal F}$, and in particular around the fixed point $\tau=i$.
If we focus on models where CP is spontaneously broken (right panel of fig. \ref{lines}) the preference for the region near the point
$\tau=i$ is more pronounced. 
Symmetric points $\tau$ and $-\bar\tau$ in CP invariant models give the same predictions, except for 
the sign of the CP violating phases. They have been considered equally successful in fig. \ref{lines} (right panel) that includes $27$ pairs of points. Two-thirds of these points fall inside the circle $|\tau-i|<0.25$, close to the self-dual point.

Such a preference may have several explanations. The statistics are limited and a small cluster of points can arise just from a fluctuation. The authors themselves might have focused their attention on a specific
part of ${\cal F}$, while scanning the $\tau$ VEVs. Or they might have selected 
only a subset of all possible representations of the modular group for the
matter multiplets.
Thus, the cluster of models around $\tau=i$ may arise from a bias of the analysis. Being aware of this possibility, here we regard the accumulation of points around $\tau=i$ as an indication of
an intrinsic property of the theory, calling for an explanation.
\begin{figure}[h!]
\centering
\includegraphics[width=0.65\linewidth]{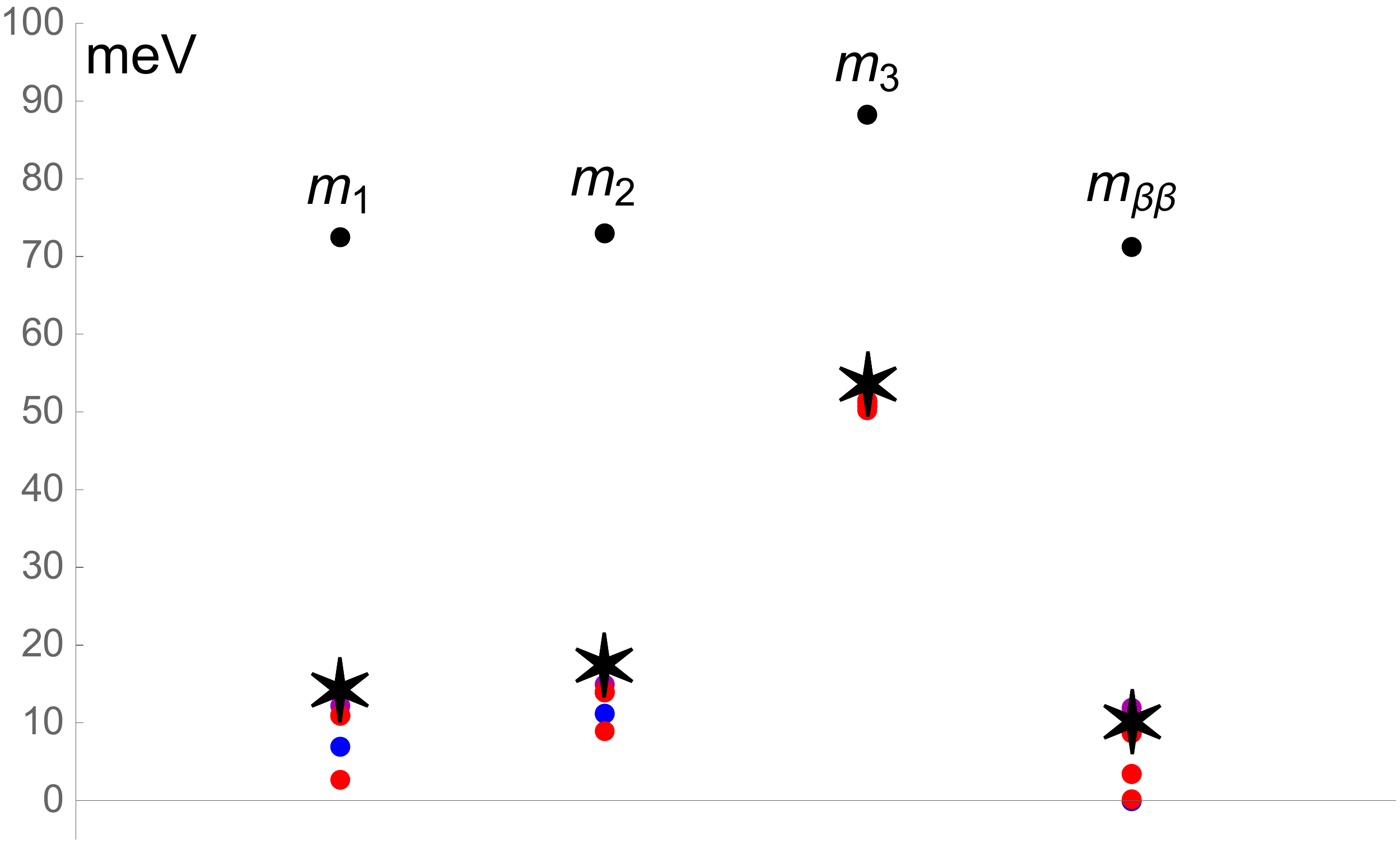}\\[20pt]
\includegraphics[width=0.75\linewidth]{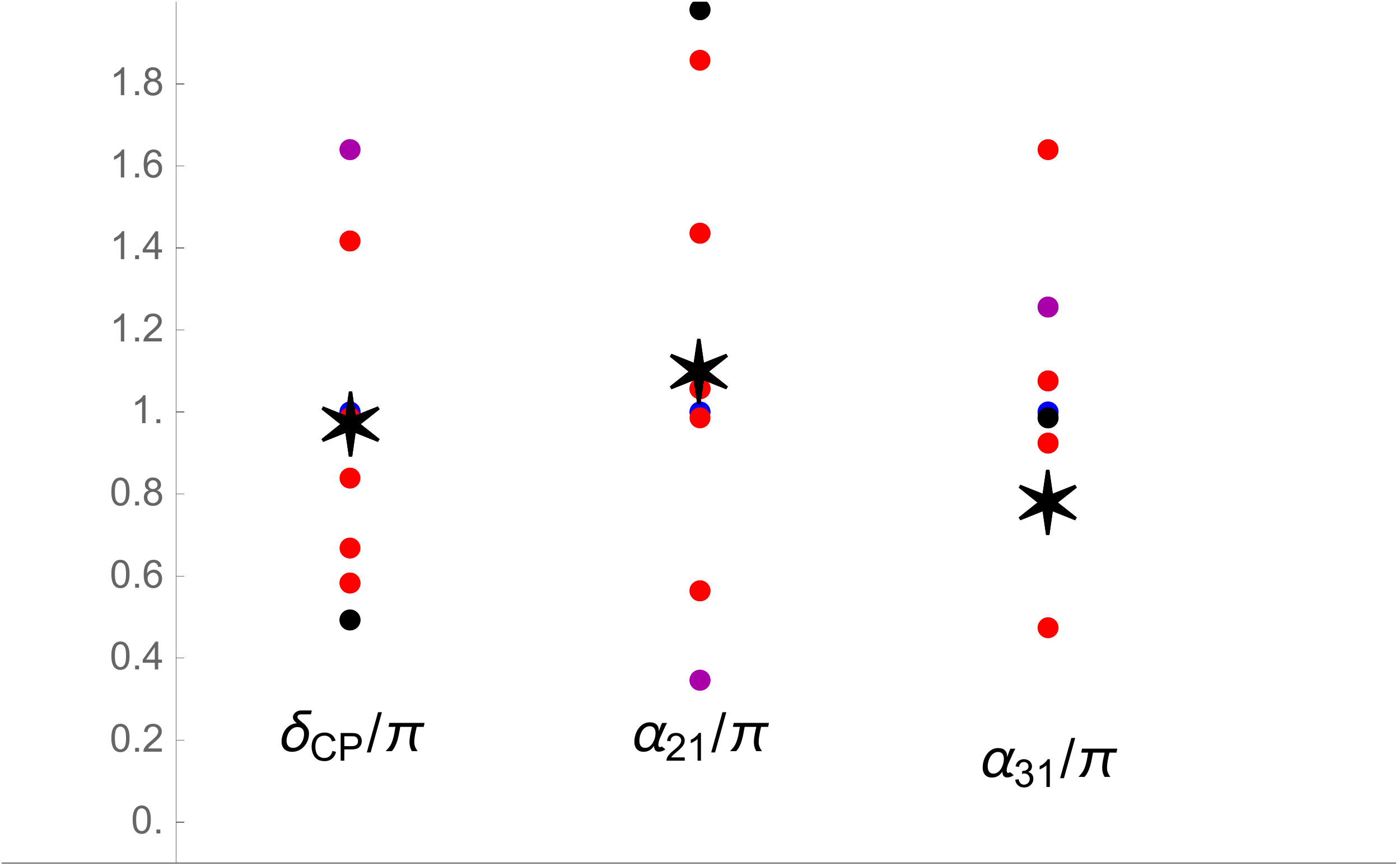}
\caption{Mass parameters and phases of 14 pairs of CP and modular invariant models featuring normal ordering and $|\tau-i|<0.25$ (see text). The full distributions of predictions
are displayed. The color code is identical to the one in fig. \ref{lines}. A star shows the average over the 14 models. CP violating phases refer to models where ${\tt Re}\tau>0$.}
\label{masses}
\end{figure}
In figs. \ref{angles} and \ref{masses} we show the predictions of a homogeneous set of $14$ pairs of CP invariant models. We discarded nine pairs by requiring $|\tau-i|<0.25$, two pairs not relying on the seesaw mechanism, one pair - the only one - predicting inverted ordering and a last pair, for which full data were not available. For some combinations of observables, the average over $13$ models of our sample is listed in eq. (\ref{ave})~\footnote{We have removed the model of ref. \cite{Wang:2021mkw}, denoted by black points in fig. \ref{masses}, for which $\rho_L$ is reducible.}.
The quoted error only illustrates the spread of the actual predictions for each combination and has no statistical meaning. All the dimensionless values are of the same order of magnitude and close to the average deviation of $\tau$ from the imaginary unit.
We will see that this behavior is a universal prediction of modular invariant models living close to $\tau=i$.
\begin{myalign}
\label{ave}
\begin{array}{lcl}
\vert \tau-i\vert=0.20\pm0.04&&|u|=|\dd\frac{\tau-i}{\tau+i}|=0.095\pm 0.015\\[10 pt]
\sum_i m_i=73.9\pm 4.6~{\rm meV}&&m_{\beta\beta}=5.5\pm4.2~{\rm meV}\\[10 pt]
\dd\frac{m_2+m_1}{2}=11.5\pm 2.2~{\rm meV}&&
\dd\frac{m_2+m_1}{2 m_3}=0.23\pm 0.04\\[10 pt]
2\dd\frac{m_2-m_1}{m_2+m_1}=0.34\pm 0.25&&
\left(\dd\frac{\Delta m^2_{sol}}{\Delta m^2_{atm}}\right)^{1/3}=0.310\pm 0.001\\[10 pt]
\sin\theta_{13}=0.148\pm 0.001&&
\left\vert \dd\frac{1}{\sqrt{2}}-\sin\theta_{12}  \right\vert=0.14\pm 0.01~.
\end{array}
\end{myalign}
\section{Modular invariance at critical points}
\label{micp}
In this Section, we extend and complete the discussion of ref. \cite{Feruglio:2022kea}. We explain why 
and in which sense at the fixed points $\otau =( i,\omega)$ of the fundamental domain ${\cal F}$, the theory
exhibit a universal behavior, independent of many details of the specific model. This behavior arises under
the only assumption that lepton doublets are assigned to an irreducible representation of the finite modular group.
The level, the modular weights and even the type of kinetic terms, which can have the most general form compatible with
modular invariance, do not affect a relevant part of the predictions. As compared to ref.~ \cite{Feruglio:2022kea},
we analyze in detail both the fixed points $\otau =( i,\omega)$ and we provide much more details of the derivation
of our results.
At $\otau =( i,\omega, i\infty)$ of the fundamental domain ${\cal F}$, both CP and the subgroup
$G_0=(\mathbb{Z}_4^S,\mathbb{Z}_2^{ST}\times\mathbb{Z}_2^{S^2},\mathbb{Z}^{T}\times\mathbb{Z}_2^{S^2})$
are preserved. However, $G_0$ is nonlinearly realized and it is preferable to move to a field basis $u(\tau)$ where
both CP and $G_0$ act linearly. We will choose a basis where $u(\tau_0)=0$, so that the group $G_0$ and CP are both
unbroken at the origin of the field space, $u=0$. This allows us to adopt $u$ as an order parameter for the breaking of 
$G_0$ and CP.
\subsection{A field redefinition}
By a field redefinition, we can move to a description where the group $G_0$ is linearly realized in the field
space. We define~\cite{Feruglio:2021dte,Novichkov:2021evw}:
\begin{myalign}
\label{fred1}
\begin{array}{lclcl}
\otau\ne i \infty:&&u\equiv\dd\frac{\tau-\otau}{~\tau-\otaus}&&
\Phi\equiv\left(1-u\right)^{k_\varphi} \varphi\\
\otau=i \infty:&&u\equiv e^{i\frac{2\pi}{N}\tau}&&\Phi\equiv\varphi~~~,
\end{array}
\end{myalign} 
where $N$ is the level of the construction. The group $G_0$ is generated by the elements $S^2$ and $\ogamma$, paired with $\otau$ in the following way: $(\ogamma,\otau) \in\{(S,i),(ST,\omega),(T, i\infty)\}$.
We show in table 2 the action of $G_0$ and CP on the new field variables.
\begin{myalign}
\nonumber
\begin{array}{lclclcl}
\hline
&&\ogamma&&S^2&& {\rm CP}\\
\hline
\hline
u&&e^{i\theta_0} u&&u&&\bar u\\
\hline
\Phi&&\Omega_\varphi(\ogamma)~\Phi&&\Omega_\varphi(S^2)~\Phi&&\bar \Phi\\
\hline
\end{array}
\end{myalign}
\vskip 0.1 cm
\noindent
{{\bf Table 2} Transformation properties of the fields $u$ and $\Phi$ under the group $G_0$ and CP. See the text for the definitions of the involved quantities.}
\vskip 0.4 cm
\noindent
The phase $\theta_0$ is given by:
\begin{myalign}
\theta_0=(\frac{1}{2},\frac{2}{3},\frac{1}{N})~2\pi~~~,
\end{myalign}
while the unitary matrices $\Omega_\varphi(\ogamma)$ and $\Omega_\varphi(S^2)$ are defined as
\begin{myalign}
\label{defOme}
\begin{array}{lccl}
\Omega_\varphi(\ogamma)\equiv j_0^{k_\varphi} \rho_\varphi(\ogamma)&&&j_0=[j(\ogamma,\otau)]^{-1}=(i,\omega,1)\\
\Omega_\varphi(S^2)\equiv (-1)^{k_\varphi} \rho_\varphi(S^2)~~~.&&&
\end{array}
\end{myalign}
The element $S^2$ of $SL(2,\mathbb{Z})$ commutes with all the other elements of the group. Thus the group
$G_0$, generated by $(\ogamma,S^2)$, is abelian.
If the representation $\rho_\varphi$ is irreducible, $\rho_\varphi(S^2)$ is proportional to the identity and
$\rho_\varphi(S^4)=\mathbb{1}$ implies 
$\Omega_\varphi(S^2)=\pm \mathbb{1}$.
The CP transformations of table 2 reproduce those of eq. (\ref{CP}), except when
the variable $u$ describes the fluctuations around the fixed point $\omega$.
In this case $u\to\bar u$ corresponds to $\tau_{CP}=-\bar\tau-1$, which becomes identical to that in eq. (\ref{CP})
after performing a modular transformation.
In conclusion, the full residual group arising from $G_0$ and CP is linearly realized on the new field variables.
At the same time, the variable $u$ vanishes at $\tau=\otau$ and, parametrizing the deviation from the fixed point,
provides a good order parameter for both $G_0$ and CP breaking.
\subsection{Lepton Lagrangian in the new field variables}
It is useful to analyze the theory expressed in the new variables.
We focus on a class of models that includes and extends those considered in Section \ref{snowplot}.
The Lagrangian for the lepton sector depends on the chiral multiplets $(\varphi=\varphi_{u,d},e^c,l)$.
\begin{myalign}
\label{starting1}
{\cal L}=\int d^2\theta d^2\bar\theta~ \sum_\varphi \varphi^\dagger {\cal K}_\varphi(\tau,\bar\tau)~ \varphi+
\left[\int d^2\theta~ {\cal W}(\tau,\varphi)+h.c.\right]~~~,
\end{myalign} 
\begin{myalign}
\label{starting2}
{\cal W}(\tau,\varphi)=-\dd\frac{1}{2\Lambda_L} (\varphi_u l)^T {\cal Y}(\tau) (\varphi_u l)-e^{cT} {\cal Y}_e(\tau) (\varphi_d l)~~~.
\end{myalign} 
An important feature of this Lagrangian is that the K\"ahler potential is not assumed to be minimal: ${\cal K}_\varphi(\tau,\bar\tau)$ are hermitian, positive definite
matrices in flavor space, with appropriate transformation properties to guarantee gauge and modular invariance.
In particular, they are not required to be flavor independent.
The functions ${\cal Y}(\tau)$ and ${\cal Y}_e(\tau)$ are linear combinations of holomorphic modular forms.
We move to the new field basis through the field redefinitions of eqs. (\ref{fred1}). In terms of the new variables 
$(\Phi=\Phi_{u,d},E^c,L)$ we have:
\begin{myalign}
\label{newL}
{\cal L}=\int d^2\theta d^2\bar\theta~ \sum_\Phi \Phi^\dagger K_\Phi(u,\bar u)~ \Phi+
\left[\int d^2\theta~ W(u,\Phi)+h.c.\right]~~~,
\end{myalign} 
\begin{myalign}
\label{newW}
W(u,\Phi)=-\dd\frac{1}{2\Lambda_L} (\Phi_u L)^T Y(u) (\Phi_u L)-E^{cT} Y_e(u) (\Phi_d L)~~~.
\end{myalign} 
We have defined:
\begin{myalign}
\begin{array}{lcl}
\otau= i \infty:&&\left\{
\begin{array}{l}
K_\Phi(u,\bar u)\equiv {\cal K}_\varphi(\tau(u),\bar\tau(u))\\
Y(u)\equiv {\cal Y}(\tau(u))\\
Y_e(u)\equiv {\cal Y}(\tau(u))
\end{array}\right.
\\
\otau\ne i \infty:&&\left\{
\begin{array}{l}
K_\Phi(u,\bar u)\equiv(1-\bar u)^{-k_\varphi}
{\cal K}_\varphi(\tau(u),\bar\tau(u))(1-u)^{-k_\varphi}\\
Y(u)\equiv(1-u)^{-2 k_u}(1-u)^{-k_l}{\cal Y}(\tau(u))(1-u)^{-k_l}\\
Y_e(u)\equiv(1-u)^{-k_d}(1-u)^{-k_{e^c}}{\cal Y}_e(\tau(u))(1-u)^{-k_l}
\end{array}\right.~~~.
\end{array}
\end{myalign} 
Under the combined $G_0$ and CP symmetries, the new field variables transform as indicated in table 3.
We also show how $K_\Phi(u,\bar u)$, $Y(u)$ and $Y_e(u)$ should transform to guarantee modular and CP
invariance of the Lagrangian.
\begin{myalign}
\nonumber
\begin{array}{lclclcl}
\hline
&&G_0&& {\rm CP}\\
\hline
\hline
\Phi_{u,d}&&\Omega_{u,d}~\Phi_{u,d}&&\bar \Phi_{u,d}\\
\hline
E^c&&\Omega_\varphi~E^c&&\bar E^c\\
\hline
L&&\Omega_\varphi~L&&\bar L\\
\hline
K_\Phi(u,\bar u)&&\Omega_\varphi ~K_\Phi(u,\bar u)~\Omega_\varphi^\dagger&&K_\Phi(u,\bar u)^* \\
\hline
Z_\Phi(u,\bar u)&&\Omega_\varphi ~Z_\Phi(u,\bar u)~\Omega_\varphi^\dagger&&Z_\Phi(u,\bar u)^* \\
\hline
Y(u)&&[\Omega_u^\dagger]^2~ \Omega_l^*~Y(u)~ \Omega_l^\dagger &&Y(u)^*\\
\hline
Y_e(u)&&\Omega_d^\dagger~\Omega_{e^c}^*~Y_e(u)~ \Omega_l^\dagger &&Y_e(u)^*\\[2pt]
\hline
\end{array}
\end{myalign}
\vskip 0.1 cm
\noindent
{{\bf Table 3} Transformation properties of the lepton supermultiplets and of the matrices $K_\Phi(u,\bar u)$, $Z_\Phi(u,\bar u)$, $Y(u)$, $Y_e(u)$ under the group $G_0$ and CP. See the text for the definitions of the involved quantities.}
\vskip 0.4 cm
\noindent
In the last four lines of table 3, the entries of the second and third columns should be read as follows:
$K_\Phi(g_0 u,g_0 \bar u)=\Omega_\varphi(g_0) ~K_\Phi(u,\bar u)~\Omega_\varphi^\dagger(g_0)$ for any element
$g_0$ of $G_0$ and $K_\Phi(\bar u,u)=K_\Phi(u,\bar u)^*$. 
The unitary matrices $\Omega_{u,d}$, $\Omega_{e^c}$ and $\Omega_l$ are defined as in eq. (\ref{defOme}),
when the group element is equal to $\ogamma$ or $S^2$. For any other element of $G_0$ these matrices are
uniquely determined by the group composition laws.
The Lagrangian of eqs. (\ref{newL}) and (\ref{newW}) has the same properties as the original one. In particular, it
is modular invariant and provides the same predictions as the Lagrangian in eqs. (\ref{starting1}) and (\ref{starting2}). In the new variables the
action of the elements $(\ogamma,S^2)$ and of the corresponding group $G_0$ is linear, which is particularly useful when working in the vicinity of $\otau$ (or $u=0$). At the same time the full group $SL(2,\mathbb{Z})$ restricts both kinetic terms and Yukawa couplings. 
Kinetic terms are made canonical by the transformation:
\begin{myalign}
\Phi\to Z_\Phi(u,\bar u)~ \Phi~~~~~~~~~~~~~K_\Phi(u,\bar u)={Z_\Phi}(u,\bar u)^{-1\dagger} {Z_\Phi}(u,\bar u)^{-1}~~~.
\end{myalign} 
The matrix $Z_\Phi(u,\bar u)$ is defined up to an arbitrary unitary matrix, which can be chosen to make
$Z_\Phi(u,\bar u)$ hermitian, a choice we adopt here. As shown in Appendix \ref{Z}, the matrix $Z_\Phi(u,\bar u)$ has the same transformation properties as $K_\Phi(u,\bar u)$, see table 3. The lepton mass matrices are 
\begin{myalign}
m_\nu(u,\bar u)=\dd\frac{v_u^2 Z_u(u,\bar u)^2}{\Lambda_L} Z_L(u,\bar u)^T Y(u)~ Z_L(u,\bar u)~~~,
\end{myalign} 
\begin{myalign}
m_e(u,\bar u)=v_d Z_d(u,\bar u)~ Z_{E^c}(u,\bar u)^T Y_e(u)~ Z_L(u,\bar u)~~~.
\end{myalign} 
From table 3, we can finally read how these matrices transform when the theory is invariant under $G_0$ and CP. We collect the results in table 4.
\begin{myalign}
\nonumber
\begin{array}{lclclcl}
\hline
&&G_0&& {\rm CP}\\
\hline
\hline
m_\nu(u,\bar u)&&\Omega^*~m_\nu(u,\bar u)~\Omega^\dagger&&m_\nu(u, \bar u)^*\\
\hline
m_\nu(u,\bar u)^{-1}&&\Omega~m_\nu(u,\bar u)^{-1}~\Omega^T&&m_\nu(u, \bar u)^{-1*}\\
\hline
m_e(u,\bar u)&&\Omega_c^*~m_e(u,\bar u)~\Omega^\dagger&&m_e(u,\bar u)^*\\
\hline
m_{\bar e e}(u,\bar u)&&\Omega~m_{\bar e e}(u,\bar u)~\Omega^\dagger&&[m_{\bar e e}(u,\bar u)]^* \\
\hline
\end{array}
\end{myalign}
\vskip 0.1 cm
\noindent
{{\bf Table 4} Transformation properties of the lepton mass matrices under the group $G_0$ and CP. We have defined:
$m_{\bar e e}(u,\bar u)\equiv m_e(u,\bar u)^\dagger m_e(u,\bar u)$.}
\vskip 0.4 cm
\noindent
The unitary matrices in table 4 read:
\begin{myalign}
\label{trans2}
\Omega\equiv \Omega_u\Omega_l~~~,~~~~~~\Omega_c\equiv \Omega_u^*\Omega_d\Omega_{e^c}~~~.
\end{myalign}
If the neutrino mass matrix arises from the seesaw mechanism,
it may occur that $m_\nu(0,0)$ is singular~\footnote{That is the limit of $m_\nu(u,\bar u)$ when $u$ goes to zero does not exist or is infinite.}. In such a case it is convenient to enforce the transformations on the inverse $[m_\nu(u,\bar u)]^{-1}$, also reported in table 4. When the theory is close to a fixed point $\otau$ the physical fermion masses, fully accounting for a possible non-holomorphic dependence coming from the K\"ahler potential, transform in a simple way under the residual symmetry.
Table 4 can be used to get the most general parametrization of $m_\nu(u,\bar u)$ and $m_{\bar ee}(u,\bar u)$
in the vicinity of $\otau$.
\subsection{Expansion around a fixed point}
We now consider the implications of the residual symmetries when $\tau$ approaches the fixed point $\otau$.
We assume that the correct vacuum is described by a point $u$ close to zero and we analyze the theory by performing a power expansion, truncated to the first few terms.
Notice that, by retaining only a few terms, we cannot enforce any more the powerful constraint
arising from the full modular group. For example, in the vicinity of $\tau_0=i$ the $T$ transformation $\tau\to \tau+1$ evaluated in terms of $u$ reads:
\begin{myalign}
u\xrightarrow{T} \dd\frac{1-(1-2i) u}{(1+2i)-u}=\frac{1-2i}{5}+\frac{4}{25}(3+4 i)u+...~~~.
\end{myalign}
This transformation is nonlinear and spoils a truncated power expansion. 
For this reason, we only analyze the constraints coming from the invariance under $G_0$ and CP.
We expand the quantities of interests, $m_\nu(u,\bar u)$ and $m_{\bar e e}(u,\bar u)$,
in powers of $u$ and $\bar u$ around $u=0$~\footnote{When $\otau=i \infty$,
the minimal K\"ahler potential reads $K_\Phi(u,\bar u)=[-i(\tau-\bar\tau)]^{-k_\varphi}=[-\frac{N}{2\pi}\log(u \bar u)]^{-k_\varphi}$. We absorb the potentially large, $G_0$ and CP invariant terms depending on $\log(u \bar u)$
in the first term of the expansion, $m_\nu^0$ and $m_{\bar e e}^0$.}:
\begin{myalign}
\label{exp}
\begin{array}{lll}
m_\nu(u,\bar u)&=&m_\nu^0+m_\nu^{1,0} u+m_\nu^{0,1} \bar u+m_\nu^{2,0} u^2+m_\nu^{0,2} {\bar u}^2+m_\nu^{1,1}u\bar u+...\\
m_{\bar e e}(u,\bar u)&=&m_{\bar e e}^0+m_{\bar e e}^{1,0} u+m_{\bar e e}^{0,1} \bar u+m_{\bar e e}^{2,0} u^2+m_{\bar e e}^{0,2} {\bar u}^2+m_{\bar e e}^{1,1}u\bar u+...
\end{array}
\end{myalign} 
where all the coefficients of the expansions are matrices in flavor space.
To respect CP invariance, from table 4 we see that all the coefficients $m_\nu^{p,q}$ and $m_{\bar e e}^{p,q}$ should be real. Since the residual symmetry $G_0$ at the fixed point is abelian, we choose a basis where the unitary matrices 
$\Omega(\ogamma)$ and $\Omega(S^2)$ are simultaneously diagonal, with elements
\begin{myalign}
\Omega(\ogamma)={\tt diag}(e^{\dd i \theta_1},e^{\dd i \theta_2},e^{\dd i \theta_3})~~~,~~~~~~~
\Omega(S^2)={\tt diag}((-1)^{\sigma_1},(-1)^{\sigma_2},(-1)^{\sigma_3})~~~.
\end{myalign} 
The charges $\sigma_i$ are integers, while $\theta_i/2\pi$ are rational numbers.
The transformation laws of table 4 determine the nonvanishing matrix elements of
$m_\nu^{p,q}$, $m_{\bar e e}^{p,q}$:
\begin{myalign}
\begin{array}{lcl}
\left(m_\nu^{p,q}\right)_{i,j}\ne 0&&(p-q)~\theta_0=-\theta_i-\theta_j~~~({\rm mod}~ 2\pi)~\&~~~~~\sigma_i+\sigma_j=0~~~({\rm mod}~ 2)\\
\left(m_{\bar e e}^{p,q}\right)_{i,j}\ne 0&&(p-q)~\theta_0=+\theta_i-\theta_j~~~({\rm mod}~ 2\pi)~\&~~~~~\sigma_i+\sigma_j=0~~~({\rm mod}~ 2)~~~.
\end{array}
\end{myalign}  
We recognize the familiar matching conditions of the Froggatt-Nielsen formalism, realized within a discrete abelian symmetry $G_0$, spontaneously broken by small order parameters $u$ and $\bar u$~\footnote{The $\mathbb{Z}_2$ component of $G_0$ generated by $S^2$ remains unbroken.}.
\subsection{Irreducible $\rho_l$}
There are models where
the lepton doublets $l$ fall in a reducible representation. Moreover in string theory compactifications, matter multiplets
often come in reducible representations of the finite modular groups. In a bottom-up approach, choosing an irreducible representation $\rho_l$ has the advantage of minimizing the number of free parameters needed to describe
$m_\nu(u,\bar u)$. For this reason this is the most frequent assignment adopted in model building and in the rest of this paper we will analyze this important case. If $\rho_l$ is irreducible, from eq. (\ref{defOme}) and (\ref{trans2})
we see that $\Omega(S^2)$ coincides with the identity, up to an overall sign. The condition $\sigma_i+\sigma_j=0$ ({\rm mod}~ 2) is always satisfied
and there is no constraint from the invariance of the theory under the $S^2$ element of the group $G_0$.

Concerning the element $\ogamma$, eq. (\ref{defOme}) and (\ref{trans2}) show that the unitary matrix constraining the pattern of both $m_\nu(u,\bar u)$ and $m_e(u,\bar u)^\dagger m_e(u,\bar u)$ is the combination
\begin{myalign}
\Omega(\ogamma)=j_0^{k_u+k_l} \rho_u(\ogamma)\rho_l(\ogamma)~~~.
\end{myalign} 
Such a matrix depends on the (integer) weights $k_u$ and $k_l$ as well as on the representations 
$\rho_u(\ogamma)$ (1-dimensional) and $\rho_l(\ogamma)$ (3-dimensional). In turn, these representations
depend on the level $N$. 
It is remarkable that, under the only assumption that $\rho_l(\ogamma)$ is irreducible,
which is commonly adopted in most concrete models, the combination $\Omega(\ogamma)$ $(\ogamma=S,ST)$
is completely fixed up to an overall phase factor:
\begin{myalign}
\label{Omega}
\Omega(S)=i^{k_S}{\tt diag}(1,-1,-1)~~~,~~~~~~~~~~~\Omega(ST)={\tt diag}(1,\omega,\omega^2)~~~,
\end{myalign} 
where $k_S$ is an integer depending on $k_u$ and $k_l$, as well as on $N$ and on the specific
3-dimensional and 1-dimensional representation $\rho_l(\ogamma)$ and $\rho_u(\ogamma)$ of $SL(2,\mathbb{Z}_N)$. Of course, $\Omega(S)$ and $\Omega(ST)$
cannot be simultaneously diagonal and the above relations are valid in two distinct bases. 
The result in eq. (\ref{Omega}) can be proved by directly inspecting all 3-dimensional irreducible representations of
of the groups $SL(2,\mathbb{Z}_N)$ that, for $N$ a power of 2 or a prime, exist only for $N=3,4,5,7,8,16$~\cite{nobs}. 
There are 33 inequivalent such representations: $1,2,2,4,8,16$ for $N=3,4,5,7,8,16$, respectively~\cite{Eholzer:1994th}.
They are collected in Appendix \ref{Ehol}. For these levels, we can also construct the
1-dimensional representations $\rho_u(\ogamma)$ of $SL(2,\mathbb{Z}_N)$, which are also
displayed in Appendix \ref{Ehol}. Making use of the prime factorization for a generic $N$, a straightforward computation leads to eq. (\ref{Omega}).
It is also possible to prove the first equality in eq. (13) without relying on explicit representations~\footnote{G.-J. Ding, private communication.},
also covering the case of other extensions of $SL(2,\mathbb{Z})$, such as the metaplectic group \cite{Liu:2020msy,Almumin:2021fbk}. In the latter case $k_S$ is multiple of an half-integer. The phenomenological interesting cases arise when $k_S$ is an integer.
A similar analysis shows that for $\Omega(T)$ the dependence on $N$ cannot be factored out as in 
eq. (\ref{Omega}). In this case, the eigenvalues of $\Omega(T)$ and their ratios depend explicitly on $N$.
\section{Lepton mass matrices for irreducible $\rho_l$}
\label{lmm}
Under the assumption that $\rho_l$ is an irreducible triplet of $SL(2,\mathbb{Z}_N)$ and exploiting the explicit form of $\Omega(\ogamma)$ in eq. (\ref{Omega}), here we 
derive the expansion in eq. (\ref{exp}) for the matrices $m_\nu(u,\bar u)$ and $m_{\bar e e}(u,\bar u)$,
to first order in $u$ and $\bar u$. We set
\begin{myalign}
u\equiv x e^{i\theta}~~~~~~~~~~~~~~~~(x>0~,~~~0\le \theta<2 \pi)~~~.
\end{myalign} 
We also discuss neutrino masses, lepton mixing angles and phases for each case. We provide
more details in Appendix \ref{diagon} and fewer details in the summary of Section \ref{summy}.
We focus on $\otau=i,\omega$, for which the results do not depend on the level $N$.
\subsection{$\ogamma=S$ $\&$ $\otau=i$}
We start by considering the vicinity to the fixed point $\otau=i$, where the symmetry group is $\mathbb{Z}_4^S$
spontaneously broken by $u$, transforming as $u\to -u$ under the generator $\gamma_0=S$. From table 4 and $\Omega(S)$ in eq. (\ref{Omega}), we find the following pattern for $m_{\bar e e}(u,\bar u)$:
\begin{myalign}
m_{\bar e e}(u,\bar u)=m_{0e}^2
\left(
\begin{array}{ccc}
{y}^0_{11}&{y}^{10}_{12}u+{y}^{01}_{12}\bar u&{y}^{10}_{13}u+{y}^{01}_{13}\bar u\\
{y}^{10}_{12}\bar u+{y}^{01}_{12}u&{y}^0_{22}&{y}^0_{23}\\
{y}^{10}_{13}\bar u+{y}^{01}_{13} u&{y}^{0}_{23}&{y}^0_{33}\\
\end{array}
\right)+...
\end{myalign} 
where dots denote higher-order terms and $m_{0e}^2$, $y_{ij}^{0}$, $y_{ij}^{10}$ and $y_{ij}^{01}$ are real to satisfy CP invariance.
We move to the basis where $m_{\bar e e}(u,\bar u)$ is diagonal:
\begin{myalign}
U_e^\dagger m_{\bar e e}(u,\bar u) U_e={\tt diag} [m_{\bar e e}(u,\bar u)]~~~.
\end{myalign} 
From Appendix \ref{diagon} we see that, up to a permutation matrix $P$ related to the ordering of the charged lepton masses, $U_e$ has the pattern:
\begin{myalign}
\label{ue}
U_e=\left(
\begin{array}{ccc}
1&x&x\\
x&1&1\\
x&1&1
\end{array}
\right)~~~.
\end{myalign}
Here unknown independent coefficients of order one for each entry have been omitted.
The neutrino mass matrices depend on the integer $k_S$. To first order in $u$ and $\bar u$ we get:
\begin{itemize}
\item[$\bullet$] {\bf $k_S$ even}
\begin{myalign}
\label{casea}
m_\nu(u,\bar u)=m_{0\nu}
\left(
\begin{array}{ccc}
x^0_{11}&x^{10}_{12}u+x^{01}_{12}\bar u&x^{10}_{13}u+x^{01}_{13}\bar u\\
\cdot&x^0_{22}&x^0_{23}\\
\cdot&\cdot&x^0_{33}\\
\end{array}
\right)+...
\end{myalign} 

\item[$\bullet$] $k_S$ odd
\begin{myalign}
\label{caseb}
m_\nu(u,\bar u)=m_{0\nu}
\left(
\begin{array}{ccc}
x^{10}_{11}u+x^{01}_{11}\bar u&x^0_{12}&x^0_{13}\\
\cdot&x^{10}_{22}u+x^{01}_{22}\bar u&x^{10}_{23}u+x^{01}_{23}\bar u\\
\cdot&\cdot&x^{10}_{33}u+x^{01}_{33}\bar u\\
\end{array}
\right)+...
\end{myalign} 
\end{itemize}
All the parameters except $u$ and $\bar u$ are real. The dimensionless ones are assumed to be of order one, 
except $x=|u|$, which is expected to be smaller than one and to provide the expansion parameter.
If $m_\nu(u,\bar u)$ is singular at $u=0$,
we consider the expansion of $[m_\nu(u,\bar u)]^{-1}$, which is identical to the one given above, with the replacement
$m_\nu(u,\bar u)\to [m_\nu(u,\bar u)]^{-1}$ and $m_{0\nu} \to m_{0\nu}^{-1}$. From eq. (\ref{ue}) we see that
in the basis where $m_{\bar e e}(u,\bar u)$ is diagonal, up to a common permutation matrix of rows and columns and up to higher-order terms in the expansion, the neutrino mass matrix maintains the same pattern shown in eqs. (\ref{casea}) and (\ref{caseb}).
To first order in $x$, the effect of the basis change can be absorbed in the coefficients $x_{ij}^{0}$, $x_{ij}^{10}$ and $x_{ij}^{01}$.
The same conclusion holds for the inverse $m_\nu(u,\bar u)^{-1}$ and, without losing generality, we discuss the neutrino mass spectrum, mixing angles and phases by directly analyzing the matrices (\ref{casea}) and (\ref{caseb}). Here we summarize the results, deferring more details to Appendix \ref{diagon}.
\subsubsection{$k_S$ even}
If $m_\nu(0,0)$ is regular, we find:
\begin{myalign}
U_\nu^T m_\nu(u,\bar u) U_\nu={\tt diag}(\tilde m_1,\tilde m_2,\tilde m_3)~~~,
\end{myalign}
where, to first order in $x$, the eigenvalues read:
\begin{myalign}
\begin{array}{l}
\tilde m_1=m_{0\nu} x_{11}^0\\
\tilde m_{2,3}=m_{0\nu} \left[\dd\frac{1}{2}(x_{22}^0+x_{33}^0)\pm\frac{1}{2}\sqrt{(x_{22}^0-x_{33}^0)^2+4 (x_{23}^0)^2}\right]~~~.
\end{array}
\end{myalign}
Up to a correct ordering, they coincide with the neutrino masses $m_{1,2,3}$.
The lepton mixing matrix is $P_e U_\nu P_\nu$, where $P_{e,\nu}$ are permutation matrices
accounting for the ordering of the mass eigenstates in the charged lepton and neutrino sectors and:
\begin{myalign}
U_\nu=
\left(
\begin{array}{ccc}
1&x_a x & x_b x\\
-(cx_a^*-s x_b^*) x&c&-s\\
-(sx_a^*+cx_b^*) x&s&c
\end{array}
\right)~~~.
\end{myalign}
where the quantities $c,s,x_a,x_b$ depend on the coefficients $x_{ij}^{0}$, $x_{ij}^{10}$, $x_{ij}^{01}$ and on the phase of $u$, but
are independent of $x$. They are explicitly given in Appendix \ref{diagon}.
By varying the parameters $x_{11}^0$, $x_{22}^0$, $x_{33}^0$, $x_{23}^0$, both ordering of neutrino masses can be accommodated. Barring cancellations, the ratio $\Delta m^2_{sol}/\Delta m^2_{atm}$ is expected to be of order one, while experimentally it is close to 0.03. When $P_{e,\nu}=\mathbb{1}$, to first order in $x$ we find:
\begin{myalign}
\sin\theta_{12}=|x_a| x~~~,~~~~~~~\sin\theta_{13}=|x_b| x~~~,~~~~~~~\sin\theta_{23}=s~~~. 
\end{myalign}
To match the experimental data we would need
$|x_b/x_a|\approx 0.27$. This suppression might originate by the approximate scaling $|x_a/x_b|\propto (\tilde m_3^2-\tilde m_1^2)/(\tilde m_2^2-\tilde m_1^2)$. Once the parameters $x_{11}^0$, $x_{22}^0$, $x_{33}^0$, $x_{23}^0$ have been adjusted to match $\Delta m^2_{sol}/\Delta m^2_{atm}$, they naturally enhance $|x_a/x_b|$.
To keep $\sin\theta_{13}={\cal O}(x)$, we can also allow $P_e=P_{23}$, the permutation between second and third rows.
This produces the exchange $\sin\theta_{23}\leftrightarrow \cos\theta_{23}$. Similarly, taking $P_\nu=P_{12}$
causes the exchange $\sin\theta_{12}\leftrightarrow \cos\theta_{12}$. In this case data requires 
$|x_b/x_a|<0.2$. Taking $P_\nu=P_{23}$
causes the exchange $\sin\theta_{12}\leftrightarrow \sin\theta_{13}$. Another set of permutations leaving
$\sin\theta_{13}={\cal O}(x)$ is $P_\nu=P_{13}$ combined with $P_e=P_{13}$ (or $P_e=P_{12}$).
In this case we end up with $\sin\theta_{12}=s$ and $\sin\theta_{23}={\cal O}(x)$ (or $\cos\theta_{23}={\cal O}(x)$),
with the result that the atmospheric angle is very far from the maximal one.
If in eq. (\ref{casea}) we make the replacement
$m_\nu(u,\bar u)\to [m_\nu(u,\bar u)]^{-1}$ and $m_{0\nu} \to m_{0\nu}^{-1}$, we see that $m_\nu(0,0)$ cannot be singular, except for the special cases where either $x^0_{11}$ or $(x^0_{23})^2-x^0_{22}x^0_{33}$ vanish.

\subsubsection{$k_S$ odd, $m_\nu(0,0)$ regular}
The neutrino masses read:
\begin{myalign} 
\begin{array}{l} 
\begin{array}{l}
m_1=m_{0\nu}h\left(1-\dd\frac{\sqrt{(k+\bar k)^2-(l-\bar l)^2}}{2h}x\right)\\
m_2=m_{0\nu}h\left(1+\dd\frac{\sqrt{(k+\bar k)^2-(l-\bar l)^2}}{2h}x\right)\\
m_3=m_{0\nu} \vert q\vert x
\end{array}\\
\\
\begin{array}{l}
\Delta m^2_{sol}=m_2^2-m_1^2= 2 m_{0\nu}^2 h\sqrt{(k+\bar k)^2-(l-\bar l)^2}x\\[8 pt]
\Delta m^2_{atm}=-m_3^2+(m_2^2+m_1^2)/2= m_{0\nu}^2 h^2\left(1+{\cal O}(x^2)\right)\\[8 pt]
r=\dd\frac{\Delta m^2_{sol}}{\Delta m^2_{atm}}=2 \dd\frac{\sqrt{(k+\bar k)^2-(l-\bar l)^2}}{h}x~~~.
\end{array}
\end{array}
\end{myalign}
The mixing matrix $U_{PMNS}$ is given in Appendix \ref{diagon}. From it we can deduce
the mixing angles:
\begin{myalign}  
\begin{array}{l}
\sin^2\theta_{12}=\frac{1}{2}\left(1-\dd\frac{l\bar k+\bar l k}{h\sqrt{(k+\bar k)^2-(l-\bar l)^2}}x\right)\\[8 pt]
\sin^2\theta_{13}=2 \dd\frac{|n|^2}{h^2} x^2\\[8 pt]
\sin^2\theta_{23}=\dd\frac{(x^0_{13})^2}{(x^0_{12})^2+(x^0_{13})^2}(1+{\cal O}(x))~~~,
\end{array}
\end{myalign}
and the CP-violating phases:
\begin{myalign}  
\begin{array}{l}
\delta_{CP}=\arg\left[\dd\frac{(c_\nu-is_\nu)^2 x^0_{12} x^0_{13}}{n}\right]+{\cal O}(x^2)\\
\alpha_{21}=\pi+{\cal O}(x)\\
\alpha_{31}=\pi-\arg(q)+\arg\left[(c_\nu-is_\nu)^2\right]+{\cal O}(x)~~~.
\end{array}
\end{myalign}
The combination $m_{\beta\beta}$ relevant to neutrino-less double beta decay is:
\begin{myalign}  
m_{\beta\beta}=m^0_\nu |x_{11}| x~~~.
\end{myalign}
The parameters $h$, $k$, $l$, $n$, $q$, $c_\nu$, $s_\nu$, $x_{11}$ are dimensionless combinations of
the coefficients $x^0_{ij}$, $x^{10}_{ij}$, $x^{01}_{ij}$ and the phase of $u$. They do not depend on $x$ and are explicitly
given in Appendix \ref{diagon}. Barring accidental cancellations they are expected to be of order one.
An inverted ordering of neutrino masses is predicted. To reproduce the observed values of $\sin^2\theta_{13}$ and $\sin^2\theta_{12}$, 
$x$ should be close to 0.15. This is in tension with the value of $x$ required by
$r=\Delta m^2_{sol}/\Delta m^2_{atm}$, experimentally close to 0.03. If we choose $x\approx 0.15$,
an additional suppression by about an order of magnitude from the combination $\sqrt{(k+\bar k)^2-(l-\bar l)^2}/h$ should be invoked.

The only acceptable permutation to keep $(U_\nu)_{e3}={\cal O}(x)$ is the one between the second and the third rows of $U_\nu$: $P=P_{23}$. All observable
remain unchanged but $\sin^2\theta_{23}\to 1-\sin^2\theta_{23}$ and $\delta_{CP}\to \pi+\delta_{CP}$ $\mod(2\pi)$.
\vskip 0.5 cm
\subsubsection{$k_S$ odd, $m_\nu(0,0)$ singular}
\label{good}
We get the following neutrino masses:
\begin{myalign}  
\label{masssing}
\begin{array}{l}
\begin{array}{l}
m_1=\dd\frac{m_{0\nu}}{h}\left(1-\dd\frac{\sqrt{(k+\bar k)^2-(l-\bar l)^2}}{2h}x\right)\\
m_2=\dd\frac{m_{0\nu}}{h}\left(1+\dd\frac{\sqrt{(k+\bar k)^2-(l-\bar l)^2}}{2h}x\right)\\
m_3=\dd\frac{m_{0\nu}}{|q| x}
\end{array}\\
\\
\begin{array}{l}
\Delta m^2_{sol}=m_2^2-m_1^2= 2 m_{0\nu}^2 \dd\frac{\sqrt{(k+\bar k)^2-(l-\bar l)^2}}{h^3}x\\[8 pt]
\Delta m^2_{atm}=m_3^2-(m_2^2+m_1^2)/2= \dd\frac{m_{0\nu}^2}{|q|^2 x^2}\left(1+{\cal O}(x^2)\right)\\[8 pt]
r=\dd\frac{\Delta m^2_{sol}}{\Delta m^2_{atm}}=2 \dd\frac{|q|^2\sqrt{(k+\bar k)^2-(l-\bar l)^2}}{h^3}x^3~~~.
\end{array}
\end{array}
\end{myalign}
The mixing matrix $U_{PMNS}$ is shown in Appendix \ref{diagon}. From it we derive
the mixing angles:
\begin{myalign}  
\label{anglessing}
\begin{array}{l}
\sin^2\theta_{12}=\frac{1}{2}\left(1+\dd\frac{l\bar k+\bar l k}{h\sqrt{(k+\bar k)^2-(l-\bar l)^2}}x\right)\\[8 pt]
\sin^2\theta_{13}=2 \dd\frac{|n|^2}{h^2} x^2\\[8 pt]
\sin^2\theta_{23}=\dd\frac{(x^0_{13})^2}{(x^0_{12})^2+(x^0_{13})^2}(1+{\cal O}(x))~~~,
\end{array}
\end{myalign}
and the CP-violating phases:
\begin{myalign}  
\begin{array}{l}
\delta_{CP}=\pi-\arg\left[\dd\frac{(c_\nu-is_\nu)^2 x^0_{12} x^0_{13}}{n}\right]+{\cal O}(x^2)\\
\alpha_{21}=\pi+{\cal O}(x)\\
\alpha_{31}=\arg(q)-\arg\left[(c_\nu-is_\nu)^2\right]+{\cal O}(x)~~~.
\end{array}
\end{myalign}
The quantity $m_{\beta\beta}$ is given by:
\begin{myalign}  
m_{\beta\beta}=m^0_\nu\dd\frac{\left\vert x_{23}^2-x_{22}x_{33}\right\vert}{ h^2|q|} x~~~.
\end{myalign}
where $x_{23}$, $x_{22}$, $x_{33}$ are complex coefficients of order one defined in Appendix \ref{diagon}.
We get a normal order of the neutrino mass spectrum. Similarly to the case of regular $m_\nu(0,0)$, to reproduce the experimental values of $\sin^2\theta_{13}$ and $\sin^2\theta_{12}$, we need $x\approx 0.15$. Unlike the previous case, now this value can adequately suppress 
$r=\Delta m^2_{sol}/\Delta m^2_{atm}$.
Of all the possibilities that can be realized at $\tau=i$, this is the only one matching all the experimental results, 
at the level of orders of magnitude, without any important adjustment of the order-one parameters. To allow a direct comparison with the averages of eqs. (\ref{ave}) we also
list an additional set of predictions. 
\begin{myalign}  
\label{othersing}
\begin{array}{l}
\dd\frac{m_1+m_2}{2}=\dd\frac{m_{0\nu}}{h}\\[8 pt]
\dd\frac{m_1+m_2}{2 m_3}=\frac{|q|}{h}x\\[8 pt]
2\dd\frac{m_2-m_1}{m_2+m_1}=\frac{\sqrt{(k+\bar k)^2-(l-\bar l)^2}}{h}x~~~.
\end{array}
\end{myalign}
Making use of eqs. (\ref{masssing}), (\ref{anglessing}) and (\ref{othersing}), we see that the averages of eqs. (\ref{ave}) 
can be reproduced by choosing
\begin{myalign} 
\label{approx} 
\begin{array}{lcl}
m_{0\nu}/h=11.5&&\\
x\approx 0.1&&\dd\frac{|q|}{h}\approx 2.3\\[8pt]
\dd\frac{\sqrt{(k+\bar k)^2-(l-\bar l)^2}}{h}\approx 3.4&&
\dd\frac{\sqrt{2}|n|}{h}\approx 1.5\\[8pt]
\dd\frac{|l\bar k+\bar l k|}{2\sqrt{2}h \sqrt{(k+\bar k)^2-(l-\bar l)^2}}\approx 1.4&&
\dd\frac{\left\vert x_{23}^2-x_{22}x_{33}\right\vert}{h |q|}\approx 4.8~~~.
\end{array}
\end{myalign}
Indeed $k_S$ is odd in all but one of the $14$ pairs of CP invariant models
discussed in Section \ref{snowplot}.
We conclude that all the dimensionless quantities in eq. (\ref{ave}) scale linearly with $x$,
with proportionality coefficients of order one. The only relatively
large coefficient is that of the combination controlling $m_{\beta\beta}$,
whose value in eq. (\ref{ave}) has the largest relative fluctuation.

Finally, concerning the effect of possible permutations, the permutation matrix $P$ from the charged lepton sector changes $U_{PMNS}$ into $PU_{PMNS}$. Since $(U_{PMNS})_{e3}={\cal O}(x)$, the only acceptable permutation is the one between the second and the third rows of $U_{PMNS}$: $P=P_{23}$. All observables remain unchanged but $\sin^2\theta_{23}\to 1-\sin^2\theta_{23}$ and $\delta_{CP}\to \pi+\delta_{CP}$ $\mod(2\pi)$.
\subsection{$\ogamma=ST$ $\&$ $\otau=\omega$}
We now move to the fixed point $\otau=\omega$, where the symmetry group is $\mathbb{Z}_3^{ST}\times \mathbb{Z}_2^{S^2}$ spontaneously broken by $u$, transforming as $u\to \omega^2 u$ under the generator $\gamma_0=ST$
and invariant under the action of $S^2$. From table 4 and $\Omega(ST)$ in eq. (\ref{Omega}), we find the following pattern for $m_{\bar e e}(u,\bar u)$:
\begin{myalign}
m_{\bar e e}(u,\bar u)=m_{0e}^2~
\left(
\begin{array}{ccc}
{y}^0_{11}&{y}^{10}_{12}u&{y}^{01}_{13}\bar u\\
{y}^{10}_{12}\bar u&{y}^0_{22}&{y}^{10}_{23} u\\
{y}^{01}_{13}u&{y}^{10}_{23} \bar u&{y}^0_{33}\\
\end{array}
\right)+...
\end{myalign} 
where dots denote higher-order terms and $m_{0e}^2$, $y_{ii}^{0}$, $y_{ij}^{10}$ and $y_{ij}^{01}$ are real to satisfy CP invariance. We move to the basis where $m_{\bar e e}(u,\bar u)$ is diagonal:
\begin{myalign}
U_e^\dagger m_{\bar e e}(u,\bar u) U_e={\tt diag} [m_{\bar e e}(u,\bar u)]~~~.
\end{myalign} 
From Appendix \ref{diagon} we see that, up to a permutation matrix $P_e$ related to the ordering of the charged lepton masses, $U_e$ is given by:
\begin{myalign}
\label{ueome}
 U_e=
\left(
\begin{array}{ccc}
1&\frac{{y}^{10}_{12}u}{y^0_{22}-y^0_{11}} & \frac{{y}^{01}_{13}\bar u}{y^0_{33}-y^0_{11}}\\
-\frac{{y}^{10}_{12}\bar u}{y^0_{22}-y^0_{11}}&1&\frac{{y}^{10}_{23}u}{y^0_{33}-y^0_{22}}\\
-\frac{{y}^{01}_{13}u}{y^0_{33}-y^0_{11}}&-\frac{{y}^{10}_{23}\bar u}{y^0_{33}-y^0_{22}}&1
\end{array}
\right)+...
\end{myalign} 
As for the neutrino mass matrix, to first order in $u$ and $\bar u$ we get:
\begin{myalign}
\label{caseome}
m_\nu(u,\bar u)=m_{0\nu}
\left(
\begin{array}{ccc}
x^0_{11}&x^{10}_{12}u&x^{01}_{13}\bar u\\
\cdot&x^{01}_{22}\bar u&x^0_{23}\\
\cdot&\cdot&x^{10}_{33}u\\
\end{array}
\right)+...
\end{myalign} 
where all the parameters except $u$ and $\bar u$ are real. All the parameters $x_{ii}^{0}$, $x_{ij}^{10}$ and $x_{ij}^{01}$ are expected to be of order one. A similar expansion holds for $m_\nu(u,\bar u)^{-1}$ but, excluding the case where either $x^0_{11}$ and/or $x^0_{23}$ vanish, $m_\nu(0,0)$ cannot be singular.
Using eq. (\ref{ueome}), we see that
in the basis where $m_{\bar e e}(u,\bar u)$ is diagonal, up to a common permutation matrix of rows and columns and up to higher-order terms in the expansion, the neutrino mass matrix maintains the same pattern shown in eq. (\ref{caseome}).
To first order in $x$, the effect of the basis change can be absorbed in the coefficients $x_{ii}^{0}$, $x_{ij}^{10}$ and $x_{ij}^{01}$.
Thus, without losing generality, we discuss the neutrino mass spectrum, mixing angles and phases by directly analyzing the matrix (\ref{caseome}). Here we report the main results, more details can be found in Appendix \ref{diagon}.
The unitary matrix that diagonalizes $m_\nu(u,\bar u)$ is:
\begin{myalign}  
\label{unu}
U_\nu=U~K_\nu~~~,
\end{myalign}
where $K_\nu$ is a diagonal unitary matrix and 
\begin{myalign}
U=
\left(
\begin{array}{ccc}
1&a(\alpha-\beta)u+b(\alpha+\beta)\bar u&-a(\alpha+\bar\beta)u+b(\alpha-\bar\beta)\bar u\\
-\sqrt{2} a \bar u&\frac{\alpha-\beta}{\sqrt{2}}&-\frac{\alpha+\bar\beta}{\sqrt{2}}\\
-\sqrt{2} b u&\frac{\alpha+\beta}{\sqrt{2}}&\frac{\alpha-\bar\beta}{\sqrt{2}}
\end{array}
\right)+...
\end{myalign}
Here $a$ and $b$ are numbers of order one, depending on $x^0_{ii}$, $x^{10}_{ij}$ and $x^{01}_{ij}$, while $\alpha$ and $\beta$ satisfy $|\alpha|^2+|\beta|^2=1$.
These coefficients are given in Appendix \ref{diagon}.
We find:
\begin{myalign}  
U_\nu^T m_\nu(u,\bar u) U_\nu={\tt diag}(\tilde m_1,\tilde m_2,\tilde m_3)~~~,
\end{myalign}
where the eigenvalues read:
\begin{myalign}  
\begin{array}{l}
\tilde m_1=m_{0\nu}~x^0_{11}\\
\tilde m_2=m_{0\nu}~|x^0_{23}|\left(1+\dd\frac{x^{10}_{33}+x^{01}_{22}}{2x^0_{23}}x\right)\\
\tilde m_3=m_{0\nu}~|x^0_{23}|\left(1-\dd\frac{x^{10}_{33}+x^{01}_{22}}{2x^0_{23}}x\right)~~~.
\end{array}
\end{myalign}
The closest eigenvalues are $\tilde m_{2,3}$, which we are led to identify with $m_{1,2}$ or $m_{2,1}$. 
If in addition we enforce the desirable property $\sin\theta_{13}={\cal O}(x)$, we find that the mixing matrix 
$U_{PMNS}$ should coincide with $P_e~U_\nu~P_\nu$ where
$P_\nu$ is equal to either $P_{13}$ or $P_{23}P_{12}$, while we have four possibilities for the permutation $P_e$: $P_{13}$, $P_{12}P_{13}$, $P_{12}$ and $P_{13}P_{12}$. By exploiting these permutations of rows and columns, related to the lepton mass ordering, we end up with:
\begin{myalign}  
\begin{array}{lclcl}
\sin\theta_{13}={\cal O}(x)&&\tan^2\theta_{23}= {\cal O}(x^2)&&\tan^2\theta_{12}= 1+{\cal O}(x)~~~,
\end{array}
\end{myalign}
where $\tan\theta_{ij}$ can be replaced also by $\cot\theta_{ij}$ in each single entry above. None of these possibilities matches the observations.
\subsection{Summary}
\label{summy}
When the modulus $\tau$ approaches one of the two fixed points $\tau=i$ and $\tau=\omega$,
many properties of the system do not depend anymore on the details of the model realization.
We can draw conclusions that are independent of the specific finite modular group $SL(2,\mathbb{Z}_N)$,
from the choice of the irreducible representation $\rho_l$ of the lepton doublets and largely independent
of the choice of the modular weights $k_\varphi$. Moreover, we are not forced to assume a minimal or 
flavor universal K\"ahler potential: the above results hold for the most general K\"ahler potential 
compatible with modular invariance~\footnote{If a flavor universal K\"ahler potential is adopted, 
the anti-holomorphic variable $\bar u$ only affects the overall scale of the mass matrices, can be absorbed in the parameter $m^0_\nu$ and drops from all dimensionless quantities.}. These are all signals that
the behavior of the system at the critical point is universal. 

In table 5 we summarize the predictions of modular invariant
models for lepton masses in the vicinity of $\tau=i$ or $\tau=\omega$, up to possible permutations affecting the
mixing matrix. In the vicinity of $\tau=\omega$, the atmospheric angle, the solar one and $(\Delta m^2_{sol}/\Delta m^2_{atm})/\sin\theta_{13}$ are not correctly described without tuning of the order-one coefficients~\footnote{
However, mass hierarchies in the charged lepton sector can be naturally generated if $\tau$ is close to $\omega$ ~\cite{Novichkov:2021evw}.}.
When $\tau$ is near the imaginary unit few possibilities can be realized, depending on the two-valued parameter $k_S$
and on the behavior (regular or singular) of $m_\nu(0,0)$. When $k_S$ is even, tuning is needed to 
reproduce $\Delta m^2_{sol}/\Delta m^2_{atm}$. Moreover $\sin^2\theta_{12}$ and $\sin^2\theta_{13}$ are expected to
be of the same order, contrary to observation. 
\begin{samepage}
\begin{myalign}
\nonumber
\begin{array}{c|lcccccc}
\hline
\tau&&\begin{array}{c}{\rm mass}\\{\rm ordering}\end{array}&\dd\frac{\Delta m^2_{sol}}{\Delta m^2_{atm}}&\sin^2\theta_{12}&\sin^2\theta_{13}&\sin^2\theta_{23}\\
\hline
\hline
\approx i&k_S~{\rm even}~~~m_\nu(0,0)~{\rm regular}&NO/IO&{\mathcal O}(1)&{\mathcal O}(x^2)&{\mathcal O}(x^2)&{\mathcal O}(1)\\[5 pt]
\hline
\approx i&k_S~{\rm odd}~~~m_\nu(0,0)~{\rm regular}&IO&{\mathcal O}(x)&\frac{1}{2}(1+{\mathcal O}(x))&{\mathcal O}(x^2)&{\mathcal O}(1)\\[5 pt]
\hline
\approx i&k_S~{\rm odd}~~~m_\nu(0,0)~{\rm singular}&NO&{\mathcal O}(x^3)&\frac{1}{2}(1+{\mathcal O}(x))&{\mathcal O}(x^2)&{\mathcal O}(1)\\[5 pt]
\hline
\hline
\approx \omega&&NO/IO&{\mathcal O}(x)&\frac{1}{2}(1+{\mathcal O}(x))&{\mathcal O}(x^2)&{\mathcal O}(x^2)\\[5 pt]
\hline
\end{array}
\end{myalign}
\vskip 0.1 cm
\noindent
{{\bf Table 5} Synopsis of predictions in modular invariant flavor models of leptons, when the modulus $\tau$ falls in the vicinity of the fixed points $\tau=i$ or $\tau=\omega$ and $\rho_l$ is an irreducible representation.}
\end{samepage}
\vskip 0.4 cm
\noindent
When $k_S$ is odd, a particularly appealing scenario occurs when $m_\nu(0,0)$ is singular,
which can be realized within the seesaw mechanism. In this case the predicted scaling of all observed quantities in terms
of the expansion parameter $x=|u|$ is compatible with observation, without requiring any tuning of the unknown
order-one parameters. A value $x$ close to $0.1$ is suggested by the data.
Out of the 27 CP-invariant models in fig. \ref{lines}, 18 satisfy $|\tau-i|<0.25$. Of these, 12 feature $k_S$ odd and $m_\nu(0,0)$ singular, through the seesaw mechanism. They present a homogeneous set of predictions with the same properties described in Section \ref{good}.
At the fixed point, $\Delta m^2_{sol}/\Delta m^2_{atm}=\sin\theta_{13}=\sin^2\theta_{12}-1/2=0$ and CP is
conserved. Nonvanishing values of these three quantities and $CP$-violating effects all originate from a small departure of $\tau$ from the critical point $\tau=i$. 

It is interesting to compare these results with the negative conclusion of ref.~\cite{Reyimuaji:2018xvs} where the authors,  
under a general set of assumptions recalled in Section \ref{survey}, prove that the only possible unbroken symmetry compatible with normal ordering requires left-handed lepton doublets to consist of three equivalent real one-dimensional representations. In this case the neutrino mass matrix $m_\nu$ is of anarchical type and both neutrino mass ratios and lepton mixing angles ensue from a favorable statistical distribution. In the case favored by our previous analysis,
the unbroken symmetry $\mathbb{Z}_4^S$ of the
fixed point $\tau=i$ offers a valid starting point to reproduce lepton mixing angles and neutrino masses with normal ordering without resorting to an anarchical scenario. There is no contradiction with the general conclusion of ref.~\cite{Reyimuaji:2018xvs}. Indeed in our case the neutrino mass matrix $m_\nu$ is singular at the
symmetric point, as signaled by $m_3\propto 1/x$. Strictly speaking, the symmetric limit cannot be applied to $m_\nu$,
it only exists for its inverse $m_\nu^{-1}$ which has a vanishing eigenvalue and is regular at $x=0$. Insisting on
$m_\nu$ as the primary object requires switching on symmetry-breaking effects to avoid the singularity. 
While these considerations 
are clearly discussed in ref.~\cite{Reyimuaji:2018xvs}, our results represent an interesting concrete example
of how the anarchical scenario can be avoided for neutrinos with normal mass ordering.

In ref.~ \cite{Feruglio:2022kea}, some of these results where anticipated. In particular, the neutrino mass spectrum
near $\otau =i$ for $k_S$ odd and $m_\nu(0,0)$ singular was explicitly reported. Here we have provided a thorough discussion of both the fixed point $\otau =i$ (including the cases  $k_S$ odd and $m_\nu(0,0)$ regular and $k_S$ even) and $\otau =\omega$. Many details are given in Appendix \ref{diagon}. Moreover we have given an explicit proof of the transformation properties of charged lepton and neutrino mass matrices (Table 4), as well as of the $x$-expansion of $m_\nu(u,\bar u)$ in the basis where
the charged lepton mass matrix is diagonal. Finally, in Appendix \ref{Ehol} we have justified the diagonal form of the matrices $\Omega(S)$ and $\Omega(ST)$, eq. (\ref{Omega}), for irreducible $\rho_l$.

\section{Paths toward criticality}
\label{road}
In all flavor models discussed here, Yukawa couplings ${\mathcal Y}(\tau)$ are field-dependent quantities of an underlying EFT. Near-criticality explored in this paper requires the $\tau$ VEV to lie close to a second-order phase transition connecting a disordered phase to an ordered one. The transition is monitored by an order parameter
$u(\tau)$. In a simple-minded discussion, we surveyed classes of models and analyzed the features of the function $u(\tau)$ required to reproduce the data, to detect possible hints of near-criticality in the Yukawa sector.
Phase transitions we are familiar with in Nature necessitate some control parameter $\xi$ to be tuned close to a critical threshold $\xi_c$. Above(below) $\xi_c$ the system is in the disordered(ordered) phase. Given the large variety of
flavor models we have neither identified the control parameters nor analyzed their role. 
In this Section we make a few comments on this important aspect, leaving a thorough discussion for future work.
We do not claim any originality, rather we collect here a few proposals that have been put forward in the literature
in similar contexts.
 
To fix the ideas we assume that the phase transition in question can be described \`a la Landau, employing a functional
$V(\xi,u(\tau))$ respecting the symmetries of the system. The order parameter $u(\tau)$ vanishes in the disordered phase and is different from zero
in the ordered one. We adopt a mean field approximation, where fluctuations of $u(\tau)$ are ignored.
The minima of $V(\xi,u(\tau))$ with respect to $\tau$, keeping 
$\xi$ fixed, represent the searched for VEVs and will be denoted by $\tau_{min}(\xi)$.
Here $\xi$ represents either some Lagrangian parameter of the theory, like the parameter $\mu^2$ of the quadratic term in the Higgs scalar potential, or some additional background field. In the Landau theory, ${\mathcal F}(\xi)\equiv V(\xi,u(\tau_{min}(\xi)))$ is the free energy, whose derivatives, and their discontinuities across the critical threshold $\xi_c$, describe the system. We can regard $V(\xi,u(\tau))$ as the energy density of the theory. 
For example, in the context of a $\mathbb{Z}_2$-symmetric model mimicking the electroweak symmetry breaking, we have:
\begin{myalign}
\label{toyew}
V(\xi,u(\tau))=\xi u(\tau)+\lambda u^2(\tau)~~~~~~~u(\tau)\equiv\dd\frac{\tau^2}{2}~~~~~~(\lambda>0)~.
\end{myalign} 

\begin{myalign}
u(\tau_{min}(\xi))=\left\{
\begin{array}{cc}
0&\xi>0\\
-\xi/2\lambda&\xi<0
\end{array}
\right.
~~~~~~~~~~~~~~~~~~~
{\mathcal F}(\xi)=\left\{
\begin{array}{cc}
0&\xi>0\\
-\dd\frac{\xi^2}{4\lambda}&\xi<0
\end{array}
\right.~~~.
\end{myalign} 
The free energy ${\mathcal F}(\xi)$ and its first derivative are continuous across the critical value $\xi_c=0$, while
the second derivative is not, signaling a second-order phase transition. 
This example can be generalized by assuming a functional $V(\xi,u(\tau))$ of the type:
\begin{myalign}
V(\xi,u(\tau))=V_0(\tau)+\xi~u(\tau)~~~.
\end{myalign} 
The minima $\tau_{min}(\xi)$ satisfy the equation:
\begin{myalign}
\dd\frac{\partial V(\xi,u(\tau))}{\partial\tau}=\dd\frac{d V_0(\tau)}{d\tau}+\xi~\dd\frac{d u(\tau)}{d\tau}=0~~~.
\end{myalign} 
A phase transition occurs when the order parameter $u(\tau_{min}(\xi))$ vanishes for $\xi>\xi_c$ and becomes different from zero for $\xi$ below $\xi_c$. The free energy reads:
\begin{myalign}
{\mathcal F}(\xi)=V_0(\tau_{min}(\xi))+\xi~u(\tau_{min}(\xi))~~~.
\end{myalign} 
The first two derivatives of the free-energy are
\begin{myalign}
\dd\frac{d{\mathcal F}(\xi)}{d\xi}=u(\tau_{min}(\xi))~~~~~~~~~~~~~~~~~~\dd\frac{d^2{\mathcal F}(\xi)}{d\xi^2}=\dd\frac{d u}{d \tau}
\dd\frac{d \tau_{min}(\xi)}{d \xi}=-2 m^2(\xi)\left[\dd\frac{d \tau_{min}(\xi)}{d \xi}\right]^2~~~,
\end{myalign} 
where
\begin{myalign}
m^2(\xi)=\dd\frac{1}{2}\left[\dd\frac{d^2 V_0(\tau)}{d\tau^2}+\xi~\dd\frac{d^2 u(\tau)}{d\tau^2}\right]_{\tau=\tau_{min}(\xi)}
\end{myalign} 
is the square mass of the field $\tau$. 
In second-order transitions $u(\tau_{min}(\xi))$ is continuous at $\xi_c$ while $d u(\tau_{min}(\xi))/d \xi$ is discontinuous, which implies the discontinuity of $d^2 \mathcal F(\xi)/d\xi^2 $. 

In this context, few possibilities can lead to near-criticality.
\begin{itemize}
\item[1.] The condition $\xi\approx\xi_c$ is accidental. The parameter $\xi$ is God-given and it happens to lie close
to the critical value $\xi_c$. The value $|\tau_{min}(\xi)|\ll 1$ is a local minimum of the energy density and the system is close to the disordered phase. In our survey of flavor models, this happens when 
the flavor symmetry group $G$ is broken, but a subgroup $H$ of $G$ is approximately conserved. 
\item[2.] The control parameter $\xi$ is a dynamical variable whose variation scans the order parameter 
$u(\tau_{min}(\xi))$, to finally deliver $\xi\approx\xi_c$ in our universe today. 
This general idea is shared by several proposals \cite{Dvali:2003br,Dvali:2004tma,Graham:2015cka,Arvanitaki:2016xds,Herraez:2016dxn,Geller:2018xvz,Giudice:2021viw}, having in common a set of key ingredients.
The control parameter $\xi$, or more generally the parameters of the theory, are promoted to dynamical variables, much as the $\theta$ parameter in the Peccei-Quinn solution of the strong CP problem. The evolution of the universe
and the presence of a non-trivial vacuum structure allows $\xi$ to vary in a wide range, possibly exploring all the relevant parameter space. Eventually, the cosmological evolution selects a vacuum $\xi\approx \xi_c$. 
A single period of inflation or eternal inflation provides the tool to populate the available vacua. Different mechanisms
can secure $\xi\approx\xi_c$ at the end of inflation. 

In the applications to the hierarchy problem, it is the quadratic parameter of the Higgs sector $\mu^2$ that is promoted to a dynamical variable. The distribution of vacua can be nearly uniform with respect to
$\mu^2$, but the cosmological evolution stops when $\mu^2$ is negative and close to zero. 
In ref. \cite{Graham:2015cka} at the beginning of inflation $\mu^2$ is large and positive and slowly rolls down. When it becomes slightly negative, the spontaneous breaking of the electroweak symmetry creates a potential barrier for $\mu^2$, preventing a further evolution of $\mu^2$. Another possibility is that the distribution of vacua strongly peaks at $\mu^2\approx 0$ \cite{Dvali:2003br,Dvali:2004tma}. In ref. \cite{Giudice:2021viw} the distribution of the control parameter,
a generic scalar field subject to a nearly flat potential, is initially random. Under general conditions, quantum fluctuations of the field during a period of inflation delete the initial conditions and lead to a stationary distribution localized at a critical point of a first-order phase transition. 

These explanations do not require New Physics at the weak scale and can avoid anthropic arguments. 
The landscape of vacua arising from string theory compactifications and the related field-dependent low-energy parameters naturally deliver the main ingredients of this scenario.
Also in our application to flavor physics the Yukawa couplings ${\mathcal Y}(\tau)$ are field-dependent quantities and the scalar fields $\tau$ span the possible vacua of the theory. The example of modular flavor symmetry is inspired 
by the string theory framework and is based on a set of inequivalent vacua arising from toroidal compactifications. \item[3.] Finally, it is worth mentioning that a critical behavior can also arise without tuning the control parameters,
as is the case of systems enjoying Self-Organized Criticality~\cite{sandpile,soc1,soc2,Eroncel:2018dkg}. These are dissipative systems
in an out-of-equilibrium regime with a slow driving force, like in the sandpile model \cite{sandpile} where sand grains are slowly added to an initial random distribution, allowing a part of them to sink and be lost. In contrast to the usual phase transitions, which require thermodynamical equilibrium, the time evolution brings the system close to a scaling behavior similar to the one exhibited in a continuous phase transition. Even though no real phase transition takes place, 
the system exhibits spatial/temporal scale invariance and power law scaling, typical of a second-order critical point.
It might be the case that also in the Yukawa sector, during some dynamical evolution, the value of $u(\tau)$ relaxes
close to a point enjoying a residual symmetry~\cite{Kofman:2004yc}, in the absence of a control parameter $\xi$.
In this case the role of $\xi$ could be played by the field $\tau$ itself, whose variation during the cosmological evolution
scans the Yukawa couplings of the theory.
\end{itemize}
We conclude this Section by mentioning the type of control parameters that can be at work in modular invariant
flavor models.
When discussing the example of modular flavor symmetries applied to the lepton sector, we saw that the neighborhood of the self-dual point
$\tau=i$ is particularly suitable to reproduce the observed pattern of neutrino masses and mixing angles.
The vicinity of the fixed point $\tau=\omega$ is instead promising to describe the hierarchy among the charged
lepton masses \cite{Novichkov:2021evw}. 

A modular invariant scalar potential, inspired by the simplest orbifold compactification, is given by:
\begin{myalign}
  V(\tau, \bar\tau) = \dd\frac{\Lambda^4}{8(\im\tau)^3 \lvert \eta \rvert^{12}}
  \left[
  \frac{4}{3} \left\lvert
  i \dd\frac{dH}{d\tau} + \frac{3}{2\pi} H \hat{G}_2
  \right\rvert^2 (\im\tau)^2 - 3 \lvert H \rvert^2
  \right]~~~,
\end{myalign}
where $\tau$ represents the overall K\"ahler modulus. By requiring no singularities within the fundamental domain, the modular-invariant function $H(\tau)$ reads:
\begin{myalign}
H(\tau)  
\,=\, \left ( j(\tau)-1728 \right )^{m/2}j(\tau)^{n/3}\mathcal{P}\left( j(\tau) \right )~~~,
\end{myalign}
where $j(\tau)$ is the Klein function, $\mathcal{P}\left( j(\tau) \right )$ is a polynomial in $j$ and $m$ and $n$ are non-negative integers. The potential $V(\tau, \bar\tau)$
depends also on $\eta(\tau)$ and $\hat{G}_2(\tau,\overline{\tau})$, the Dedekind function and the non-holomorphic Eisenstein function:
\begin{myalign}
\hat{G}_2(\tau,\overline{\tau}) = {G}_2(\tau) - \dd\frac{\pi}{\im\tau}~~~,~~~~~~~~~~~~\frac{\eta'(\tau)}{\eta(\tau)} = \dd\frac{i}{4\pi} G_2(\tau)~~~.
\end{myalign}
Choosing $\mathcal{P}=1$, which renders $V(\tau, \bar\tau)$ CP-invariant, few cases have been studied in the literature \cite{Font:1990nt,Cvetic:1991qm}. When $(m,n) = (0,0), (1,1), (0,3)$ the global minima lie at $\tau \simeq 1.2\, i$ (imaginary axis), $\tau \simeq \pm 0.24+0.97\,i$ (two equivalent minima on the unit arc) and $\tau = i$, respectively. The first two cases lead to $|u|\approx 0.1$, the value preferred in the class of models for lepton masses we have inspected. 
The three minima at $(m,n) = (0,0), (1,1), (0,3)$ are all CP-conserving. Indeed, in ref.~\cite{Cvetic:1991qm}, it was conjectured that all extrema of $V(\tau, \bar\tau)$ are at CP-conserving values of $\tau$, that is either on the boundary of the fundamental domain or on the imaginary $\tau$ axis. Only recently, CP-violating minima have also been found \cite{Novichkov:2022wvg,Leedom:2022zdm}. They lie in the vicinity of
the fixed point $\tau=\omega$, at a typical distance $|\tau-\omega|$ of order 0.01, and require $n=0$ and $m\ne 0$.
Both the minima close to $\tau=i$ and those near $\tau=\omega$ correspond to a negative cosmological constant but, depending on the integers $(m,n)$, variants of $V(\tau, \bar\tau)$ can exhibit local minima with a positive energy density \cite{Leedom:2022zdm}. Turning on additional moduli and/or fluxes leads to a richer variety of possibilities.
Distributions of $\tau$ VEVs in the presence of quantized  three-form fluxes in Type IIB string theory have been investigated in refs.~\cite{Ishiguro:2020tmo,Ishiguro:2022pde}.

Apart from the overall constant $\Lambda^4$ providing the correct dimension, in the simple example of $V(\tau, \bar\tau)$ 
with $\mathcal{P}=1$ we have no continuous parameters and the small deviations of the minima at $(m,n) = (0,0), (1,1), (m\ne 0,0)$ from the corresponding fixed points all arise from the special properties of the Dedekind and Klein functions. Unlike the case of the Higgs potential, $V(\tau, \bar\tau)\vert_{\mathcal{P}=1}$ depends
only on the two discrete variables $m$ and $n$, which act as control parameters. 
Other continuous parameters can be provided by a nontrivial polynomial $\mathcal{P}$.
Since they parametrize physically distinct systems, much as the case of $\theta$ vacua in QCD, $m$ and $n$ and possibly additional continuous parameters are not dynamical variables and transitions between different vacua cannot occur at this stage.
To explore the landscape of vacua during the evolution of our universe, we would need to promote one or more
of these control parameters to dynamical variables.
\section{Conclusions}
Solutions to the hierarchy problem based on symmetry arguments alone, like for example low-energy supersymmetry,
require New Physics near the TeV scale, an expectation that has not found confirmation so far. By reinterpreting the hierarchy problem
as the closeness of the SM parameters to the critical point separating the unbroken electroweak phase from the broken one, new solutions have been put forward. Much as in the case of the Peccei-Quinn solution of the strong CP problem, the parameters of the theory are promoted to dynamical variables that can approach a critical value during the
evolution of our universe. In almost all the attempts to explain the observed pattern of fermion masses and mixing angles, the Yukawa couplings are also promoted to dynamical variables. String theory has no free parameters apart from a fundamental length scale, and Yukawa couplings are completely determined by the fields describing the background over which the string propagates. Similarly, in all models based on flavor symmetries,
the Yukawa couplings depend on a set of fields $\tau$ responsible for the necessary symmetry breaking.

It is natural to ask whether also the Yukawa sector exhibits hints of near-criticality,
pointing to a common interpretation of most of the parameters describing our fundamental interactions.
In this paper, we made the very first steps to explore such a possibility. Due to the complexity of the Yukawa
sector, even simply detecting clues of near-criticality is not straightforward, and we had to make several 
assumptions and drastic simplifications. 
First of all, unlike in the Higgs system, there is no theory of fermion masses and mixing angles
but rather a large variety of models, able to shed light only on a part of the problem. Most of them are based on flavor symmetries and rely on complicated symmetry-breaking sectors, preventing a classification of the relevant
control parameters. To simplify our task we assumed that the phase transition relevant to the Yukawa sector 
is of the second order, between an ordered phase and a disordered one. If this is the case a necessary condition for near-criticality is the closeness to zero of the order parameter $u(\tau)$, a convenient function of the fields responsible for symmetry breaking.  

This indicator is not without ambiguities, since by construction many flavor models are realized as small perturbations
around a symmetric scheme, as we recalled in a short survey. However, the discussion of the lepton sector revealed that
closeness to a symmetric phase is not a general property, if nontrivial outputs are requested in the unbroken limit.
Many models of lepton masses make use of small symmetry-breaking order parameters, but mass ratios and mixing parameters are undefined at the symmetric point, a feature that makes these models dependent on vacuum alignment. Moreover, under rather general assumptions, the only possible unbroken symmetry compatible with normal ordering requires left-handed lepton doublets to consist of three trivial singlets. No symmetry-breaking sector is required and the system is always in the unbroken phase, where both neutrino masses and lepton mixing angles are completely unconstrained. 

Furthermore, good candidates where the closeness to a symmetric point can truly be interpreted as a hint of near-criticality
are models where the flavor symmetry is nonlinearly realized. Unlike the models with a linear realization,
these models are not conceived as perturbations of a symmetric system.
Everywhere in the moduli space parametrized by $\tau$, the flavor symmetry is always broken and there is no prejudice about the value of $\tau$ providing a successful description of the fermion spectrum. In a given model of this type, data may or may not drive $\tau$ close to a point of residual symmetry. 

We have analyzed modular invariant models of lepton masses, where $\tau$ parametrizes the inequivalent tori associated with the compactification of two extra dimensions. An important feature of this class of models is that the choice of the flavor symmetry group is mandatory. Modular invariance is a built-in property, to remove the redundancy arising when $\tau$ is left to vary in the upper half complex plane. Modular transformations are nonlinear and the size
of $\tau$ has no invariant meaning in these models. There are no expectations about the value of $\tau$ that better describes lepton masses and $\tau$ is treated as a free parameter in a fit. In a bottom-up approach other free parameters are the choice of the level $N$ and the modular weights of the matter multiplets, leading to a virtually infinite number of
possible realizations. For these reasons, the fact that in most of the existing modular invariant and CP invariant models the preferred value of $\tau$ falls close to the self-dual point $\tau=i$ is very remarkable. 
The dimensionless parameter $\tau$ can be interpreted as the VEV of a scalar field in units of some fundamental scale
and the typical deviation of $\tau$ from the self-dual point is of order ten percent, a huge value, if compared with the electroweak VEV expressed in units of the Planck mass. Nevertheless, the preference for a neighborhood of $\tau=i$ is impressive
and we regard it as the indication of an intrinsic property of the theory, deserving an explanation. 

We have identified
a "kinematical" justification. Close to the fixed points $\otau =( i,\omega)$ the theory enjoys an approximate symmetry under CP and the subgroups $(\mathbb{Z}_4^S,\mathbb{Z}_2^{ST}\times\mathbb{Z}_2^{S^2})$.
By expanding the lepton mass matrices in powers of the symmetry-breaking parameter around
these fixed points we have classified neutrino masses and mixing angles. Among the few 
possibilities, a neutrino mass matrix singular at $\tau=i$ provides the best zeroth-order approximation
for a successful fit to the data. The singularity at the symmetric point arises when a right-handed neutrino
becomes massless in the see-saw mechanism and allows to circumvent the no-go theorem on
unbroken symmetries quoted above.
Moreover, in the vicinity of this point, the models exhibit a universal behavior, independent of the level $N$ of the construction, the modular weights of the matter multiplets and even the form of their kinetic terms. 
The neutrino spectrum is normally ordered.
Physical quantities, such as mass ratios and mixing angles
satisfy common scaling laws in terms of the symmetry-breaking parameter, a feature reminiscent of systems belonging
to the same universality class in second-order phase transitions.

A major advancement would be to find a "dynamical" explanation. This aspect goes well beyond the scope of this work.
At a very preliminary level, we have revisited an example of a modular invariant scalar potential for $\tau$. In the simplest case, the potential depends only on a set of discrete parameters and, interestingly enough,
a choice exists leading to $|\tau-i|\approx 0.1$, within the desired range. 
It would be interesting to see whether the parameters of the scalar potential can be promoted to 
dynamical variables, allowing our universe to explore the landscape of vacua.
The possibility that the vacuum selected by the cosmological evolution could explain
at the same time the closeness of our universe to the electroweak phase transition, the observed pattern of fermion masses and mixing angles, and perhaps even the smallness of the cosmological constant, is a fascinating one.

\section*{Acknowledgements}
I thank Gianguido Dall'Agata, Sergey Ketov, Jacob Michael Leedom, Nicole Righi and Alexander Westphal 
for useful correspondence. I warmly thank Gian Giudice for reading the manuscript and for his encouraging comments. I am very grateful to Arsenii Titov for very helpful comments and for invaluable assistance during the final revisions of this work. Finally, I thank Odd Magne \O greid, organizer of Discrete 2020-2021, 
Bergen (Norway), November $29^{th}$ - December $3^{rd}$ 2021, for recalling the work of Wilson Bentley and stimulating my curiosity about snowflakes and their properties. This work was supported by the INFN.
\vskip 0.2 cm
\noindent
\newpage


\newpage
\appendix
\addcontentsline{toc}{section}{Appendices}
\section*{Appendices}
\section{General aspects of near-criticality and universality}
\label{fmcu}
The vacua of the theory are parametrized by the fields $\tau$ belonging to a space ${\cal M}$ on which a symmetry group $G$ acts. Here we will mainly focus on linearly realized symmetries, though in Section~\ref{modularsec} we explain how
to trace back the case of nonlinearly realized symmetries to the one discussed here.
We assume $\tau$ to be small dimensionless
quantities, close to a point of residual symmetry under a subgroup $H$ of $G$, and the Lagrangian parameters $\lambda$ to be generic numbers of order one. 
In this scenario we expect that an observable $A(\lambda;\tau)$ enjoys an approximate leading order approximation of the type:
\begin{myalign}
A(\lambda;\tau)\approx c_A(\lambda) \Pi(\tau)~~~,
\end{myalign}
where $\Pi(\tau)$ is a suitable expansion in powers of $\tau$ and the dependence on $\lambda$ is carried by the function $c_A(\lambda)$. As we have seen, the phenomenologically preferred vacuum may involve widely different values among the components of $\tau$, causing $\Pi(\tau)$ to have a nontrivial structure, with both positive and negative powers of $\tau$, see eq.~(\ref{uni1}). We would like to understand how, in general, $\Pi(\tau)$ can be accounted for by
the critical behavior of the system and whether it is possible to extract 
some information that does not depend on the function $c_A(\lambda)$.

It is useful to recall some properties of the $n$-dimensional space ${\cal M}$ spanned by the fields $\tau$,
carrying a representation of the group $G$.
Along any orbit of the group passing through $\tau$, that is the set $\large\{g\tau,~g\in G\large\}$, the mass spectrum does not change and the variables $\tau$ provide a redundant description of the system. It is convenient to remove this degeneracy by projecting $\tau$
onto the quotient space obtained by identifying points lying along the same group orbits.
This projection can be realized by removing from $\tau$ as many components as possible, by performing the most general $G$-transformation. 
Starting from a multiplet $\tau$ possessing $n$ real components, we look for a subset ${\cal F}$ of points $\hat\tau$ such that
\begin{itemize}
\item[{\it i)}] any point $\tau$ in ${\cal M}$ can be reached from a point $\hat\tau$ in ${\cal F}$ by means of a suitable 
$G$-transformation; \item[{\it ii)}] no two points in the interior of ${\cal F}$ are related by a $G$-transformation.
\end{itemize}
The subset ${\cal F}$ is the fundamental domain of the group $G$ in ${\cal M}$ and parametrizes the inequivalent vacua of the system. It provides a concrete realization of the coset ${\cal M}/G$.
The projections $\hat\tau$ and the space ${\cal F}$ are not uniquely defined: if we perform a rigid $G$ transformation
on all points of ${\cal F}$, we end up with another fundamental domain. The new domain and the original one
have the same number $n_I$ of independent components.
If the group is continuous, we expect $n_I<n$, the number $n_I$ depending not only on the dimension of the group $G$ 
but also on the transformation properties of $\tau$. 
When $G$ is discrete, a general $G$ transformation will not reduce the number of relevant components of $\tau$,
but allows to restrict the fields $\tau$ to some subspace of $\mathbb{R}^n$, with the same dimension $n$: $n_I=n$. 
For example, if $n=1$ a discrete translation $\tau\to \tau+1$ allows to restrict the field $\tau$ between zero and one~\footnote
{When the symmetry is linearly realized, the notion of fundamental domain ${\cal F}$ is intimately related to that of orbit space.}.
\begin{figure}[h!]
\centering
\includegraphics[width=0.5\linewidth]{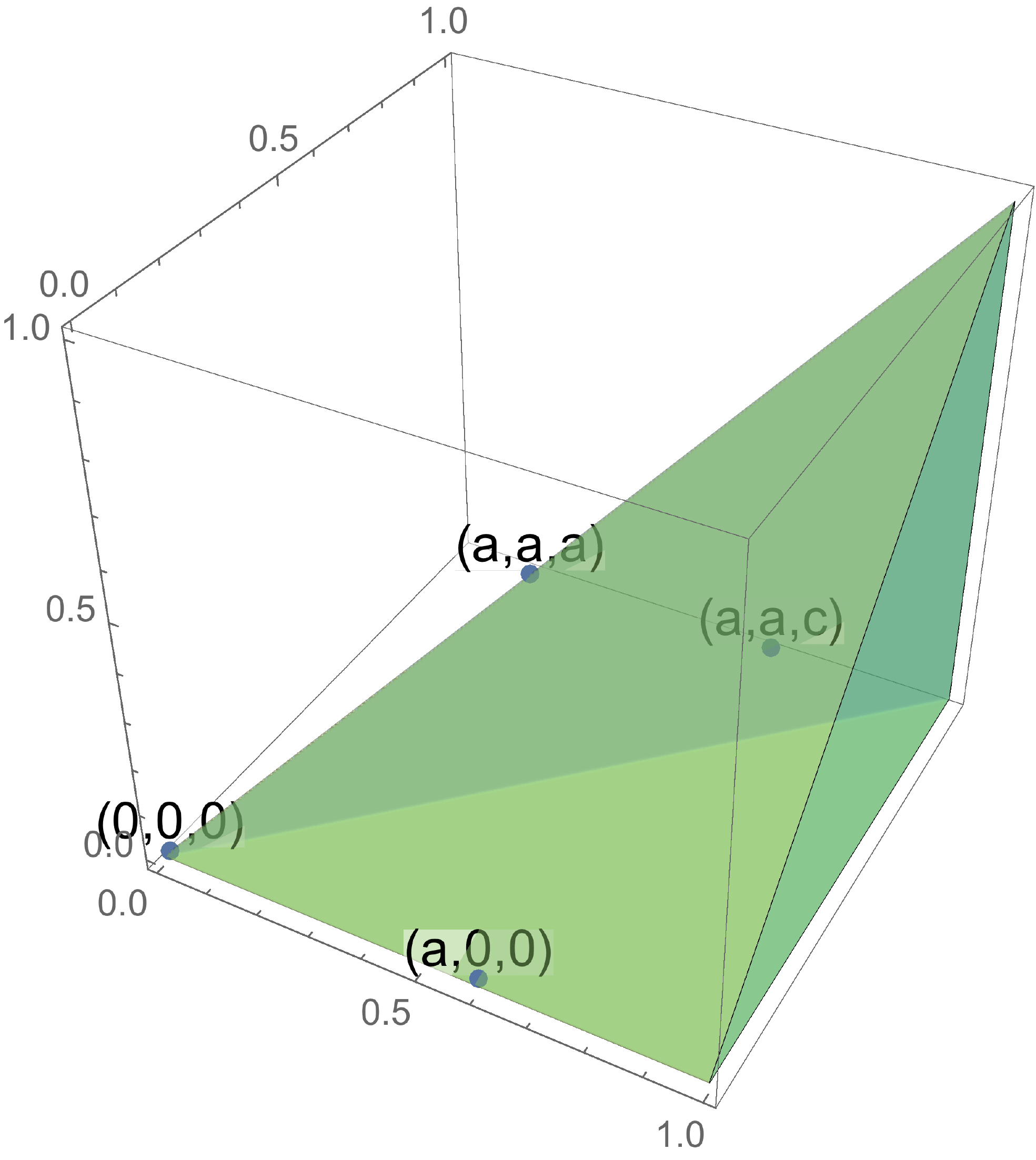}
\caption{Portion of the fundamental domain of the group $O(3)\times O(3)$ in ${\cal M}=\mathbb{R}^9$ spanned by the $(3,3)$
representation $\tau$. The variables $a$, $b$, $c$ have been restricted in the interval between 0 and 1. Points enjoying residual symmetries have been highlighted.}
\label{lines1}
\end{figure}
\vskip 0.2 cm
\noindent
As an example, consider the set ${\cal M}$ of $3\times 3$ real matrices $\tau$, transforming in the $(3,3)$
representation of $G=O(3)\times O(3)$:
\begin{myalign}
\tau\to U^T \tau V~~~,
\end{myalign}
$U$ and $V$ being two independent 3$\times$3 orthogonal matrices. 
Through a $G$ transformation, it is always possible to map a generic element $\tau$ into a diagonal non-negative
and ordered matrix $\hat\tau$ (only the diagonal elements are shown here):
\begin{myalign}
\hat\tau=(a,b,c)~~~~~~~~~~(a\ge b\ge c\ge 0)~~~.
\end{myalign}
We can choose this set as fundamental domain ${\cal F}$ of $O(3)\times O(3)$ in ${\cal M}$, see fig.~\ref{lines1}.
The domain ${\cal F}$ has three two-dimensional boundaries, that intersect along three one-dimensional boundaries, 
which finally join at the origin.
In the interior of ${\cal F}$, where $a$, $b$, $c$ are generic, $G$ is broken down to the minimal residual symmetry, $(Z_2)^3=Z_2(a)\times Z_2(b)\times Z_2(c)$~\footnote{By $Z_2(i)$ we denote the parity symmetry related to 
$U$ and $V$ transformations both operating a change of sign of the element $i$.}. 
The two-dimensional boundary are $\hat\tau=(a,a,c)$, $\hat\tau=(a,b,b)$ and $\hat\tau=(a,b,0)$.
Points lying on these planes enjoy the residual symmetry $O(2)\times Z_2(c)$, $O(2)\times Z_2(a)$ and $(Z_2)^4$, respectively.
The three one-dimensional boundaries are $\hat\tau=(a,0,0)$, $\hat\tau=(a,a,0)$ and $\hat\tau=(a,a,a)$.
They leave unbroken the subgroups $O(2)\times O(2)'\times Z_2(a)$, $O(2)\times (Z_2)^2$ and $O(3)$, respectively.
Finally, at the origin $\hat\tau=(0,0,0)$ the whole
group $G$ is preserved.

There are many possible breaking chains, listed in table 6, that depend on the relative size of $a$, $b$, $c$.
We omitted from the list the trivial chain $G\to (Z_2)^3$ realized when $a$, $b$, $c$ are generic numbers
of order one. Thus a breaking chain is described by a point $\hat\tau$ lying near one of the boundaries of ${\cal F}$.
The hierarchy among the different components of $\hat\tau$ is important to qualify a breaking chain.
Indeed, given a generic set $(a,b,c)$ it is always possible to parametrize it in a way that mimics a given chain.
For instance $(a,b,c)=(c+\delta+\epsilon,c+\delta,c)$, if we define $\delta=b-c$ and $\epsilon=a-b$. However,
$(a,b,c)$ does not represent the breaking chain $G\to O(3)\to O(2)\times Z_2 \to (Z_2)^3$ unless
$a-b\ll b-c\ll c$. In general, given a group $G$ acting linearly on scalar fields $\tau$, we can distinguish the maximal residual symmetry,
coinciding with $G$, and the minimal residual symmetry $H_{\rm min}$, not necessarily trivial.
Starting from $G$ we can reach $H_{\rm min}$ through different paths:
\begin{myalign}
\label{bc}
G\equiv H_0\to H_1\to \cdot\cdot\cdot \to H_p\equiv H_{\rm min}~~~,
\end{myalign}
each consisting of a certain number $p$ of steps.
These paths are sequences of proper subgroups starting from $G$ and ending in $H_{\rm min}$.
We say that the multiplet $\hat\tau$ identifies the breaking chain ${\cal C}$ associated to the path (\ref{bc}) if 
\begin{itemize}
\item[{\it i)}]
it decomposes as the sum of $p$ representations: 
\begin{myalign}
\label{decomposition}
\hat\tau=\hat\tau_1\oplus \hat\tau_2 \oplus\cdot\cdot\cdot \oplus \hat\tau_p~~~,
\end{myalign}
where $\hat\tau_k$ is a singlet under $H_{k}$ and transforms non trivially under $H_{k-1}$ $(k=1,...,p)$. 
\item[{\it ii)}] the following inequalities hold:
\begin{myalign}
\label{hie}
1>|\hat\tau_1|\gg |\hat\tau_2|\gg \cdot\cdot\cdot \gg |\hat\tau_p|~~~.
\end{myalign}
\end{itemize}
The number $p$ of representations occurring in a chain is smaller or equal to the number $n_I$ of components of $\hat\tau$. Hence, in the present notation, a single representation $\hat \tau_i$ may involve several components $\tau_{ia}$. 
For example, along the chain $G\to O(2)\times O(2)'\times Z_2\to (Z_2)^3$ we have $\hat\tau_1=(a,0,0)$,
$\hat\tau_2=(0,\delta_1,\delta_2)$, $1>a\gg \delta_1, \delta_2$. 
\begin{myalign}
\nonumber
\begin{array}{l|l|l}
\hline
~~~~~~~~~\hat\tau&&~~~~~~~~~~~~~{\rm breaking~ chain}\\
\hline
\hline
(\delta_1,\delta_2,\delta_3)&\delta_i\ll1&G\to (Z_2)^3\\
\hline
\hline
(a,\delta_1,\delta_2)&\delta_i\ll a<1&G\to O(2)\times O(2)'\times Z_2\to (Z_2)^3\\
\hline
(a,\delta+\epsilon,\delta)&\epsilon\ll\delta\ll a<1&G\to O(2)\times O(2)'\times Z_2\to O(2)\times Z_2\to (Z_2)^3\\
\hline
(a,\delta,\epsilon)&\epsilon\ll\delta\ll a<1&G\to O(2)\times O(2)'\times Z_2\to (Z_2)^4\to (Z_2)^3\\
\hline
\hline
(a+\delta_1,a,\delta_2)&\delta_i\ll a<1&G\to O(2)\times (Z_2)^2\to (Z_2)^3\\
\hline
(a+\delta,a,\epsilon)&\epsilon\ll\delta\ll a<1&G\to O(2)\times (Z_2)^2\to (Z_2)^4\to (Z_2)^3\\
\hline
(a+\epsilon,a,\delta)&\epsilon\ll\delta\ll a<1&G\to O(2)\times (Z_2)^2\to O(2)\times (Z_2)\to (Z_2)^3\\
\hline
\hline
(a+\delta_1,a+\delta_2,a)&\delta_i\ll a<1&G\to O(3)\to (Z_2)^3\\
\hline
(a+\delta,a+\epsilon,a)&\epsilon\ll\delta\ll a<1&G\to O(3)\to O(2)\times Z_2\to(Z_2)^3\\
\hline
(a+\delta+\epsilon,a+\delta,a)&\epsilon\ll\delta\ll a<1&G\to O(3)\to O(2)\times Z_2\to(Z_2)^3\\
\hline
\hline
(a+\delta,a,c)&\delta\ll a-c<1&G\to O(2)\times Z_2\to (Z_2)^3\\
\hline
(a,b+\delta,b)&\delta\ll b<1&G\to O(2)\times Z_2\to (Z_2)^3\\
\hline
(a,b,\delta)&\delta\ll b<1&G\to (Z_2)^4\to (Z_2)^3\\
\hline
\hline
\end{array}
\end{myalign}
\vskip 0.2 cm
\noindent
{{\bf Table 6} Possible breaking chains of the group $G=O(3)\times O(3)$ acting on a $(3,3)$
representation, described by $3\times 3$ real matrices $\tau$. }
\subsection{Near-criticality}
Physical quantities $A(\lambda;\tau)$ of our interest depend on $\tau$ only through the representative $\hat\tau$:
\begin{myalign}
\label{pq}
A(\lambda;\tau)=f_A(\lambda;\hat\tau)~~~.
\end{myalign}
We assume that the comparison between data and model predictions selects (a range of) values $\hat\tau$ reproducing the observed pattern of fermion masses and mixing angles. Since no exact residual symmetries are expected
in the fermion spectrum, $\hat\tau$ will lie in the interior of the fundamental domain ${\cal F}$, where $G$ is maximally
broken down to $H_{\rm min}$. 
We say that near-criticality occurs when:
\begin{itemize}
\item[1.]
The value of $\hat\tau$ preferred by the data identifies a breaking chain ${\cal C}$.
\item[2.]
Within each representation $\hat\tau_i$, the components $\hat\tau_{ia}$ are of the same order of magnitude.
\end{itemize}
The symmetry $G$ is completely broken down to $H_{\rm min}$ that, in model building, usually consists of
the trivial subgroup. However, the system lies close to a critical point $\hat\tau_0$. Indeed, a reasonable approximation
of the preferred value $\hat\tau$ is obtained by setting to zero the component  
$\hat\tau_p$ of $\hat\tau$ in eq. (\ref{decomposition}): 
\begin{myalign}
\label{t0}
\hat\tau\approx\hat\tau_0=\hat\tau_1\oplus \hat\tau_2 \oplus\cdot\cdot\cdot \oplus \hat\tau_{p-1}\oplus 0~~~.
\end{myalign}
In $\hat\tau_0$ the residual symmetry is enhanced from $H_{\rm min}$ to a non-minimal group $H_{p-1}$, proper subgroup of $G$. 
The above definition of near-criticality not only requires that the value of $\hat\tau$ reproducing the observed pattern of masses and mixing angles lies close to a critical point $\hat\tau_0$, but also that a hierarchy among the different components of $\hat\tau$ reflects a specific pattern of increasing symmetry of the system. 
The degree of criticality of the system can be estimated from the relations (\ref{hie}): the stronger the inequalities, the higher the degree of criticality.
\subsection{Universality}
For near-critical models, we can investigate universality by inspecting the behavior of the system in a neighborhood of the critical point and by looking for properties
that are independent of the details of the model, such as order-one Lagrangian coefficients. 
To this purpose, we need to characterize the asymptotic behavior of the functions $f_A(\lambda;\hat\tau)$
representing the physical quantities in eq. (\ref{pq}) in the vicinity of the relevant breaking chain:
\begin{myalign}
f_A(\lambda;\hat\tau)\approx c_A(\lambda) \Pi(\hat\tau)~~~,
\end{myalign}
$\Pi(\hat\tau)$ being a suitable expansion in powers of $\hat\tau$. From the examples discussed in the previous Section
we recognize that a Taylor expansion of the type:
\begin{myalign}
\begin{array}{rl}
f_A(\lambda;\hat\tau)=&\sum_{n_1\cdot\cdot\cdot n_p} c_{n_1\cdot\cdot\cdot n_p}(\lambda)(\hat\tau_1)^{n_1}\cdot\cdot\cdot (\hat\tau_p)^{n_p}\\
\approx& c_{\alpha_1\cdot\cdot\cdot \alpha_p}(\lambda)(\hat\tau_1)^{\alpha_1}\cdot\cdot\cdot (\hat\tau_p)^{\alpha_p}~~~,
\end{array}
\end{myalign}
where sums run over non-negative $n_i$ and the last line represent the leading term, is too restrictive to
capture the behavior of the function $f_A(\lambda;\hat\tau)$. In general, this function is 
not analytic in the vicinity of the chain and we should resort to some other representation.
We start by assuming that the function $f_A(\lambda;\hat\tau)$
has a non-singular limit when $\hat\tau_p/\hat\tau_{p-1}$ approaches zero, keeping fixed the values of the other
independent variables $\hat \tau_i$, $(i=1,...,p-1)$. Assuming a power-like leading dependence on $\hat \tau_p/\hat\tau_{p-1}$, when $\hat\tau_p\ll \hat\tau_{p-1}$ we get:
\begin{myalign}
f_A(\lambda;\hat\tau)\approx f^{(p-1)}_A(\lambda;\hat\tau_1,...,\hat\tau_{p-1}) \left(\dd\frac{\hat \tau_p}{\hat\tau_{p-1}}\right)^{\alpha^A_{p}}~~~~~~~~~~~~(\alpha^A_{p}\ge 0)~~~.
\end{myalign}
Now we reiterate the ansatz, analyzing the function $f^{(p-1)}_A(\lambda;\hat\tau_1,...,\hat\tau_{p-1})$ when
$\hat\tau_{p-1}\ll \hat\tau_{p-2}$:
\begin{myalign}
f_A(\lambda;\hat\tau)\approx f^{(p-2)}_A(\lambda;\hat\tau_1,...,\hat\tau_{p-2}) 
\left(\dd\frac{\hat \tau_{p-1}}{\hat\tau_{p-2}}\right)^{\alpha^A_{p-1}}
\left(\dd\frac{\hat \tau_p}{\hat\tau_{p-1}}\right)^{\alpha^A_{p}}~~~~~~~~~~~~(\alpha^A_{p-1},\alpha^A_{p}\ge 0)~~~.
\end{myalign}
After repeating similar steps we get:
\begin{myalign}
\label{scaling}
f_A(\lambda;\hat\tau)\approx c_A(\lambda)~ \prod_{i=1}^p \xi_{i}^{\alpha^A_{i}}~~~~~~~~~~~~~~
\xi_{i}\equiv \dd\frac{\hat\tau_{i}}{\hat\tau_{i-1}}~~~(i=1,...,p;~\hat\tau_0=1)~~~.
\end{myalign}
with non-negative exponents $\alpha^A_i$. 
For instance, if $p=n_I=3$,
\begin{myalign}
f_A(\lambda;\hat\tau)\approx c_A(\lambda)~\hat\tau_1^{\alpha^A_1}~
\left(\dd\frac{\hat\tau_2}{\hat\tau_1}\right)^{\alpha^A_{2}}
~\left(\dd\frac{\hat\tau_3}{\hat\tau_2}\right)^{\alpha^A_{3}}~~~~~~~~(\alpha^A_1,\alpha^A_{2},\alpha^A_{3}\ge0)~~~.
\end{myalign}
Eq. (\ref{scaling}) reproduces accurately the case $p=n_I$. When $p<n_I$ and one or more representations
$\hat\tau_i$ have different components $\hat\tau_{ia}$, we have a more complicated expansion in terms of the variables $\xi_{iab}\equiv\hat\tau_{ia}/\hat\tau_{i-1b}$.
The dependence of the observables $A$ on the Lagrangian parameters is included in $c_A(\lambda)$.
The critical exponents $\alpha^A_{i}$ do not depend on $\lambda$ and
describe the behavior of the system along the breaking chain. 
Moreover, assuming that hierarchies are due to $\hat\tau$ and not to $\lambda$, we have the order-of-magnitude
relations:
\begin{myalign}
\label{neareq}
A(\lambda;\tau)\approx f_A(1;\hat\tau)~~~.
\end{myalign}
If we start from $n_A>n_I$ observables, we can eliminate the $n_I$ parameters $\hat\tau$ from
eq. (\ref{neareq}) and obtain $n_A-n_I$ order-of-magnitude relations between observables:
\begin{myalign}
\label{oom}
R_\alpha[A]\approx 0~~~~~~(\alpha=1,n_A-n_I)~~~.
\end{myalign}
The set of exponents $\alpha^A_i$ characterizing the scaling properties of eq. (\ref{scaling}) represents universal properties of the system. The degree of universality is provided by the number $n_A-n_I$ of order-of-magnitude relations in eq. (\ref{oom}).

Universality should not be confused with predictability. Predictability requires 
the number of input parameters to be smaller than the number of observables
and output quantities
matching a given experimental precision. In general, neither the scaling relations of eq. (\ref{scaling}), nor the order-of-magnitude relations
in eq. (\ref{oom}) allow predictions to the required level of accuracy. Hence a model can possess a high degree of
universality and, at the same time, a low level of predictability. This is the case of FN type of models, where
$n_I=1$ and we have the greatest possible number of relations, $n_A-1$. Nevertheless, observables
can only be predicted up to a large number of unknown input parameters. Universality is meant to capture the 
essentials of the symmetry-breaking pattern suggested by the data, not its detailed implementation.

\vskip 0.5 cm

\section{Square root of an hermitian positive definite matrix $K$}
\label{Z}
Given an hermitian positive definite matrix $K$, there is a unique positive definite square root $K^{1/2}$, that can be represented as
\begin{myalign}
\label{wiki}
K^{1/2}=||K||^{1/2}\left[\mathbb{1}-\sum_{n=1}^\infty \left\vert \left(\begin{array}{c}1/2\\n\end{array}\right)\right|\left(
\mathbb{1}-\dd\frac{K}{||K||}\right)^n\right]~~~.
\end{myalign}
Here $||K||$ is the norm of the matrix $K$ and $\left(\begin{array}{c}1/2\\n\end{array}\right)$ is the generalized binomial:
\begin{myalign}
\left(\begin{array}{c}1/2\\n\end{array}\right)=\dd\frac{1/2\times\cdot\cdot\cdot\times (1/2-n+1)}{n!}~~~.
\end{myalign}
When $K$ undergoes the transformation:
\begin{myalign}
K\to \Omega~ K~ \Omega^\dagger~~~,
\end{myalign}
$\Omega$ being a unitary matrix, from eq. (\ref{wiki}) we see that its square root $K^{1/2}$ transforms in the same way:
\begin{myalign}
K^{1/2}\to \Omega~ K^{1/2}~ \Omega^\dagger~~~.
\end{myalign}
\newpage
\section{Irreducible representations of $SL(2,\mathbb{Z}_N)$}
\label{Ehol}
The models discussed in Section \ref{modularsec} assign lepton doublets to irreducible 3-dimensional representations 
of the finite modular groups $SL(2,\mathbb{Z}_N)$. The independent such representations can all be inferred from
those of the levels $N=3,4,5,7,8,16$. Here we follow the discussion of ref~\cite{Eholzer:1994th} (see Appendix A). To build the representations for a generic level $N$, we exploit the prime decomposition:
\begin{myalign}
N=\dd\prod_p p^{\lambda^p}~~~,
\end{myalign}
and the fact that the group $SL(2,\mathbb{Z}_{N})$ factorizes as:
\begin{myalign}
\label{fafa}
SL(2,\mathbb{Z}_{N})=\dd\prod_p SL(2,\mathbb{Z}_{p^{\lambda^p}})
\end{myalign}
The three-dimensional representations of these product groups are constructed by using the three-dimensional representations of one of the groups and one-dimensional representations of all the others. 
Thus we only need to discuss one and three-dimensional representations of groups of the type $SL(2,\mathbb{Z}_{p^{\lambda^p}})$, with $p$ prime. Moreover, if the level $N$ is of the type $N=p^\lambda$, where $p$ is a prime and $\lambda>1$ an integer, all representations of the groups $SL(2,\mathbb{Z}_{p^{\bar\lambda}})$, with $1\le {\bar\lambda}< \lambda$, are also representations of $SL(2,\mathbb{Z}_{p^\lambda})$. 
The independent one-dimensional representations for $N=p^\lambda$ are listed in Table 7.
\noindent
\begin{align}
\nonumber
\begin{array}{lclclcl}
\hline
 \mathrm{level~ N} &&\rho(S) &&\rho(T)&&\rho(ST)\\
 \hline
 2&&-1&&-1&&+1\\
 \hline
 3&&+1&&\omega&&\omega\\
 &&+1&&\omega^2&&\omega^2\\
\hline
 4&&+i&&-i&&+1\\
 &&-i&&+i&&+1\\
\hline
\end{array}
\end{align}
\vskip 0.2 cm
\noindent
{{\bf Table 7} 
Independent one-dimensional representations for $N=p^\lambda$, where $p$ is a prime, $\lambda$ an integer
and $\omega=e^{i 2\pi/3}$. The trivial representation $(\rho(S),\rho(T))=(+1,+1)$ is common to all levels.}
\vskip 0.5 cm
\noindent
Table 7 should be read as follows. The trivial representation $(\rho(S),\rho(T))=(+1,+1)$ is common to all levels.
In addition, level $N=2$ admits the representation listed in the table, and level $N=4$ admits the representations
corresponding to the rows $N=2,4$. For higher levels, $N=2^\lambda$, $\lambda>2$, the representations are those
of level $N=4$. Level $N=3$ admits only the representations listed in the table and the trivial one.
For higher levels, $N=3^\lambda$, $\lambda>1$, the representations are those of level $N=3$.
Finally, when $N=p^\lambda$, $p>3$ and $\lambda\ge 1$, the only allowed representation is the trivial one.
In all cases $\rho(S)$ is a power of $i$ and $\rho(ST)$ is a power of $\omega$, which follows from the properties
$S^4=(ST)^3=1$.

Table 8 collects all independent three-dimensional irreducible representations for $N=p^\lambda$.
To correctly count their number, we should take into account that those corresponding to the levels
$N=4,8,16$ are obtained by multiplying the elements explicitly listed in table 8 by each of the four 
one-dimensional representations at level $N=4$. We have $1,2,2$ representations at the level
$N=3,5,7$, respectively. At levels $N=p^\lambda$, $p=3,5,7$ and $\lambda>1$, the representations are those of
$N=3,5,7$, respectively. At level $N=4$ we have 4 representations. At level $N=8$ we have those of
level $N=4$ plus 8 new ones. At level $N=16$ we have those of
levels $N=4,8$ plus 16 new ones. At levels $N=2^\lambda$, $\lambda>4$, the representations
are those of $N=16$.
Finally, at levels $N=p^\lambda$, $p>7$, there are no irreducible three-dimensional representations.
\begin{smalign}
\nonumber
\begin{array}{clccl}
\hline
 \mathrm{level} & \rho(S) &&& \log(\rho(T))/(2\pi i)\\
 \hline
 3 & \frac13\begin{pmatrix} -1 &2 &2 \\ 2 &-1 &2 \\ 2 &2 &-1 \end{pmatrix}
 & & &\diag( \frac13, \frac23, 0)\\
\hline
5 &
\frac2{\sqrt{5}}\begin{pmatrix}
       \frac12 &\frac1{\sqrt2}   &\frac1{\sqrt2} \\
    \frac1{\sqrt2} &-s_1 &s_2 \\
    \frac1{\sqrt2} & s_2 &-s_1
    \end{pmatrix}&s_k=\cos(k \pi/5)
& &\diag(0,\frac15, \frac45 )\\
\hline
5 &
\frac2{\sqrt{5}}\begin{pmatrix}
       -\frac12 &-\frac1{\sqrt2}   &-\frac1{\sqrt2} \\
    -\frac1{\sqrt2} &-s_2  &s_1 \\
    -\frac1{\sqrt2} & s_1  &-s_2
    \end{pmatrix}&s_k=\cos(k \pi/5)
&&\diag( 0, \frac25, \frac35 )\\
\hline
7 &
\frac2{\sqrt{7}} \begin{pmatrix}
    s_1 & s_2 & s_3 \\
   s_2 & -s_3 & s_1 \\
   s_3 &  s_1 & -s_2
   \end{pmatrix}&s_k=\sin(k \pi/7)
&&\diag( \frac27, \frac17, \frac47 )~~~~~\mathrm{or}~~~~~\diag( \frac57, \frac67, \frac37 )\\
\hline
4 &
-\frac{i}2\begin{pmatrix} 0    &\sqrt{2} &\sqrt{2} \\
              \sqrt{2} &-1    &1    \\
              \sqrt{2} &1     &-1
    \end{pmatrix}
&&&\diag(\frac14, \frac24, 0 )\\
\hline
8&
-\frac{i}2\begin{pmatrix} 0    &\sqrt{2} &\sqrt{2} \\
              \sqrt{2} &1    &-1    \\
              \sqrt{2} &-1    &1
    \end{pmatrix}
&&&\diag( \frac48, \frac58, \frac18 )\\
\hline
8&
\frac{i}2\begin{pmatrix} 0    &\sqrt{2} &\sqrt{2} \\
              \sqrt{2} &1    &-1    \\
              \sqrt{2} &-1    &1
    \end{pmatrix}
&&&\diag( \frac48, \frac78,\frac38 )\\
\hline
16 &
-\frac{i}2\begin{pmatrix} 0    &\sqrt{2} &\sqrt{2} \\
             \sqrt{2} &1    &-1    \\
             \sqrt{2} &-1    &1
    \end{pmatrix}
 &&&\diag(\frac{10}{16}, \frac{1}{16}, \frac{9}{16} )~~~~~\mathrm{or}~~~~~\diag( \frac{2}{16}, \frac{5}{16}, \frac{26}{16} )\\
 \hline
16&
-\frac{i}2\begin{pmatrix} 0    &\sqrt{2} &\sqrt{2} \\
             \sqrt{2} &-1    &1    \\
             \sqrt{2} &1    &-1
    \end{pmatrix}
 &&& \diag( \frac{14}{16}, \frac{3}{16}, \frac{11}{16} )~~~~~\mathrm{or}~~~~~\diag( \frac{6}{16}, \frac{15}{16}, \frac{7}{16} )\\
\hline
\end{array}
\end{smalign}
\vskip 0.2 cm
\noindent
{{\bf Table 8} Three-dimensional irreducible representations of $SL(2,\mathbb{Z}_N)$. To obtain the full set of 
independent representations at level $N=4,8,16$, we should multiply each of those listed here by the four one-dimensional representations given in table 7. }
\vskip 0.5 cm
\noindent
From the two previous tables, a straightforward computation shows that the eigenvalues of the three-dimensional representations $\rho(S)$ and $\rho(ST)$ for the levels $N=3,4,5,7,8,16$ are those listed in table 9.
By taking into account also the contribution of one-dimensional representations, we finally get
the result of eq. (\ref{Omega}), which applies to any level $N$:
\begin{myalign}
\Omega(S)=i^{k}{\tt diag}(1,-1,-1)~~~~~(k~{\rm integer})~~~,~~~~~~~~~~~\Omega(ST)={\tt diag}(1,\omega,\omega^2)~~~
\end{myalign} 
\begin{align}
\nonumber
\begin{array}{ccc}
\hline
 \mathrm{level} & {\mathrm Eig}[\rho(S)] & {\mathrm Eig}[\rho(ST))]\\
 \hline
 3,5,7 & 
(+1,-1,-1)&(+1,\omega,\omega^2)\\
\hline
4,8,16 &
\pm(+1,-1,-1)&(+1,\omega,\omega^2)\\
&\pm i(+1,-1,-1)&(+1,\omega,\omega^2)\\
\hline
\end{array}
\end{align}
\vskip 0.2 cm
\noindent
{{\bf Table 9} Eigenvalues of the three-dimensional representations $\rho(S)$ and $\rho(ST)$.}
\vskip 0.5 cm
\noindent
\newpage
\section{Diagonalization of mass matrices}
\label{diagon}
In this Appendix we provide a detailed derivation of the results presented in Section~\ref{lmm}.
\subsection{Case $\tau\approx i$}
\subsubsection{Charged lepton mass matrix}
\begin{myalign}
m_e^\dagger m_e=m_{0e}^2~Y(u,\bar u)~~~.
\end{myalign}
\begin{myalign}
\label{caseb3}
Y(u,\bar u)=
\left(
\begin{array}{ccc}
y^0_{11} &y_{12} x&y_{13} x\\
y_{12}^* x&y^0_{22}&y^0_{23} \\
y_{13}^* x&y^0_{23}&y^0_{33}\\
\end{array}
\right)+...~~~~~~~~~~~~~~~~~~~~~~~~~~~~~~~
\end{myalign} 
\begin{myalign} 
\begin{array}{ll} 
y^{10}_{ij}u+y^{01}_{ij}\bar u=(y^{10}_{ij}e^{\dd i \theta}+y^{01}_{ij}e^{\dd -i \theta})x\equiv y_{ij} x&~~~.
\end{array}
\end{myalign}
If the theory is CP invariant, the case we will discuss here, the parameters $y^0_{ij}$, $y^{10}_{ij}$ and $y^{01}_{ij}$ are all real. We define:
\begin{myalign}  
\begin{array}{ll} 
U_{e1}=
\left(
\begin{array}{ccc}
1&0&0\\
0&c&-s\\
0&s&c
\end{array}
\right)&~~~~~~~~~~~~~\dd\frac{2cs}{c^2-s^2}=\dd\frac{2 y_{23}^0}{y_{22}^0-y_{33}^0}\\ 
&\\
 U_{e2}=
\left(
\begin{array}{ccc}
1&y_a x & y_b x\\
-y_a^* x&1&0\\
-y_b^* x&0&1
\end{array}
\right)&~~~~~~~~
\begin{array}{l}
y_a=\dd\frac{c y_{12}+s y_{13}}{Y_2-Y_1}\\
y_b=\dd\frac{-s y_{12}+c y_{13}}{Y_3-Y_1}
\end{array}
\end{array}~~~,
\end{myalign} 
\begin{myalign}
U_e=U_{e1} U_{e2}=
\left(
\begin{array}{ccc}
1&y_a x & y_b x\\
-(cy_a^*-s y_b^*) x&c&-s\\
-(sy_a^*+cy_b^*) x&s&c
\end{array}
\right)~~~.
\end{myalign}
To first order in $x$, we find:
\begin{myalign}
U_e^\dagger Y(u,\bar u) U_e={\tt diag}(Y_1,Y_2,Y_3)~~~,
\end{myalign}
where the eigenvalues are 
\begin{myalign}
\begin{array}{l}
Y_1=y_{11}^0\\
Y_{2,3}=\dd\frac{1}{2}(y_{22}^0+y_{33}^0)\pm\frac{1}{2}\sqrt{(y_{22}^0-y_{33}^0)^2+4 (y_{23}^0)^2}~~~.
\end{array}
\end{myalign}
The contribution from the charged lepton sector to the lepton mixing is $U_e P$, where $P$ is a permutation matrix
accounting for the ordering of the mass eigenstates.

\subsubsection{Neutrino mass matrix}
\noindent
{\large\bf $k_S$ even}
\vskip 0.5cm

\noindent
The neutrino mass matrix $m_\nu=m_{0\nu}X(u,\bar u)$ reads:
\begin{myalign}
\label{caseax}
X(u,\bar u)=
\left(
\begin{array}{ccc}
x^0_{11}&x_{12} x&x_{13} x\\
\cdot&x^0_{22}&x^0_{23}\\
\cdot&\cdot&x^0_{33}\\
\end{array}
\right)+...
\end{myalign}
\begin{myalign} 
\begin{array}{ll} 
u=x~ e^{\dd i \theta}~~~,~~~~~\bar u=x~ e^{\dd- i \theta}&(x>0,2 \pi> \theta\ge 0)\\
x^{10}_{ij}u+x^{01}_{ij}\bar u=(x^{10}_{ij}e^{\dd i \theta}+x^{01}_{ij}e^{\dd -i \theta})x\equiv x_{ij} x&~~~.
\end{array}
\end{myalign}
We move to the basis where the charged lepton mass matrix is diagonal, by transforming the whole lepton doublet
through the unitary matrix $U_e$ (the effect of the permutation $P$ is discussed later).
In the new basis, $X(u,\bar u)$ does not change form. The only effect is a redefinition of parameters
$x_{ij}$ and $x_{11}^0,x_{22}^0,x_{23}^0,x_{33}^0$,
which we continue to denote with the same symbol. We define:
\begin{myalign}  
\begin{array}{ll} 
U_{\nu1}=
\left(
\begin{array}{ccc}
1&0&0\\
0&c&-s\\
0&s&c
\end{array}
\right)&~~~~~~~~~~~~~\dd\frac{2cs}{c^2-s^2}=\dd\frac{2 x_{23}^0}{x_{22}^0-x_{33}^0}\\ 
&\\
 U_{\nu 2}=
\left(
\begin{array}{ccc}
1&x_a x & x_b x\\
-x_a^* x&1&0\\
-x_b^* x&0&1
\end{array}
\right)&~~~~~~~~
\begin{array}{l}
x_a=\dd\frac{c(X_1 x_{12}+X_2 x_{12}^*)+s(X_1 x_{13}+X_2 x_{13}^*)}{X_2^2-X_1^2}\\[10 pt]
x_b=\dd\frac{-s(X_1 x_{12}+X_3 x_{12}^*)+c(X_1 x_{13}+X_3 x_{13}^*)}{X_3^2-X_1^2}
\end{array}
\end{array}~~~,
\end{myalign} 
\begin{myalign}
U_\nu=U_{\nu1} U_{\nu2}=
\left(
\begin{array}{ccc}
1&x_a x & x_b x\\
-(cx_a^*-s x_b^*) x&c&-s\\
-(sx_a^*+cx_b^*) x&s&c
\end{array}
\right)~~~.
\end{myalign}
To first order in $x$, we find:
\begin{myalign}
U_\nu^T X(u,\bar u) U_\nu={\tt diag}(X_1,X_2,X_3)~~~,
\end{myalign}
where the eigenvalues read:
\begin{myalign}
\begin{array}{l}
X_1=x_{11}^0\\
X_{2,3}=\dd\frac{1}{2}(x_{22}^0+x_{33}^0)\pm\frac{1}{2}\sqrt{(x_{22}^0-x_{33}^0)^2+4 (x_{23}^0)^2}~~~.
\end{array}
\end{myalign}
The lepton mixing matrix is $P_e U_\nu P_\nu$, where $P_{e,\nu}$ are permutation matrices
accounting for the ordering of the mass eigenstates in the charged lepton and neutrino sectors.

By adapting the parameters $x_{11}^0$, $x_{22}^0$, $x_{33}^0$, $x_{23}^0$, both ordering of neutrino masses can be accommodated. Barring cancellations, the ratio $\Delta m^2_{sol}/\Delta m^2_{atm}$ is expected to be of order one, while experimentally is close to 0.03. When $P_{e,\nu}=\mathbb{1}$, to first order in $x$ we find:
\begin{myalign}
\sin\theta_{12}=|x_a| x~~~,~~~~~~~\sin\theta_{13}=|x_b| x~~~,~~~~~~~\sin\theta_{23}=s~~~. 
\end{myalign}
To match the experimental data we would need
$|x_b/x_a|\approx 0.27$. This suppression might mostly originate from the approximate scaling $|x_a/x_b|\propto (X_3^2-X_1^2)/(X_2^2-X_1^2)$. Once the parameters $x_{11}^0$, $x_{22}^0$, $x_{33}^0$, $x_{23}^0$ have been adjusted to match $\Delta m^2_{sol}/\Delta m^2_{atm}$, they naturally enhance $|x_a/x_b|$.
To keep $\sin\theta_{13}={\cal O}(x)$, we can allow $P_e=P_{23}$, the permutation between second and third rows.
This produces the exchange $\sin\theta_{23}\leftrightarrow \cos\theta_{23}$. Similarly, taking $P_\nu=P_{12}$
causes the exchange $\sin\theta_{12}\leftrightarrow \cos\theta_{12}$. Now data requires 
$|x_b/x_a|<0.2$. Taking $P_\nu=P_{23}$
causes the exchange $\sin\theta_{12}\leftrightarrow \sin\theta_{13}$. Another set of permutations leaving
$\sin\theta_{13}={\cal O}(x)$ is $P_\nu=P_{13}$ combined with $P_e=P_{13}$ (or $P_e=P_{12}$).
In this case we end up with $\sin\theta_{12}=s$ and $\sin\theta_{23}={\cal O}(x)$ (or $\cos\theta_{23}={\cal O}(x)$),
with the result that the atmospheric angle is very far from the maximal one.
\vskip 0.5cm
\noindent
{\large\bf $k_S$ odd}
\vskip 0.5cm

\noindent
The two cases we are interested in are: $m_\nu=m_{0\nu}X(u,\bar u)$ and $m_\nu^{-1}=m_{0\nu}^{-1}X(u,\bar u)$
where:
 \begin{myalign}
\label{caseb2}
X(u,\bar u)=
\left(
\begin{array}{ccc}
x_{11} x&x^0_{12}&x^0_{13}\\
\cdot&x_{22} x&x_{23} x\\
\cdot&\cdot&x_{33} x\\
\end{array}
\right)+...~~~~~~~~~~~~~~~~~~~~~~~~~~~~~~~
\end{myalign} 

\begin{myalign} 
\begin{array}{ll} 
u=x~ e^{\dd i \theta}~~~,~~~~~\bar u=x~ e^{\dd- i \theta}&(x>0,2 \pi> \theta\ge 0)\\
x^{10}_{ij}u+x^{01}_{ij}\bar u=(x^{10}_{ij}e^{\dd i \theta}+x^{01}_{ij}e^{\dd -i \theta})x\equiv x_{ij} x&~~~.
\end{array}
\end{myalign}
We move to the basis where the charged lepton mass matrix is diagonal, by transforming the whole lepton doublet
through the unitary matrix $U_e$ (the effect of the permutation $P$ is discussed later).
In the new basis, $X(u,\bar u)$ does not change form. The only effect is a redefinition of parameters
$x_{ij}$ and $x_{12(3)}^0$,
which we continue to denote with the same symbol. We define three unitary matrices
\begin{myalign}  
\begin{array}{ll} 
U_{\nu1}= \dd\frac{1}{\sqrt{2}h}
\left(
\begin{array}{ccc}
+h&-h&0\\
x^0_{12}&x^0_{12}&-\sqrt{2}x^0_{13}\\
x^0_{13}&x^0_{13}&\sqrt{2}x^0_{12}
\end{array}
\right)&h=\sqrt{(x^0_{12})^2+(x^0_{13})^2}\\ 
&\\
 U_{\nu2}=
\left(
\begin{array}{ccc}
1&0&-\frac{n}{h}x\\
0&1&+\frac{n}{h}x\\
+\frac{\bar n}{h}x&-\frac{\bar n}{h}x&1
\end{array}
\right)&\\
&\\
U_{\nu3}=
\left(
\begin{array}{ccc}
\alpha&-\bar \beta&0\\
\beta&\bar\alpha&0\\
0&0&1
\end{array}
\right)&\begin{array}{l}
\alpha=c_\nu\\
\beta=s_\nu e^{\dd -i \varphi_\nu}~~~~~c_\nu^2+s_\nu^2=1~~~,
\end{array}
\end{array}
\end{myalign} 
where $c_\nu$, $s_\nu$ and $\varphi_\nu$ are the solution of:
\begin{myalign}  
-2 c_\nu s_\nu h \cos \varphi_\nu-2 i c_\nu s_\nu \sin\varphi_\nu k x+(c_\nu^2- s_\nu^2) l x=0~~~.
\end{myalign}
If $l$ is real a solution is:
\begin{myalign}  
\dd\frac{2 c_\nu s_\nu}{c_\nu^2- s_\nu^2}=\dd\frac{l}{h} x~~~~~~~~~~~~~~~\varphi_\nu=0~~~.
\end{myalign}
This solution is relevant when $u$ is real.
When $l$ is complex a solution is:
\begin{myalign}  
\sin\varphi_\nu=-1+{\cal O}(x^2)~~~~~~~~~\cos\varphi_\nu=-\dd\frac{i}{h}\left(\dd\frac{l\bar k+\bar l k}{l-\bar l}\right)x
~~~~~~~~~~\dd\frac{2 c_\nu s_\nu}{c_\nu^2- s_\nu^2}=i\dd\frac{l-\bar l}{k+\bar k}+{\cal O}(x^2)
\end{myalign}
This solution requires $i(l-\bar l)$ to be of order 1. The parameters $k$, $l$, $n$ and $q$ read
\begin{myalign}  
\begin{array}{l}
k=+\dd\frac{x_{11}}{2}+\dd\frac{1}{2h^2}
\left((x^0_{12})^2 x_{22}+(x^0_{13})^2 x_{33}+2 x^0_{12} x^0_{13} x_{23}\right)\\
l=-\dd\frac{x_{11}}{2}+\dd\frac{1}{2h^2}
\left((x^0_{12})^2 x_{22}+(x^0_{13})^2 x_{33}+2 x^0_{12} x^0_{13} x_{23}\right)\\
n=\dd\frac{1}{2h^2}\left[\left(x^0_{12})^2 - (x^0_{13})^2 \right) x_{23} + x^0_{12} x^0_{13} \left(-x_{22} + 
x_{33}\right)\right]\\
q=\dd\frac{1}{h^2}\left( (x^0_{13})^2 x_{22}+(x^0_{12})^2 x_{33}-2 x^0_{12} x^0_{13} x_{23}\right)~~~.
\end{array}
\end{myalign}
We define a matrix $K_\nu$ to remove phases from the eigenvalues of $X(u,\bar u)$
\begin{myalign}
K_\nu=
\left(
\begin{array}{ccc}
e^{\dd -i \varphi_1/2}&0&0\\
0&e^{\dd -i \varphi_2/2}&0\\
0&0&e^{\dd -i \varphi_3/2}
\end{array}
\right)~~~.
\end{myalign}
When $l\ne \bar l$:
\begin{myalign}  
\begin{array}{l}
\varphi_1=\left[\dd\frac{(l+\bar l)}{2}\sqrt{(k+\bar k)^2-(l-\bar l)^2}-(l\bar k+\bar l k)
\right]\dd\frac{x}{i h(l-\bar l)}\\
\varphi_2=\pi-\left[\dd\frac{(l+\bar l)}{2}\sqrt{(k+\bar k)^2-(l-\bar l)^2}-(l\bar k+\bar l k)
\right]\dd\frac{x}{i h(l-\bar l)}\\
\varphi_3=\arg~q~~~.
\end{array}
\end{myalign}
Finally, we combine the unitary matrices $U_{\nu1}$, $U_{\nu2}$, $U_{\nu3}$ and $K_\nu$ into a mixing matrix $U_\nu$
\begin{myalign}  
\label{unu1}
U_\nu=U_{\nu1}U_{\nu2}U_{\nu3}K_\nu~~~,
\end{myalign}
that diagonalizes $X(u,\bar u)$
\begin{myalign}  
U_\nu^T X(u,\bar u) U_\nu={\tt diag}(X_1,X_2,X_3)~~~.
\end{myalign}
The eigenvalues are given by
\begin{myalign}  
\begin{array}{l}
X_1=h\left(1+\dd\frac{\sqrt{(k+\bar k)^2-(l-\bar l)^2}}{2h}x\right)\\
X_2=h\left(1-\dd\frac{\sqrt{(k+\bar k)^2-(l-\bar l)^2}}{2h}x\right)\\
X_3=\vert q\vert x
\end{array}~~~.
\end{myalign}
\subsection*{Normal Ordering}
Normal ordering occurs when $[m_\nu(u,\bar u)]^{-1}=m_{0\nu}^{-1} X(u,\bar u)$. Neutrino masses read:
\begin{myalign}  
\begin{array}{l}
m_1= \dd\frac{m_{0\nu}}{X_1}=\dd\frac{m_{0\nu}}{h}\left(1-\dd\frac{\sqrt{(k+\bar k)^2-(l-\bar l)^2}}{2h}x\right)\\
m_2= \dd\frac{m_{0\nu}}{X_2}=\dd\frac{m_{0\nu}}{h}\left(1+\dd\frac{\sqrt{(k+\bar k)^2-(l-\bar l)^2}}{2h}x\right)\\
m_3= \dd\frac{m_{0\nu}}{X_3}=\dd\frac{m_{0\nu}}{|q| x}
\end{array}~~~.
\end{myalign}
Other combinations of interest are
\begin{myalign}  
\begin{array}{l}
\dd\frac{m_1+m_2}{2}=\dd\frac{m_{0\nu}}{h}\\[8 pt]
\dd\frac{m_1+m_2}{2 m_3}=\frac{|q|}{h}x\\[8 pt]
2\dd\frac{m_2-m_1}{m_2+m_1}=\frac{\sqrt{(k+\bar k)^2-(l-\bar l)^2}}{h}x
\end{array}~~~,
\end{myalign}
and
\begin{myalign}  
\begin{array}{l}
\Delta m^2_{sol}=m_2^2-m_1^2= 2 m_{0\nu}^2 \dd\frac{\sqrt{(k+\bar k)^2-(l-\bar l)^2}}{h^3}x\\[8 pt]
\Delta m^2_{atm}=m_3^2-(m_2^2+m_1^2)/2= \dd\frac{m_{0\nu}^2}{|q|^2 x^2}\left(1+{\cal O}(x^2)\right)\\[8 pt]
r=\dd\frac{\Delta m^2_{sol}}{\Delta m^2_{atm}}=2 \dd\frac{|q|^2\sqrt{(k+\bar k)^2-(l-\bar l)^2}}{h^3}x^3
\end{array}~~~.
\end{myalign}
The mixing matrix $U_{PMNS}$ is $U_\nu^*$ of eq. (\ref{unu1}):
\begin{myalign}  
U_{PMNS}=U_\nu^*~~~.
\end{myalign}
The mixing angles are:
\begin{myalign}  
\begin{array}{l}
\sin^2\theta_{12}=\frac{1}{2}\left(1+\dd\frac{l\bar k+\bar l k}{h\sqrt{(k+\bar k)^2-(l-\bar l)^2}}x\right)\\[8 pt]
\sin^2\theta_{13}=2 \dd\frac{|n|^2}{h^2} x^2\\[8 pt]
\sin^2\theta_{23}=\dd\frac{(x^0_{13})^2}{h^2}(1+{\cal O}(x))
\end{array}~~~.
\end{myalign}
The CP-violating phases are:
\begin{myalign}  
\begin{array}{l}
\delta_{CP}=\pi-\arg\left[\dd\frac{(c_\nu-is_\nu)^2 x^0_{12} x^0_{13}}{n}\right]+{\cal O}(x^2)\\
\alpha_{21}=\pi+{\cal O}(x)\\
\alpha_{31}=\arg(q)-\arg\left[(c_\nu-is_\nu)^2\right]+{\cal O}(x)
\end{array}~~~.
\end{myalign}
The quantity $m_{\beta\beta}$ is given by:
\begin{myalign}  
m_{\beta\beta}=m^0_\nu\dd\frac{\left\vert x_{23}^2-x_{22}x_{33}\right\vert}{h^2|q|} x~~~.
\end{myalign}
The permutation matrix $P$ from the charged lepton sector changes $U_\nu^*$ into $PU_\nu^*$. Since $(U_\nu^*)_{e3}={\cal O}(x)$, the only acceptable permutation is the one between the second and the third rows of $U_\nu^*$: $P=P_{23}$. All observable
remain unchanged but $\sin^2\theta_{23}\to 1-\sin^2\theta_{23}$ and $\delta_{CP}\to \pi+\delta_{CP}$ $\mod(2\pi)$.
\subsection*{Inverted Ordering}
Inverted ordering occurs when $m_\nu(u,\bar u)=m_{0\nu} X(u,\bar u)$. Neutrino masses read:
\begin{myalign}  
\begin{array}{l}
m_1= m_{0\nu}X_2=m_{0\nu}h\left(1-\dd\frac{\sqrt{(k+\bar k)^2-(l-\bar l)^2}}{2h}x\right)\\
m_2=m_{0\nu}X_1=m_{0\nu}h\left(1+\dd\frac{\sqrt{(k+\bar k)^2-(l-\bar l)^2}}{2h}x\right)\\
m_3= m_{0\nu}X_3=m_{0\nu} \vert q\vert x
\end{array}~~~.
\end{myalign}
Other relevant combinations are
\begin{myalign}  
\begin{array}{l}
\dd\frac{m_1+m_2}{2}=m_{0\nu} h\\[8 pt]
\dd\frac{m_1+m_2}{2 m_3}=\frac{h}{|q|x}\\[8 pt]
2\dd\frac{m_2-m_1}{m_2+m_1}=\frac{\sqrt{(k+\bar k)^2-(l-\bar l)^2}}{h}x
\end{array}~~~,
\end{myalign}
and
\begin{myalign}  
\begin{array}{l}
\Delta m^2_{sol}=m_2^2-m_1^2= 2 m_{0\nu}^2 h\sqrt{(k+\bar k)^2-(l-\bar l)^2}x\\[8 pt]
\Delta m^2_{atm}=-m_3^2+(m_2^2+m_1^2)/2= m_{0\nu}^2 h^2\left(1+{\cal O}(x^2)\right)\\[8 pt]
r=\dd\frac{\Delta m^2_{sol}}{\Delta m^2_{atm}}=2 \dd\frac{\sqrt{(k+\bar k)^2-(l-\bar l)^2}}{h}x
\end{array}~~~.
\end{myalign}
The mixing matrix $U_{PMNS}$ is the one of eq. (\ref{unu1}), after permuting the first and second columns:
\begin{myalign}  
U_{PMNS}=U_\nu P_{12}~~~.
\end{myalign}
The mixing angles are:
\begin{myalign}  
\begin{array}{l}
\sin^2\theta_{12}=\frac{1}{2}\left(1-\dd\frac{l\bar k+\bar l k}{h\sqrt{(k+\bar k)^2-(l-\bar l)^2}}x\right)\\[8 pt]
\sin^2\theta_{13}=2 \dd\frac{|n|^2}{h^2} x^2\\[8 pt]
\sin^2\theta_{23}=\dd\frac{(x^0_{13})^2}{h^2}(1+{\cal O}(x))
\end{array}~~~.
\end{myalign}
The CP-violating phases are:
\begin{myalign}  
\begin{array}{l}
\delta_{CP}=\arg\left[\dd\frac{(c_\nu-is_\nu)^2 x^0_{12} x^0_{13}}{n}\right]+{\cal O}(x^2)\\
\alpha_{21}=\pi+{\cal O}(x)\\
\alpha_{31}=\pi-\arg(q)+\arg\left[(c_\nu-is_\nu)^2\right]+{\cal O}(x)
\end{array}~~~.
\end{myalign}
The quantity $m_{\beta\beta}$ is given by:
\begin{myalign}  
m_{\beta\beta}=m^0_\nu |x_{11}| x~~~.
\end{myalign}
The permutation matrix $P$ from the charged lepton sector changes $U_\nu^*$ into $PU_\nu^*$. Since $(U_\nu^*)_{e3}={\cal O}(x)$, the only acceptable permutation is the one between the second and the third rows of $U_\nu^*$: $P=P_{23}$. All observable
remain unchanged but $\sin^2\theta_{23}\to 1-\sin^2\theta_{23}$ and $\delta_{CP}\to \pi+\delta_{CP}$ $\mod(2\pi)$.
\subsection{Case $\tau\approx \omega$}
We start from the charged lepton mass matrix:
\begin{myalign}
m_e^\dagger m_e=m_{0e}^2~Y(u,\bar u)~~~.
\end{myalign}
\begin{myalign}
\label{caseb1}
Y(u,\bar u)=
\left(
\begin{array}{ccc}
{y}^0_{11}&{y}^{10}_{12}u&{y}^{01}_{13}\bar u\\
{y}^{10*}_{12}\bar u&{y}^0_{22}&{y}^{10}_{23} u\\
{y}^{01*}_{13}u&{y}^{10*}_{23} \bar u&{y}^0_{33}\\
\end{array}
\right)+...~~~~~~~~~~~~~~~~~~~~~~~~~~~~~~~
\end{myalign} 
If the theory is CP invariant, the case we will discuss here, the parameters $y^0_{ij}$, $y^{10}_{ij}$ and $y^{01}_{ij}$ are all real.
\begin{myalign}
 U_e=
\left(
\begin{array}{ccc}
1&\frac{{y}^{10}_{12}u}{y^0_{22}-y^0_{11}} & \frac{{y}^{01}_{13}\bar u}{y^0_{33}-y^0_{11}}\\
-\frac{{y}^{10}_{12}\bar u}{y^0_{22}-y^0_{11}}&1&\frac{{y}^{10}_{23}u}{y^0_{33}-y^0_{22}}\\
-\frac{{y}^{01}_{13}u}{y^0_{33}-y^0_{11}}&-\frac{{y}^{10}_{23}\bar u}{y^0_{33}-y^0_{22}}&1
\end{array}
\right)~~~.
\end{myalign} 
Eigenvalues, up to terms quadratic in $u$ and/or $\bar u$:
\begin{myalign}
Y_1={y}^0_{11}~~~,~~~~~~Y_2={y}^0_{22}~~~,~~~~~~Y_3={y}^0_{33}~~~.
\end{myalign}
The contribution from the charged lepton sector to the lepton mixing is $U_e P$, where $P$ is a permutation matrix
accounting for the ordering of the mass eigenstates.

\vskip 0.3cm
We now discuss the neutrino mass matrix. We start from the case $m_\nu=m_{0\nu}X(u,\bar u)$, where the matrix 
$X(u,\bar u)$ reads:
\begin{myalign}
X(u,\bar u)=
\left(
\begin{array}{ccc}
x^0_{11}&x^{10}_{12}u&x^{01}_{13}\bar u\\
\cdot&x^{01}_{22}\bar u&x^0_{23}\\
\cdot&\cdot&x^{10}_{33}u\\
\end{array}
\right)+...
\end{myalign} 
We move to the basis where the charged lepton mass matrix is diagonal, by transforming the whole lepton doublet
through the unitary matrix $U_e$ (the effect of the permutation $P$ is discussed later).
In the new basis, $X(u,\bar u)$ does not change form. The only effect is a redefinition of parameters
$x^{10}_{12,33}$, and $x^{01}_{13,22}$,
which we continue to denote with the same symbol. We define the three unitary matrices:
\begin{myalign}  
\begin{array}{ll} 
U_{\nu1}= \dd\frac{1}{\sqrt{2}}
\left(
\begin{array}{ccc}
\sqrt{2}&0&0\\
0&1&-1\\
0&1&1
\end{array}
\right)\\ 
&\\
 U_{\nu2}=
\left(
\begin{array}{ccc}
1&(a u + b \bar u)&(-a u + b \bar u)\\
-(b u + a \bar u)&1&0\\
-(b u -a \bar u)&0&1
\end{array}
\right)
&
\begin{array}{l}
a=-\frac{x^0_{11} x^{10}_{12} + x^{01}_{13} x^0_{23}}{\sqrt{2} ((x^0_{11})^2 - (x^0_{23})^2)}\\
b=-\frac{x^0_{11} x^{01}_{13} + x^{10}_{12} x^0_{23}}{\sqrt{2} ((x^0_{11})^2 - (x^0_{23})^2)}
\end{array}
\\
&\\
U_{\nu3}=
\left(
\begin{array}{ccc}
1&0&0\\
0&\alpha&-\bar \beta\\
0&\beta&\bar\alpha
\end{array}
\right)&\begin{array}{l}
\alpha=c_\nu\\
\beta=s_\nu e^{\dd -i \varphi_\nu}~~~~~c_\nu^2+s_\nu^2=1~~~,
\end{array}
\end{array}
\end{myalign} 
where $c_\nu$, $s_\nu$ and $\varphi_\nu$ are the solution of:
\begin{myalign}  
\begin{array}{l}
-2 c_\nu s_\nu h \cos \varphi_\nu-2 i c_\nu s_\nu \sin\varphi_\nu k+(c_\nu^2- s_\nu^2) l=0\\
h=x^0_{23}\\
k=\frac{1}{2}(x^{10}_{33}u+x^{01}_{22}\bar u)\\
l=\frac{1}{2}(x^{10}_{33}u-x^{01}_{22}\bar u)
\end{array}
\end{myalign}
If $l$ is real a solution is:
\begin{myalign}  
\dd\frac{2 c_\nu s_\nu}{c_\nu^2- s_\nu^2}=\dd\frac{l}{h}~~~~~~~~~~~~~~~\varphi_\nu=0~~~.
\end{myalign}
This solution is relevant when $u$ is real.
When $l$ is complex and $u=x e^{\dd i\theta}$ $(x>0)$, a solution is:
\begin{myalign}  
\sin\varphi_\nu=-1+{\cal O}(x^2)~~~~~~~~~\cos\varphi_\nu=-\dd\frac{(x^{10}_{33}-x^{01}_{22})}{2x^0_{23}\sin\theta}x~~~~~~~~~~\dd\frac{2 c_\nu s_\nu}{c_\nu^2- s_\nu^2}=-\tan\theta+{\cal O}(x^2)~~~.
\end{myalign}
We choose the solution $2 c_\nu s_\nu\approx-\sin\theta$, $c_\nu^2- s_\nu^2\approx\cos\theta$. We define:
\begin{myalign}
K_\nu=
\left(
\begin{array}{ccc}
e^{\dd -i \varphi_1/2}&0&0\\
0&e^{\dd -i \varphi_2/2}&0\\
0&0&e^{\dd -i \varphi_3/2}
\end{array}
\right)~~~.
\end{myalign}
When $l$ is complex:
\begin{myalign}  
\begin{array}{l}
\varphi_1=\dd\frac{1-{\tt sign}(x^0_{11})}{2}\pi\\
\varphi_2=\dd\frac{1-{\tt sign}(x^0_{23})}{2}\pi+\tan\left(\dd\frac{\theta}{2}\right)\dd\frac{x^{10}_{33}-x^{01}_{22}}{2x^0_{23}}x\\
\varphi_3=\dd\frac{1+{\tt sign}(x^0_{23})}{2}\pi-\tan\left(\dd\frac{\theta}{2}\right)\dd\frac{x^{10}_{33}-x^{01}_{22}}{2x^0_{23}}x~~~.\\
\end{array}
\end{myalign}
Up to permutations related to the mass ordering of charged leptons and neutrinos, the lepton mixing matrix reads:
\begin{myalign}  
\label{unu2}
U_\nu=U_{\nu1}U_{\nu2}U_{\nu3}K_\nu~~~.
\end{myalign}
It satisfies
\begin{myalign}  
U_\nu^T X(u,\bar u) U_\nu={\tt diag}(X_1,X_2,X_3)~~~,
\end{myalign}
where the eigenvalues are given by
\begin{myalign}  
\begin{array}{l}
X_1=x^0_{11}\\
X_2=|x^0_{23}|\left(1+\dd\frac{x^{10}_{33}+x^{01}_{22}}{2x^0_{23}}x\right)\\
X_3=|x^0_{23}|\left(1-\dd\frac{x^{10}_{33}+x^{01}_{22}}{2x^0_{23}}x\right)
\end{array}~~~.
\end{myalign}
The product $U_{\nu1}U_{\nu2}U_{\nu3}$ reads
\begin{myalign}
U_{\nu1}U_{\nu2}U_{\nu3}\approx
\left(
\begin{array}{ccc}
1&a(\alpha-\beta)u+b(\alpha+\beta)\bar u&-a(\alpha+\bar\beta)u+b(\alpha-\bar\beta)\bar u\\
-\sqrt{2} a \bar u&\frac{\alpha-\beta}{\sqrt{2}}&-\frac{\alpha+\bar\beta}{\sqrt{2}}\\
-\sqrt{2} b u&\frac{\alpha+\beta}{\sqrt{2}}&\frac{\alpha-\bar\beta}{\sqrt{2}}
\end{array}
\right)~~~.
\end{myalign}
The smallest entries are those linear in $x=|u|$. To reproduce $\sin^2\theta_{13}$, we need $x\approx 0.15$.
By exploiting the possible permutations of rows and columns, related to the lepton mass ordering, we end up with:
\begin{myalign}  
\begin{array}{lclcl}
{\rm either}&&\tan^2\theta_{12}\approx {\cal O}(x^2)&&\tan^2\theta_{23}\approx 1+{\cal O}(x)\\
{\rm or}&&\tan^2\theta_{23}\approx {\cal O}(x^2)&&\tan^2\theta_{12}\approx 1+{\cal O}(x)~~~,
\end{array}
\end{myalign}
where $\tan\theta_{ij}$ can be replaced also by $\cot\theta_{ij}$ in each singe entry above. 

\end{document}